\documentclass[
reprint,
a4paper,
australian,
amsmath,
amssymb,
aps,
prd,
twocolumn,
superscriptaddress,
preprintnumbers
]{revtex4-2}
\usepackage{braket}
\usepackage{cancel}
\usepackage{lipsum}
\usepackage{tikz, pgf}
\usetikzlibrary{positioning}
\usepackage{graphicx}
\usepackage{amssymb}
\usepackage{amsmath}
\usepackage{color,xcolor}
\definecolor{darkgreen}{rgb}{0.1,0.4,0.0}
\usepackage{hyperref}
\usepackage{times}
\usepackage{mathrsfs}
\usepackage[utf8]{inputenc}
\usepackage[capitalise]{cleveref}
\usepackage[normalem]{ulem}
\crefrangelabelformat{equation}{(#3#1#4--#5#2#6)}
\graphicspath{{./figs/}}
\newcommand{\rcite}[1]{\cite{#1}}
\newcommand{\rref}[1]{\rcite{#1}}
\newcommand{\refref}[1]{Ref.~\rcite{#1}}
\newcommand{\eref}[1]{Eq.~(\ref{#1})}
\newcommand{\sref}[1]{Sec.~\ref{#1}}
\newcommand{\tref}[1]{Table~\ref{#1}}
\newcommand{\fref}[1]{Fig.~\ref{#1}}

\newcommand{\dkt}{d^3 k\,}

\renewcommand{\bar}[1]{\mkern 1.8mu\overline{\mkern-1.8mu#1\mkern-1.8mu}\mkern 1.8mu}

\usepackage{bm} 
\newcommand{\vect}[1]{\bm{#1}}

\newcommand{\CSSM}{Special Research Centre for the Subatomic Structure
  of Matter (CSSM), Department of Physics, University of
  Adelaide, Adelaide, South Australia 5005, Australia}
\newcommand{\UCAS}{School of Physical Sciences, University of Chinese Academy of Sciences (UCAS), Beijing 100049, China}
\newcommand{\SCNT}{Southern Center for Nuclear-Science Theory (SCNT), Institute of Modern Physics, Chinese Academy of Sciences, Huizhou 516000, Guangdong Province, China}
\newcommand{\LanzhouSchool}{School of Physical Science and Technology, Lanzhou University, Lanzhou 730000, China}
\newcommand{\LanzhouHCSR}{Research Centre for Hadron and CSR Physics, Lanzhou University and Institute of Modern Physics of CAS, Lanzhou 730000, China}
\newcommand{\LanzhouTheory}{Lanzhou Center for Theoretical Physics, Key Laboratory of Theoretical Physics of Gansu Province, Key Laboratory of Quantum Theory and Applications of MoE, and Frontiers Science Center for Rare Isotopes, Lanzhou University, Lanzhou 730000, China}
\begin{document}
\preprint{ADP-23-17-T1226}
%
%
\title{Low-lying odd-parity nucleon resonances as quark-model like states}
\author{Curtis D. Abell}
\email[Corresponding author: ]{curtis.abell@adelaide.edu.au}
\affiliation{\CSSM}
\author{Derek B. Leinweber}
\affiliation{\CSSM}
\author{Zhan-Wei Liu}
\affiliation{\LanzhouSchool}
\affiliation{\LanzhouHCSR}
\affiliation{\LanzhouTheory}
\author{Anthony W. Thomas}
\affiliation{\CSSM}
\author{Jia-Jun Wu}
\affiliation{\UCAS}
\affiliation{\SCNT}
%
%
\begin{abstract}
  Recent lattice QCD results for the low-lying odd-parity excitations of the nucleon near the $N^{*}(1535)$ and $N^{*}(1650)$ resonance positions have revealed that the lattice QCD states have magnetic moments consistent with predictions from a constituent-quark-model.
  Using Hamiltonian Effective Field Theory (HEFT) to describe pion-nucleon scattering in the $I(J^{P}) = \frac{1}{2}(\frac{1}{2}^{-})$ channel, we represent these two quark-model like states as two single-particle bare basis states, dressed and mixed by meson-baryon scattering channels.
  By constraining the free parameters of the Hamiltonian with $S_{11}$ pion-nucleon scattering data, we perform the first calculation of the finite-volume spectrum using two bare-baryon basis states.
  By comparing this spectrum to contemporary lattice QCD results at three lattice volumes, we analyse the eigenvectors of the Hamiltonian to gain insight into the structure and composition of these two low-lying resonances.
  We find that an interpretation of the two low-lying nucleon resonances as quark-model like states dressed by meson-baryon interactions is consistent with both the $S_{11}$ scattering data and lattice QCD.
  We introduce a novel HEFT formalism for estimating scattering-state contaminations in lattice QCD correlation functions constructed with standard three-quark operators.
  Not only are historical lattice QCD results described with excellent accuracy, but correlation functions with large scattering-state contaminations are identified.
\end{abstract}
\maketitle
%
%
\section{Introduction}
An analysis of the nature of pion-nucleon resonances is a vital component of the quest to understand the nature of non-perturbative QCD.
The low-lying odd-parity nucleon resonances, the $N^{*}(1535)$ and $N^{*}(1650)$, are a subject of particular interest, as the $N^{*}(1535)$ sits above the first positive-parity excitation of the nucleon, the $N^{*}(1440)$ (Roper resonance), contrary to simple quark-model predictions.
There is now evidence for the Roper resonance as primarily being dynamically generated by strong $\pi N$ and $\pi\pi N$ re-scattering, with only a small bare state contribution \rref{Kiratidis:2015vpa,Liu:2016uzk,Lang:2016hnn,Wu:2017qve}.
The nature of the odd-parity nucleons however is less clear.
Both interpretations as being dynamically generated \rref{Bruns:2010sv,Bruns:2019fwi}, and as being primarily a three-quark state dressed by $\pi N$ and $\eta N$ interactions \rref{Liu:2015ktc} have been argued.

Lattice QCD offers an alternate source of insight into the nature of these resonances, providing a first-principles approach to the nuances of hadron spectroscopy.
In particular, a recent lattice QCD study \rref{Stokes:2019zdd} of the odd-parity nucleon states near these resonances found their magnetic moments resemble constituent-quark-model predictions.
As such, a consideration of the $N^{*}(1535)$ and $N^{*}(1650)$ as single-particle three-quark states dressed by meson-baryon interactions is now well-motivated.

Lattice QCD calculations are performed in a finite-volume under the evolution of Euclidean time, preventing the direct calculation of resonance properties such as the particle width, or scattering quantities such as the phase shifts and inelasticities.
L\"uscher's method \rref{Luscher:1985dn,Luscher:1986pf,Luscher:1990ux} has proven capable of bridging the finite-volume energy eigenstates of lattice QCD with infinite-volume scattering observables, however generalisations of L\"uscher's method to multiple channels \rref{He:2005ey,Lage:2009zv,Bernard:2010fp,Guo:2012hv,Hu:2016shf,Li:2012bi,Hansen:2012bj} and three particles \rref{Doring:2018xxx,Hansen:2019nir,Blanton:2019vdk} require parametrisations of the scattering observables, and present additional technical difficulties.

As an alternative to traditional implementations of L\"uscher's method, Hamiltonian Effective Field Theory (HEFT) is a non-perturbative extension of chiral effective field theory also incorporating L\"uscher's formalism.
Here, the Hamiltonian is parametrised to describe scattering interactions.
Though demonstrated to be equivalent to L\"uscher's formalism up to exponentially suppressed terms in $m_{\pi}L$ \rref{Wu:2014vma}, HEFT is readily generalisable to include multiple two-particle scattering channels, as well as quark-model like single-particle states referred to as bare states.

By constraining the Hamiltonian with infinite-volume scattering data, one can bring this information to finite-volume, where the eigenvalue equation for the Hamiltonian is solved to predict the energy eigenstates of lattice QCD.
Most importantly for developing an understanding into the nature of states formed through QCD interactions, HEFT also provides insight into the composition of these eigenstates through an analysis of the Hamiltonian's eigenvectors.

Previous studies \rref{Hall:2013qba,Hall:2014uca,Wu:2014vma,Liu:2015ktc,Liu:2016uzk,Wu:2017qve,Li:2019qvh,Liu:2020foc,Li:2021mob,Abell:2021awi,Guo:2022hud} have utilised HEFT for a variety of resonances, however these have all been limited to containing a single bare basis state in the Hamiltonian.
Only recently have two bare basis states been considered in HEFT.
While one study \rref{Yang:2021tvc} focused on exotic meson resonances, another study examined the interplay of two bare baryon states in an exploratory manner \rref{Abell:2023qgj}.
This is the first quantitative analysis of a baryon system describing two nearby single-particle basis states.

In \sref{sec:HEFT}, we begin by constructing a Hamiltonian with two bare basis states, representing the three-quark cores of the odd-parity nucleons, dressed by interactions with $\pi N$, $\eta N$, and $K\Lambda$ scattering states.
From here, a brief overview of both the infinite-volume and finite-volume formalisms is provided.
In \sref{sec:inf}, we formulate the coupled-channel scattering equations for this Hamiltonian, constraining the free parameters of the Hamiltonian with $S_{11}$ scattering data, and predicting the positions of poles in the scattering amplitude.

\sref{sec:3fm} makes a connection with lattice QCD at $L\sim 3$ fm, where the pion mass dependence of the bare basis states is constrained.
This allows us to study the structure of energy eigenstates observed in lattice QCD calculations.
By associating the lattice eigenstates with HEFT energy eigenstates, we are able to analyse their eigenvector composition and gain insight into their structure.

In \sref{sec:2fm}, predictions are made for the finite-volume energy spectrum at $L\sim 2$ fm, using constraints of the $L\sim 3$ fm analysis.
An eigenvector analysis is performed for the states to illustrate their composition.
An analysis is performed for a $L\sim 4$ fm lattice in \sref{sec:4fm}, where recent lattice QCD results from the CLS consortium \rref{Bulava:2022vpq} are compared with HEFT.
Remarkably, the lattice QCD results are described with excellent precision for the lattice results at both 2 and 4 fm.

\sref{sec:contamination} introduces a novel method for simulating the scattering state contaminations in lattice QCD correlation functions constructed with standard three-quark operators.
The contamination functions are constructed with both HEFT eigenvectors, and lattice QCD correlation matrix eigenvectors, with remarkable agreement between them.
We also consider the interplay between contamination due to two-particle scattering-state contributions and nearby eigenstates with significant single-particle components.

Finally, \sref{sec:conclusion} concludes the results presented herein.
%
%

\section{Hamiltonian Effective Field Theory}\label{sec:HEFT}
\subsection{Hamiltonian Model} \label{sec:HamMod}
\begin{figure}
  \centering
  \includegraphics[width=0.48\textwidth]{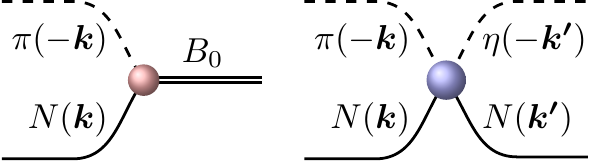}
  \caption{Diagrammatic representations of the interactions $G_{\pi N}^{B_{0}}(\vect{k})$ (left) and $V_{\pi N\eta N}(\vect{k},\vect{k'})$ (right).
  Time flows from left to right or vice versa to remain in the rest frame. }
  \label{fig:Sigma_diagrams}
\end{figure}

In the centre-of-mass frame, the Hamiltonian for an interacting system can be constructed as
\begin{equation}
  H = H_0 + H_{\text{I}} \, ,
\end{equation}
where \( H_0 \) is the free, non-interacting Hamiltonian, and \( H_{\text{I}}
\) is the interaction Hamiltonian.  In the HEFT formalism we allow for
the presence of single-particle bare-baryon basis states \( \ket{B_0} \), which may be thought of as quark model states (states in the $P$-space in the notation of \refref{Thomas:1982kv}).
With coupled two-particle channels \( \ket{\alpha} \), the free Hamiltonian \( H_0 \) can be expressed as
\begin{align}
  H_0 &= \sum_{B_0}\, \ket{B_0} m_{B_0} \bra{B_0} + \sum_{\alpha} \int\dkt \nonumber\\
      &\quad \times \ket{\alpha(\vect k)} \left[ \sqrt{m_{\text{B}_{\alpha}}^2 + k^2} + \sqrt{m_{\text{M}_{\alpha}}^2 + k^2} \right] \bra{\alpha(\vect k)} \, ,
\label{eq:H0}
\end{align}
where \( m_{\text{B}_{\alpha}} \) and \( m_{\text{M}_{\alpha}} \) are the baryon and meson masses respectively in channel \( \alpha\,, \) and \( m_{B_0} \) is the mass of each bare basis state.
For this study, the two-particle channels considered are $\pi N$, $\eta N$, and $K\Lambda$.
In general, \( H_{\text{I}} \) is governed by two types of interactions, examples of which are given in \fref{fig:Sigma_diagrams}.  The
first, which is denoted by \( g\,, \) represents the vertex
interaction between the bare state \( B_0 \), and the
two-particle basis states \( \alpha\,,\)
\begin{align}
  g = \sum_{\alpha, B_0}\, \int\dkt &\Bigl\{ \ket{B_0} G_{\alpha}^{B_0}(\vect k) \bra{\alpha(\vect k)}  \Bigr. \nonumber\\
                               &\qquad + \Bigl. \ket{\alpha(\vect k)} \left.G_{\alpha}^{B_0}\right.^{\dagger}(\vect k) \bra{B_0} \Bigr\}
\,,
\label{eq:Hg}
\end{align}
where \( G_{\alpha}^{B_0} \) is the momentum-dependent strength of the interaction between a bare state and each two-particle state.
The momentum-dependence of these couplings is selected to
reproduce the established vertex functions of chiral perturbation theory (\( \chi \)PT).
The second type of interaction represents the coupling between two different two-particle basis
states \( \alpha \) and \( \beta \) with momentum-dependent interaction strength
\( V_{\alpha\beta}\,, \) and is given by
\begin{equation}
  v = \sum_{\alpha\,\beta}\, \int\dkt\int\dkt^{'} \ket{\alpha(\vect k)} V_{\alpha\beta}(\vect k,\vect{k'}) \bra{\beta(\vect{k'})} \, .
\label{eq:Hv}
\end{equation}
The interaction Hamiltonian is therefore given by
\begin{equation}
  H_{\text{I}} = g + v\,. \label{eq:HI}
\end{equation}

\subsection{Finite-Range Regularisation}
In order to work within a finite Hilbert space, we require a renormalisation scheme.
One such renormalisation scheme is finite-range regularisation (FRR), which has been shown to reproduce
other schemes, such as dimensional regularisation, while in the power-counting regime (PCR) of
$\chi$PT (\( m_{\pi} \sim m_{\text{phys}} \)) \rref{Young:2002ib}.

Finite-range regularisation introduces a regulator, \( u(k,\Lambda)\,,
\) a function which cuts off the UV
contributions at a rate governed by the
regulator parameter \( \Lambda\,. \) While in
principle, regulators such as a sharp cutoff can be used, it is
desirable to have a smooth regulator which phenomenologically
respects the shape of the source.
For this study, a dipole regulator of the form
\begin{equation}
  u(k,\Lambda) = \left( 1 + \frac{k^2}{\Lambda^2} \right)^{-2} \, ,
  \label{eq:dip_reg}
\end{equation}
is considered.  As illustrated in \refref{Abell:2021awi}, both dipole and Gaussian functional forms were able to describe similar ranges of HEFT systems.

The FRR expansion contains a resummation of higher-order terms that come into play as one works
beyond the PCR, extending the range of
utility~\rref{Young:2002ib,Leinweber:2003dg,Leinweber:2005cm}.  The resummation ensures the FRR
loop-integral contributions are smooth and approach zero for large pion masses, providing a natural
explanation for the slow variation with increasing quark mass observed in lattice QCD results.  FRR
provides a mechanism to exactly preserve the leading nonanalytic terms of chiral perturbation
theory, including the values of the model-independent coefficients of the nonanalytic terms, even
when working beyond the PCR.  As one addresses larger quark masses, $\Lambda$ can take on a
physical role modelling the physical size of the particles \rref{Leinweber:2003dg}.
%
%

\subsection{Infinite-Volume Framework} \label{sec:infVolScat}
In order to constrain bare state masses and potential coupling strengths, we can fit the scattering phase shifts and inelasticities calculated via the \( T \)-matrix.
This can be obtained by solving the coupled-channel integral equations,
\begin{align}
  T_{\alpha\beta}(k,k';E) &= \tilde{V}_{\alpha\beta}(k,k',E) \nonumber\\
                          &+ \sum_{\gamma}\int dq\,q^2\, \frac{\tilde V_{\alpha\gamma}(k,q,E)\, T_{\gamma\beta}(q,k';E)}{E - \omega_\gamma(q)  + i\epsilon}\,, \label{eq:BS}
\end{align}
where \( \omega_{\gamma}(q) = \sqrt{q^2 + m_{\text{M}_{\gamma}}^2} + \sqrt{q^2 + m_{\text{B}_{\gamma}}^2}\,. \)
We have also defined the coupled-channel potential \( \tilde V_{\alpha\beta} \), which considers all bare states \( B_0 \) as
\begin{equation}
  \tilde{V}_{\alpha\beta}(k,k',E) = \sum_{B_0} \frac{G_{\alpha}^{B_0\,\dagger}(k)\,G_{\beta}^{B_0}(k')}{E - m_{B_0}} + V_{\alpha\beta}(k,k') \, .
\label{eq:ccV}
\end{equation}
The phase shifts and inelasticity however are extracted from the unitary \( S \)-matrix, which is related to the \( T \)-matrix by
\begin{equation}
  S_{\alpha\beta}(E) = \delta_{\alpha\beta} - 2i\pi \sqrt{\rho_{\alpha}\,\rho_{\beta}}\, T_{\alpha\beta}(k_{\text{on},\alpha}, k_{\text{on},\beta}; E) \, ,
\label{eq:Smat}
\end{equation}
where \( k_{\text{on},\alpha} \) is the on-shell momentum in channel \( \alpha\,, \)
and \( \rho_\alpha \) is the density of states, given by
\begin{equation}
  \rho_{\alpha} = \frac{\sqrt{k_{\text{on},\alpha}^2 + m_{\text{M}_{\alpha}}^2}\, \sqrt{k_{\text{on},\alpha}^2 + m_{\text{B}_{\alpha}}^2}}{E}\, k_{\text{on},\alpha} \, .
\label{eq:rho}
\end{equation}
The inelasticity, \( \eta_\alpha \), and phase shift, \( \delta_\alpha \), are then calculated from
\begin{equation}
  S_{\alpha\alpha}(E) = \eta_\alpha \exp(2i \delta_\alpha)\,.
\end{equation}

In order to search for poles in the $T$-matrix, with a negative imaginary component corresponding to a resonance, we search for zeroes in the dressed propagator
\begin{equation}
  A_{B_0,B_0'}(E) = \left[ \delta_{B_0,B_0'}\left( E - m_{B_0} \right)  - \bar{\Sigma}_{B_0,B_0'}(E) \right]^{-1}.
\end{equation}
Here, $\bar{\Sigma}_{B_0,B_0'}(E)$ is the sum of all self-energy contributions, such as those in \fref{fig:Sigma_diagrams}.
In evaluating these self-energy contributions, integrals over all $k$-space are rotated by $k \rightarrow k\,e^{i\theta}$, where $\theta$ is chosen to be approximately $-70^{\circ}$ for all scattering channels, such that all poles are found in the correct Riemann sheet.
Poles in the $T$-matrix of complex energy $E_{\text{pole}}$ are therefore found such $\det\left( A_{B_0,B_0'}(E_{\text{pole}})^{-1} \right) = 0$.
%
%
%

\subsection{Finite-Volume Matrix Method} \label{sec:finVol_method}
On a three-dimensional, cubic lattice of volume \( L^3\,, \) the allowed momentum is discretised to
\begin{equation}
  \vect{k_{\vect n}} = \frac{2\pi}{L}\, \vect{n}\,, \quad \vect{n}=(n_x\,,n_y\,,n_z) \,,
\label{eq:mom_disc}
\end{equation}
where \( n_x\,, n_y\,, \) and \( n_z \) can take any integer values, which for $S$-wave scattering give $\vect{k}_{\min} = \vect{0}$.
As a result of this, the integrals over momentum in \eref{eq:H0} to \eref{eq:Hv} undergo discretisation of the form
\begin{equation}
  \int \dkt \rightarrow \sum_{\vect n\in \mathbb{Z}^3}\, \left(\frac{2\pi}{L}\right)^3 \,. \label{eq:int_disc}
\end{equation}
For a sufficiently large lattice extent \( L \) however, we can approximate spherical symmetry and consider only the degenerate momentum states, where the effect of this approximation was discussed in \refref{Li:2019qvh}.
These degenerate momentum states are labelled \( k_n\,, \) where we have defined the integer \( n = n_x^2 + n_y^2 + n_z^2\,. \)
The degeneracy of these states is given by the function \( C_3(n)\,, \) which counts the number of ways the squared integers \( n_x^2\,,n_y^2\,, \) and \( n_z^2 \) can sum to each \( n\,. \)
Some example values of this function are \( C_3(2) = 12\,, \) and \( C_3(7) = 0\,, \) as there are no combinations of square integers that sum to 7.
Using this definition in \eref{eq:int_disc}, we therefore have the total transformation
\begin{equation}
  4\pi \int k^2\, dk = \int d^{3}k \rightarrow \left( \frac{2\pi}{L} \right)^3 \sum_{n \in \mathbb{N}} C_3(n) \,. \label{eq:C3n_trans}
\end{equation}

As our regulator parameter \( \Lambda \) provides a momentum cutoff, the Hamiltonian matrix will have a finite extent.
Defining $k_{n_{\max}}$ as the maximum momentum allowed in the Hamiltonian, this value must be sufficiently high compared to the regulator mass such that variation of $k_{n_{\max}}$ does not change the Hamiltonian solution.
Such a momentum is found as the solution of $u(k_{n_{\max}}, \Lambda) = u_{\min}$ for a given regulator form factor and regulator parameter, where $u_{\min}$ is chosen as the regulator value which satisfies this criteria.
The value of $u_{\min}$ is tuned such that the size of the matrix Hamiltonian is minimised to reduce computational requirements, while also ensuring there are a sufficient quantity of basis states such that the eigenvalues of the Hamiltonian converge to fixed values.
A value of \( u_{\min} = 10^{-2} \) is selected to balance these two requirements, and an exploration of this choice is presented in \refref{Abell:2021awi}.

Inserting \( u_{\min} \) into \eref{eq:dip_reg} and solving for the maximum momentum gives $k_{n_{\max}} = \Lambda \sqrt{u_{\min}^{-\frac{1}{2}} - 1}$.
Given the quantisation condition from \eref{eq:mom_disc}, the size of the finite Hamiltonian matrix is therefore given as
\begin{equation}
  n_{\max} = \left( \frac{k_{n_{\max}}\,L}{2\pi} \right)^2\,.
\end{equation}
Given the maximum allowed momentum for this system, in $S$-wave the free Hamiltonian for this system takes the finite matrix form of
\begin{align}
  H_0 = \text{diag}\, &\left[ m_{N_1}^{(0)},\, m_{N_2} ^{(0)},\, \omega_{\pi N}(0),\, \omega_{\eta N}(0),\, \omega_{K\Lambda}(0),\, \right. \nonumber\\
  &\, \left. \omega_{\pi N}(k_1),\, \omega_{\eta N}(k_1),\, \omega_{K\Lambda}(k_1),\, \cdots,\, \omega_{K\Lambda}(k_{n_{\max}})\, \right]\,.
\end{align}

Additionally, the potentials in \eref{eq:Hg} and \eref{eq:Hv} undergo a scaling due to finite-volume factors.
These finite-volume potentials are labelled as \( \bar{G}_{\alpha}^{B_0}(k) \) and \(\bar{V}_{\alpha\beta}(k,k')\,, \) given by
\begin{align}
  \bar G_{\alpha}^{N_i}(k_n) &= \sqrt{\frac{C_3(n)}{4\pi}}\, \left( \frac{2\pi}{L} \right)^{\frac{3}{2}}\, G_{\alpha}^{N_i}(k_n)\,, \nonumber\\
  \bar V_{\alpha\beta}(k_n, k_{m}) &= \sqrt{\frac{C_3(n)}{4\pi}}\, \sqrt{\frac{C_3(m)}{4\pi}}\, \left( \frac{2\pi}{L} \right)^{3} V_{\alpha\beta}(k_n, k_m) \,.
\end{align}
In matrix form, the interaction Hamiltonian is therefore written as
\begin{widetext}
\begin{equation}
  H_{\text{I}} =
  \begin{pmatrix}
    0 & 0 & \bar G_{\pi N}^{N_1}(0) & \bar G_{\eta N}^{N_1}(0) & \bar G_{K\Lambda}^{N_1}(0) & \bar G_{\pi N}^{N_1}(k_1) & \cdots \\
    0 & 0 & \bar G_{\pi N}^{N_2}(0) & \bar G_{\eta N}^{N_2}(0) & \bar G_{K\Lambda}^{N_2}(0) & \bar G_{\pi N}^{N_2}(k_1) & \cdots \\
    \bar G_{\pi N}^{N_1}(0) & \bar G_{\pi N}^{N_2}(0)  & \bar V_{\pi N\pi N}(0,0) & \bar V_{\pi N\eta N}(0,0) & \bar V_{\pi NK\Lambda}(0,0) & \bar V_{\pi N\pi N}(0,k_1) & \cdots \\
    \bar G_{\eta N}^{N_1}(0) & \bar G_{\eta N}^{N_1}(0) & \bar V_{\eta N\pi N}(0,0) & \bar V_{\eta N\eta N}(0,0) & \bar V_{\eta NK\Lambda}(0,0) & \bar V_{\eta N\pi N}(0,k_1) & \cdots \\
    \bar G_{K\Lambda}^{N_1}(0) & \bar G_{K\Lambda}^{N_1}(0) & \bar V_{K\Lambda\pi N}(0,0) & \bar V_{K\Lambda\eta N}(0,0) & \bar V_{K\Lambda K\Lambda}(0,0) & \bar V_{K\Lambda \pi N}(0,k_1) & \cdots\\
    \bar G_{\pi N}^{N_1}(k_1) & \bar G_{\pi N}^{N_2}(k_1) & \bar V_{\pi N\pi N}(k_1,0) & \bar V_{\eta N\pi N}(k_1,0) & \bar V_{K\Lambda \pi N}(k_1,0) & \bar V_{\pi N\pi N}(k_1,k_1) & \cdots \\
    \vdots & \vdots & \vdots & \vdots & \vdots & \vdots & \ddots \\
  \end{pmatrix}\,.
\end{equation}
\end{widetext}

Considering the full Hamiltonian $H = H_0 + H_{\text{I}}$, we may solve the eigenvalue equation $\text{det}\left( H - E_i\,  \mathbb{I}  \right) = 0$ for energies $E_i$.
Associated with each energy $E_i$, we may solve for the eigenvectors of the Hamiltonian, labelled $\braket{B_j | E_i}$.
These eigenvectors provide the contribution from each basis state $\ket{B_j}$ to the interacting eigenstate $\ket{E_i}$, providing insight into the structure of each finite-volume eigenstate.
%
%
%
%
%
%

\section{Infinite-Volume Scattering} \label{sec:inf}
To describe the interactions between the basis states in this system, we use standard $S$-wave parametrisations for the two potentials.
For some channel $\alpha$ and bare state $N_i$, the interaction $\braket{\alpha(k) | g | N_i}$ takes the heavy-baryon $\chi$PT-motivated form \rref{Wu:2014vma}
\begin{equation}
  G_{\alpha}^{N_i}(k) = \frac{\sqrt{3}\, g_{\alpha}^{N_i}}{2\pi f_{\pi}}\, \sqrt{\omega_{\text{M}_{\alpha}}(k)}\, u(k)\,,
\end{equation}
where the label $\text{M}_{\alpha}$ refers to the meson in channel $\alpha$, giving $\omega_{\text{M}_{\alpha}}(k) = \sqrt{k^2 + m_{\text{M}_{\alpha}}^2}$, $f_{\pi} = 92.4$ MeV, and $g_{\alpha}^{N_i}$ is the dimensionless coupling strength of this interaction.
The interaction $\braket{\beta(k') | v | \alpha(k)}$ takes the form
\begin{equation}
  V_{\alpha\beta}(k,k') = \frac{3\, v_{\alpha\beta}}{4\pi^2 f_{\pi}^2}\, \tilde{u}(k)\, \tilde{u}(k')\,,
\end{equation}
with coupling strength $v_{\alpha\beta}$.
For all interactions between scattering channels, the regulator gains a low-energy enhancement in order to better fit the low-energy phase-shifts \rref{Liu:2015ktc},
\begin{equation}
  \tilde{u}(k) = \frac{\omega_{\pi}(k) + m_{\pi}^{\text{phys}}}{\omega_{\pi}(k)}\, u(k)\,.
\end{equation}

Using standard numerical techniques, we are able to fit the $\pi N$ phase shifts and inelasticites solved from the $T$-matrix formalism in \sref{sec:infVolScat} to the $S_{11}$ WI08 solution
from \refref{site:SAID,Workman:2012hx}.
As the $N^{*}(1535)$ lies approximately 100 MeV below the $K\Lambda$ threshold, the coupling $g_{K\Lambda}^{N_1}$ was held fixed at zero.
With the remaining coupling strengths, bare state masses, and regulator parameters, there are a total of 21 free parameters present in this system.
A $\chi^2$ may be calculated by comparing the HEFT phase shifts and inelasticities with the WI08 solution.
Using Powell's derivative-free optimisation procedure \rref{doi:10.1093/comjnl/7.2.155} to minimise the $\chi^2$, the resultant set of parameters is presented in \tref{tab:2b3c_params}.

\begin{figure*}
  \centering
  \includegraphics[width=0.48\textwidth]{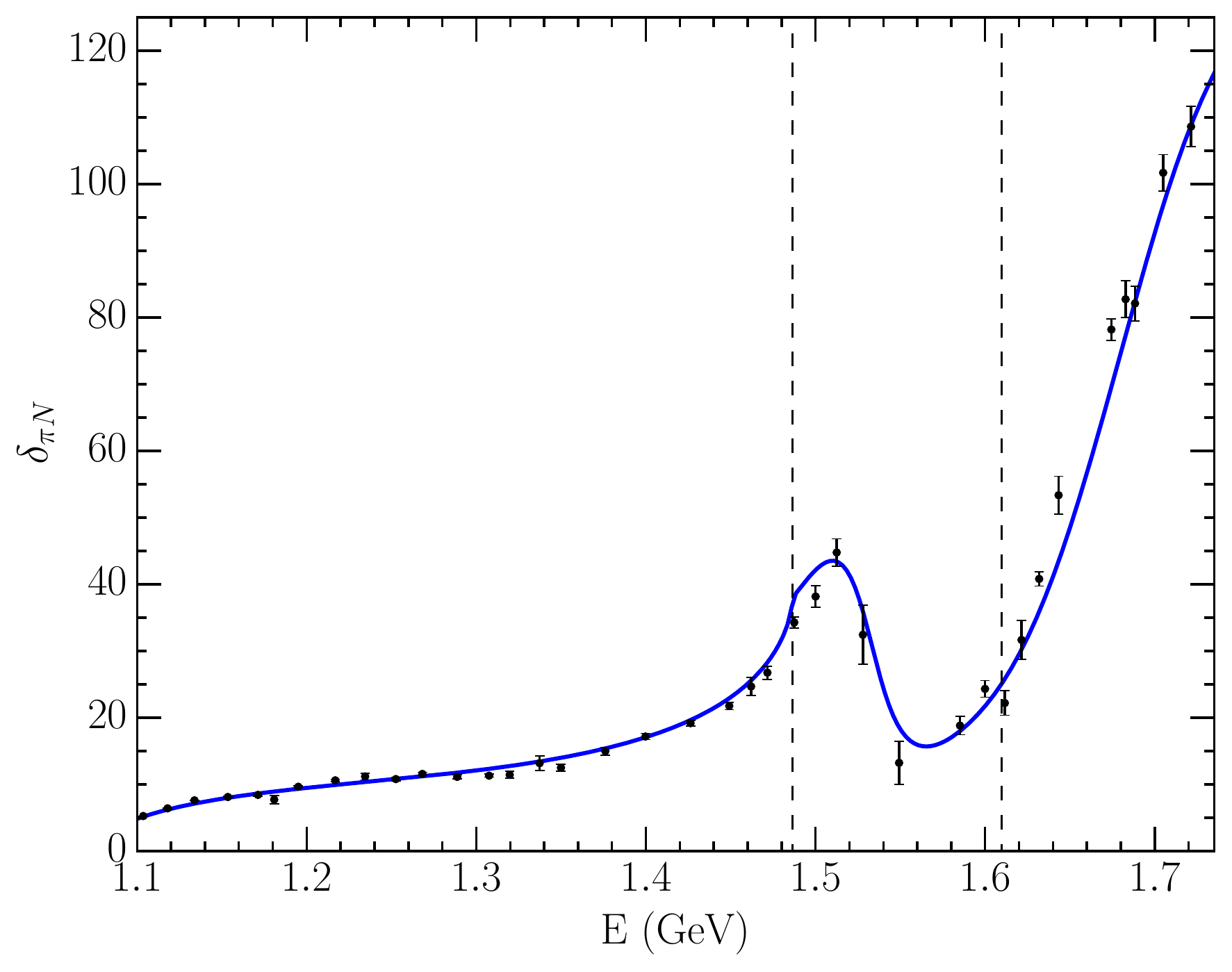}~
  \includegraphics[width=0.48\textwidth]{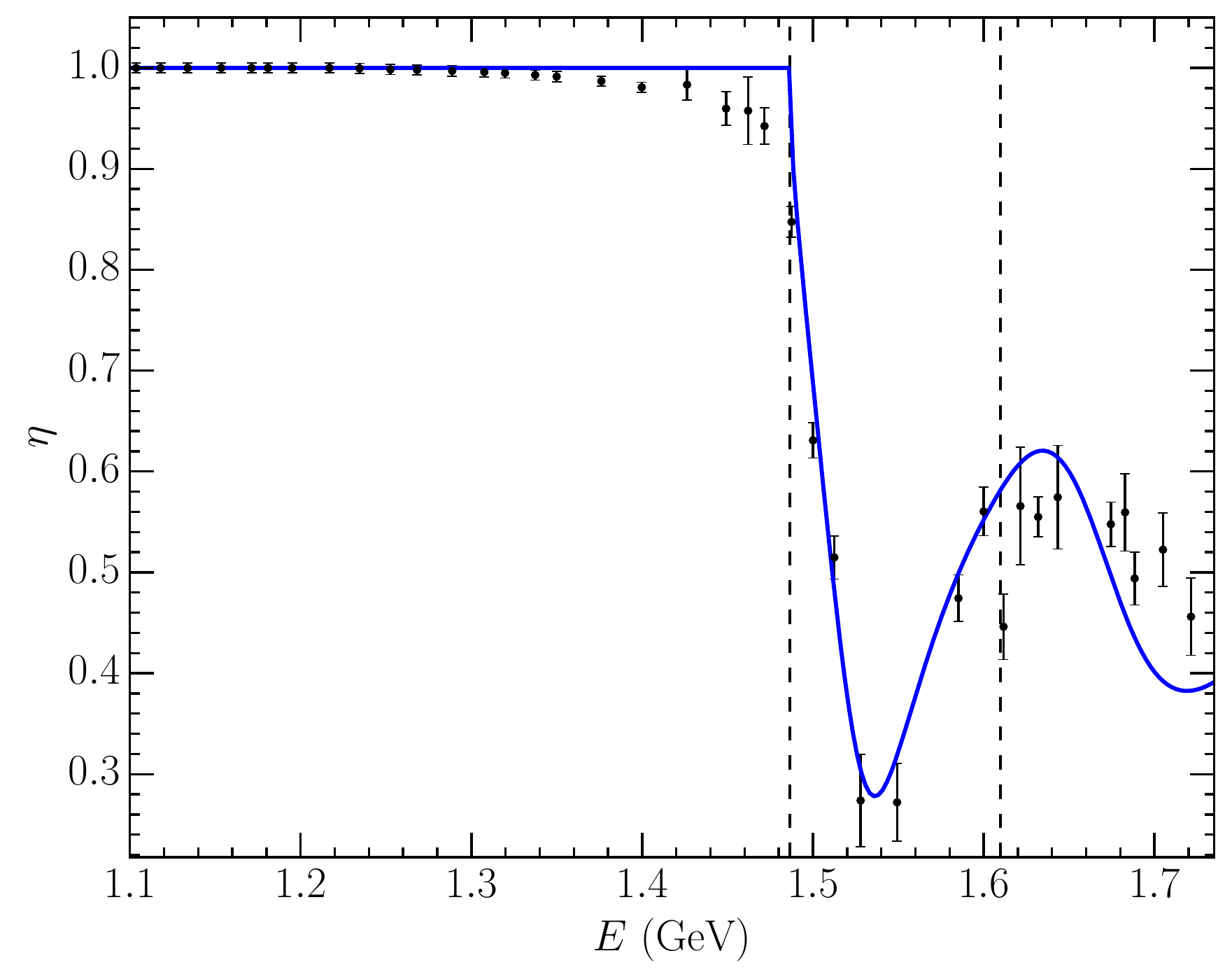}
  \caption{Phase shift and inelasticity for the parameters in \tref{tab:2b3c_params}. The solid (blue) lines are the theoretical calculations from HEFT, while the data points are the SAID WI08 solution \rref{site:SAID,Workman:2012hx}. The dashed vertical lines denote the $\eta N$ and $K\Lambda$ thresholds.}
  \label{fig:2b3c_phase}
\end{figure*}

The resultant $S_{11}$ phase shift and inelasticity are illustrated in \fref{fig:2b3c_phase}.
Using this parameter set, we are able to characterise the $S_{11}$ phase shifts in the energy range considered.
This fit results in a $\chi^2$ of 604, and with $78-21=57$ degrees-of-freedom (d.o.f.s), a $\chi^2/{\text{d.o.f.}}$ of 10.6.
While this $\chi^2/{\text{d.o.f.}}$ is large, it can be attributed to missing three-particle $\pi\pi N$ threshold effects, as can be seen in the tension in the inelasticity predictions near 1.4 GeV.
Difficulties describing the inelasticities above the $N^{*}(1650)$ region may also be attributed to a sizeable contribution from $\pi\pi N$ states, or additional hyperon channels such as $K\Sigma$.
While this $\chi^2/{\text{d.o.f.}}$ is not directly comparable with similar studies of $S_{11}$ scattering \rref{Bruns:2010sv} due to the use of the WI08 solution versus single-energy values, visually the fit of \refref{Bruns:2010sv} and that presented here produce a similar quality of fit.
Alternatively, quantities such as the positions of poles corresponding with the two odd-parity resonances may prove to be a better source of comparison.

%
%
\begin{table}
  \centering
  \caption{HEFT fit parameters constrained by the WI08 solution \rref{site:SAID,Workman:2012hx} for $S_{11}$ scattering, up to 1.75 GeV.}
    \begin{ruledtabular}
      \begin{tabular}{cc|cc}
        Parameter & Value & Parameter & Value \\
        \hline
        $m_{N_1}^{(0)}$ / GeV & 1.6301 & $m_{N_2}^{(0)}$ / GeV & 1.8612 \\
        $g_{\pi N}^{N_1}$ & 0.0898   & $g_{\pi N}^{N_2}$ & 0.2181 \\
        $g_{\eta N}^{N_1}$ & 0.1525  & $g_{\eta N}^{N_2}$ & 0.0009 \\
        $g_{K\Lambda}^{N_1}$ & 0.0000 & $g_{K\Lambda}^{N_2}$ & -0.2367 \\
        $\Lambda_{\pi N}^{N_1}$ / GeV & 1.2335  & $\Lambda_{\pi N}^{N_2}$ / GeV & 1.4000 \\
        $\Lambda_{\eta N}^{N_1}$ / GeV & 1.2642 & $\Lambda_{\eta N}^{N_2}$ / GeV & 0.9521 \\
        $\Lambda_{K\Lambda}^{N_1}$ / GeV & $\cdots$     &$\Lambda_{K\Lambda}^{N_2}$ / GeV & 0.7283 \\
        $v_{\pi N, \pi N}$ & -0.0655 & $v_{\eta N,\eta N}$ & -0.0245 \\
        $v_{\pi N, \eta N}$ & 0.0388 & $v_{\eta N, K\Lambda}$ & 0.0320 \\
        $v_{\pi N, K\Lambda}$ & -0.0757 & $v_{K\Lambda,K\Lambda}$ & 0.1371 \\
        $\Lambda_{v,\pi N}$ / GeV & 0.6000 & $\Lambda_{v,\eta N}$ / GeV & 0.9036 \\
        $\Lambda_{v,K\Lambda}$ / GeV & 0.6060 \\
      \end{tabular}
    \end{ruledtabular}
  \label{tab:2b3c_params}
\end{table}

In the Particle Data Group (PDG) tables \rref{ParticleDataGroup:2022pth}, the poles for the two low-lying odd-parity nucleon resonances are given as
\begin{align}
  E_{N^{*}(1535)} &= 1510\pm 10 - (65\pm10)i \text{ MeV}\,, \nonumber\\
  E_{N^{*}(1650)} &= 1655\pm 15 - (67\pm18)i \text{ MeV}\,.
\end{align}
With the set of parameters in \tref{tab:2b3c_params}, and searching in the second Riemann sheet, using HEFT two poles are found at energies
\begin{align}
  E_{1} &= 1500 - 50i \text{ MeV}\,, \nonumber\\
  E_{2} &= 1658 - 56i \text{ MeV}\,,
\end{align}
in excellent agreement with the PDG pole positions.
These were found by searching for solutions of $\det\left( A(E)^{-1} \right) = 0$, as described in \sref{sec:infVolScat}.

By comparing phase shifts and inelasticities calculated in HEFT with those from resources such as SAID, and $T$-matrix poles with PDG values, it is clear that an interpretation of the low-lying odd-parity nucleon resonances as quark-model like states is consistent with experiment.
By moving to a finite-volume and comparing with results from lattice QCD however, we are able to gain a larger degree of understanding, and further test this interpretation.
%
%

\section{Finite-Volume HEFT at 3 \lowercase{fm}} \label{sec:3fm}
\subsection{Pion Mass Dependence}
By varying the pion mass $m_{\pi}$, and the lattice extent $L$, one can solve for the eigenvalues and eigenvectors of the Hamiltonian to obtain the finite-volume energy spectrum, the results of which can be compared with lattice QCD.
As the pion mass is increased, the masses of the other hadrons are also increased proportionally, as to match the hadron masses calculated by PACS-CS~\rref{PACS-CS:2008bkb}.
As the pion-mass extrapolations for the bare states are unknown, we give them a simple expansion of the form
\begin{equation} \label{eq:hadron_mass}
  m_{N_i}(m_{\pi}^2) = m_{N_i}^{(0)} + \alpha_{N_i} \left( m_{\pi}^2 - \left.m_{\pi}^2 \right|_{\text{phys}} \right)\,,
\end{equation}
where the mass-slopes $\alpha_{N_i}$ are varied to fit 10 lattice QCD data points at $L \sim 3$ fm, and a pion mass varying from 169 to 623 MeV in the Sommer scheme.
It was found that as the bare mass slope only has an impact at significantly larger than physical pion masses, fitting to the lattice QCD energies at the lightest pion mass had little effect on the mass slope.
As such, the fitting procedure focused on minimising the distance between the lattice QCD data at the three heaviest pion masses, and HEFT energy eigenvalues.
While there is also precise data available at $L\sim 2$ fm which could also be used for the fitting procedure, it is desirable to confront the spectrum at 2 fm as a prediction from the 3 fm analysis.
As the parameters of the Hamiltonian are constrained by experiment, the key input from the 3 fm analysis is the quark-mass slope of the bare masses, $\alpha_{N_1}$ and $\alpha_{N_2}$.

The 3 fm fitting procedure gives mass slopes
\begin{align}
  \label{eq:mass_slopes}
  \alpha_{N_1} &= 0.944 \,\text{GeV}^{-1}\,, \nonumber\\
  \alpha_{N_2} &= 0.611 \,\text{GeV}^{-1}\,.
\end{align}
The differences in the slope parameters are in accord with quark model expectations.
The lower state is dominated by hyperfine attraction in spin-1/2 components of the wave function.
The strength of the hyperfine attraction is inversely proportional to the product of the constituent quark masses.
Thus, as the constituent quark mass increases, the hyperfine attraction is lost and the baryon mass increases rapidly.
On the other hand, the second state is dominated by spin-3/2 components contributing to hyperfine repulsion.
For the second state, repulsion is lost as the constituent quark masses increase and thus the baryon mass rises more slowly.
%
%

\subsection{Finite-Volume Energy Spectrum}
\begin{figure}
  \centering
  \includegraphics[width=0.48\textwidth]{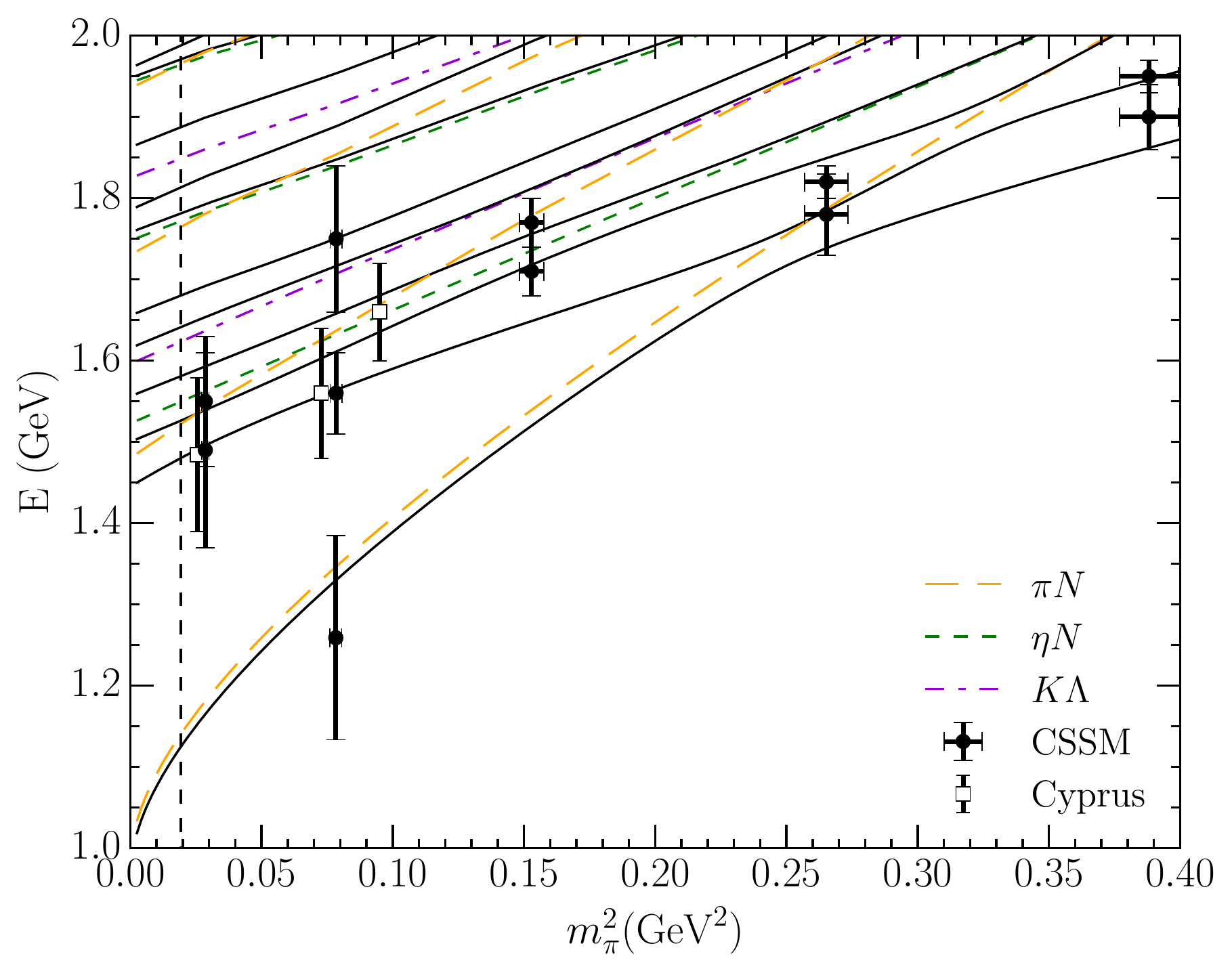}
  \caption{Finite-volume energy spectrum for $L\sim 3$ fm.
    The vertical dashed line represents the physical points, while the remaining dashed lines correspond with non-interacting basis states for each channel.
    The solid curves are the finite-volume eigenenergies calculated in HEFT.
    Lattice QCD data from CSSM \rref{Stokes:2019zdd,Mahbub:2013ala,Kiratidis:2015vpa} and the Cyprus collaboration \rref{Alexandrou:2014mka} is overlaid for comparison.}
  \label{fig:2b3c_3fm_spectrum}
\end{figure}
We are now able to calculate the full finite-volume energy spectrum for this system.
In the Sommer scheme, the physical volume varies with the quark mass.
At the physical point, the lattice extent is 2.99 fm, corresponding with the lattice size at the lightest lattice QCD point.
As the pion mass is increased, the lattice size is linearly interpolated between each lattice QCD point, giving a final lattice size of 3.27 fm.
The results of this can be seen in \fref{fig:2b3c_3fm_spectrum}, where the non-interacting basis states have been displayed as dashed lines, and the interacting energies displayed as solid lines.
Here we observe a significant shift from the non-interacting states, as well as the presence of many avoided level crossings in the excited states of the spectrum, demonstrating the complexity of the system.
The HEFT spectrum is compared to available lattice QCD results at approximately 3 fm.
We find that all eigenstates from lattice QCD correspond with at least one HEFT energy eigenvalue.

\begin{figure*}
  \centering
  \includegraphics[width=0.32\textwidth]{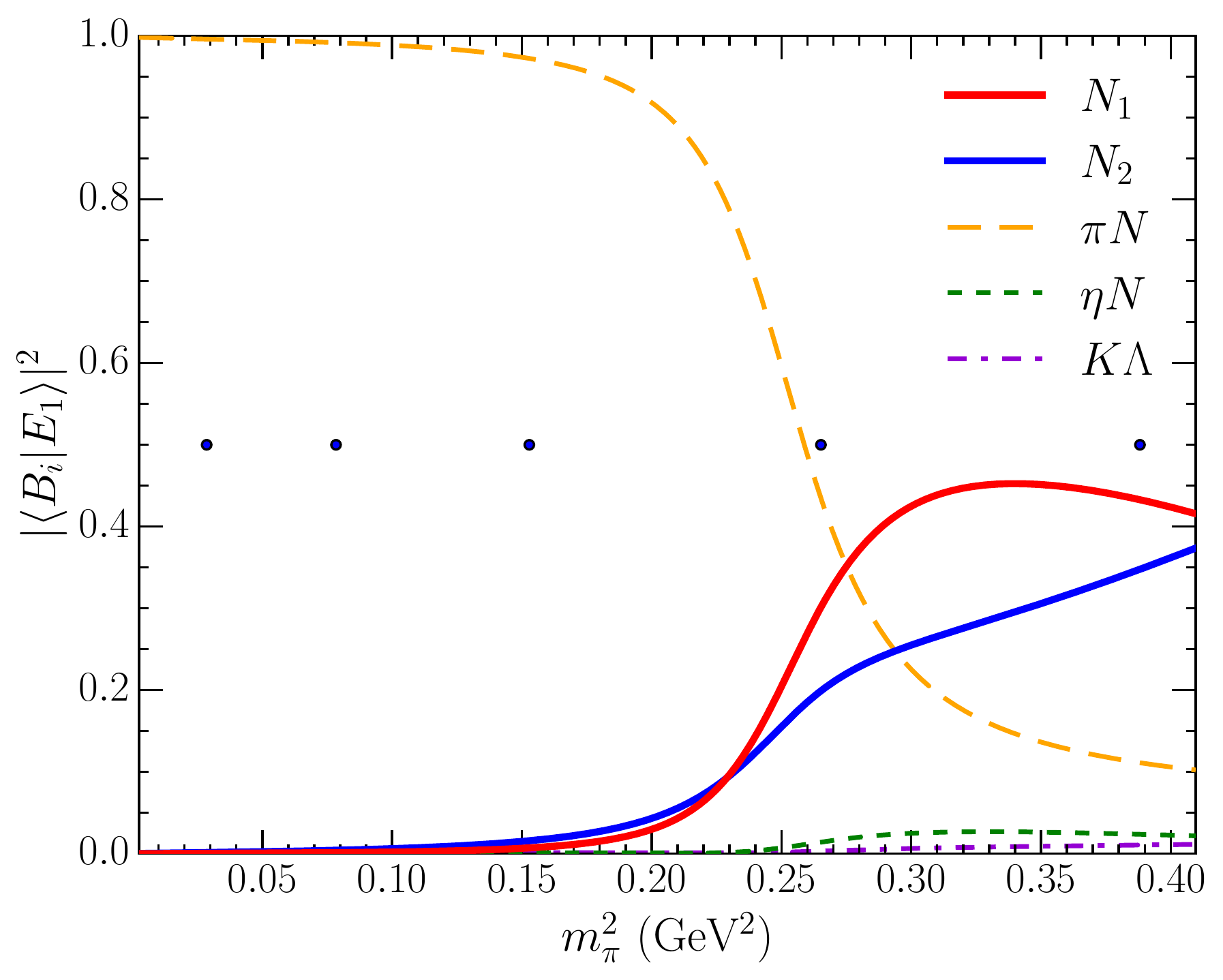}~
  \includegraphics[width=0.32\textwidth]{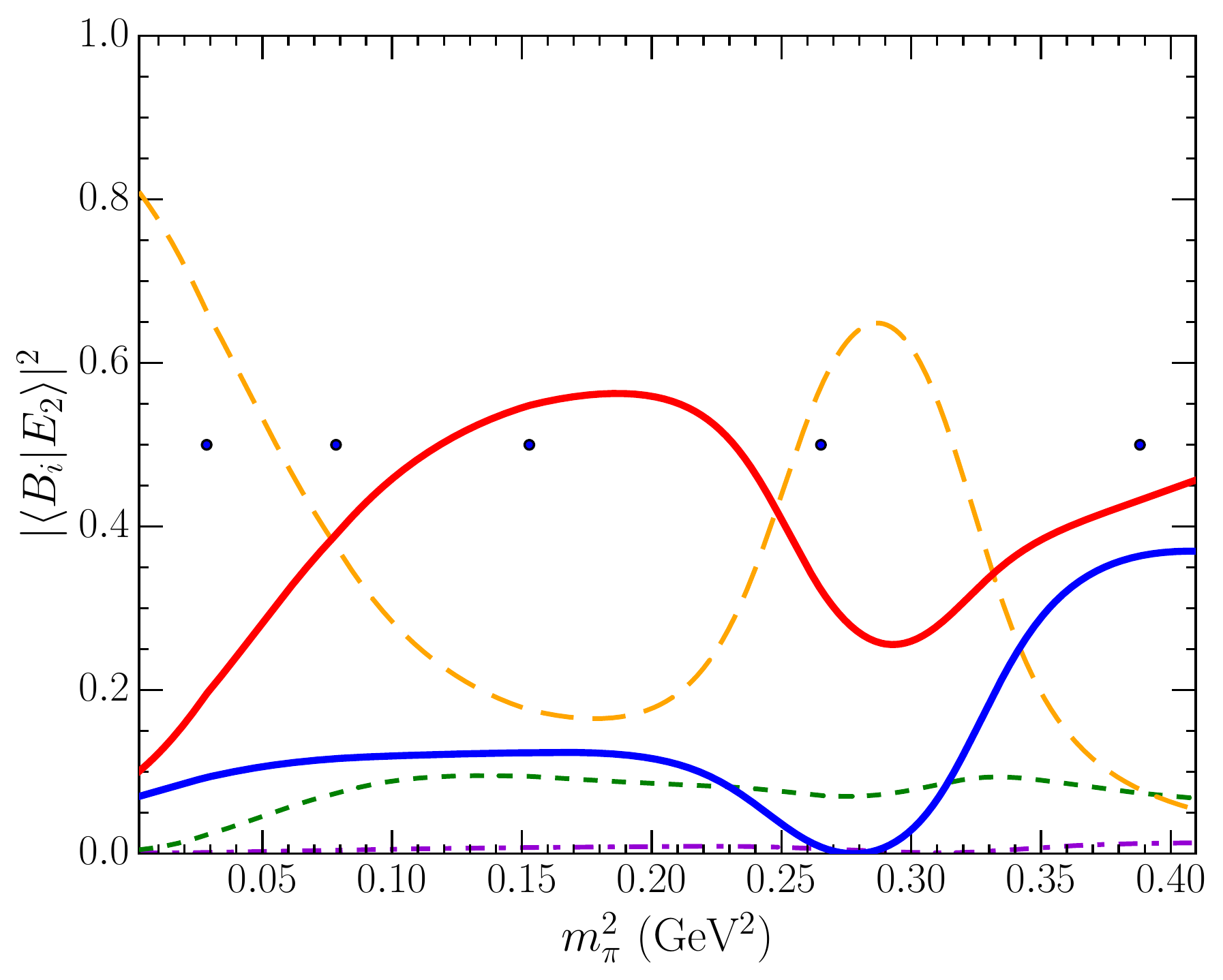}~
  \includegraphics[width=0.32\textwidth]{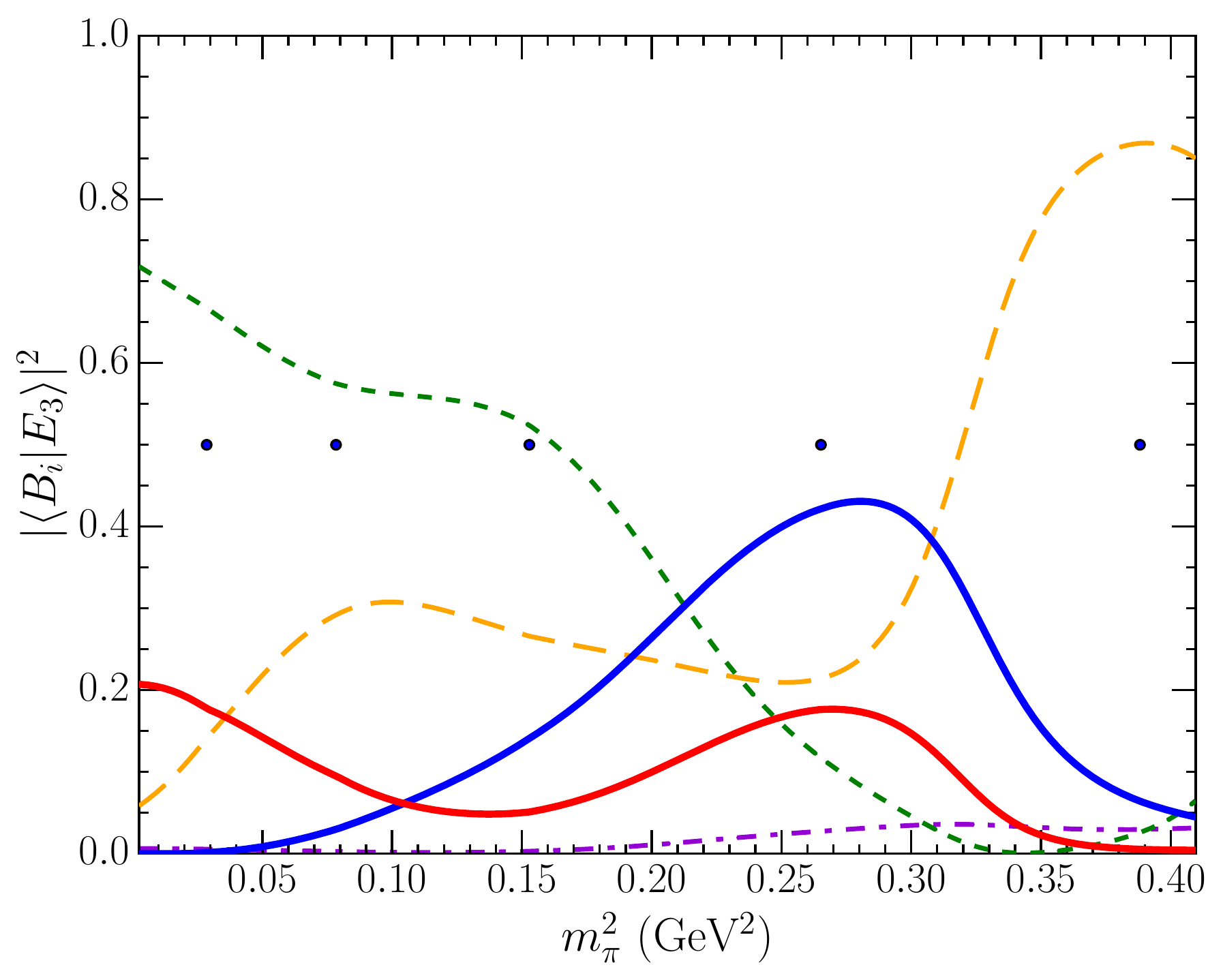}~\\
  \includegraphics[width=0.32\textwidth]{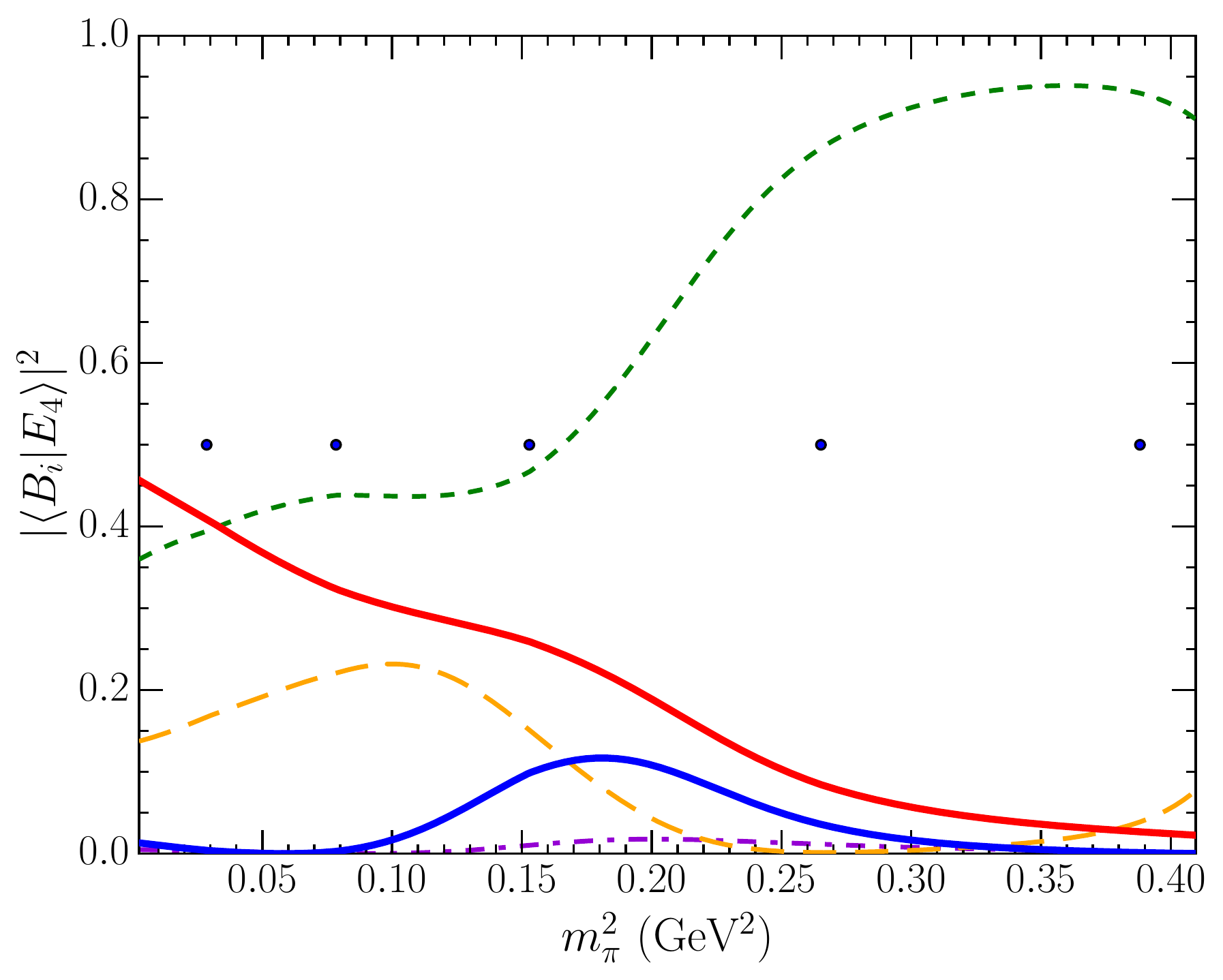}~
  \includegraphics[width=0.32\textwidth]{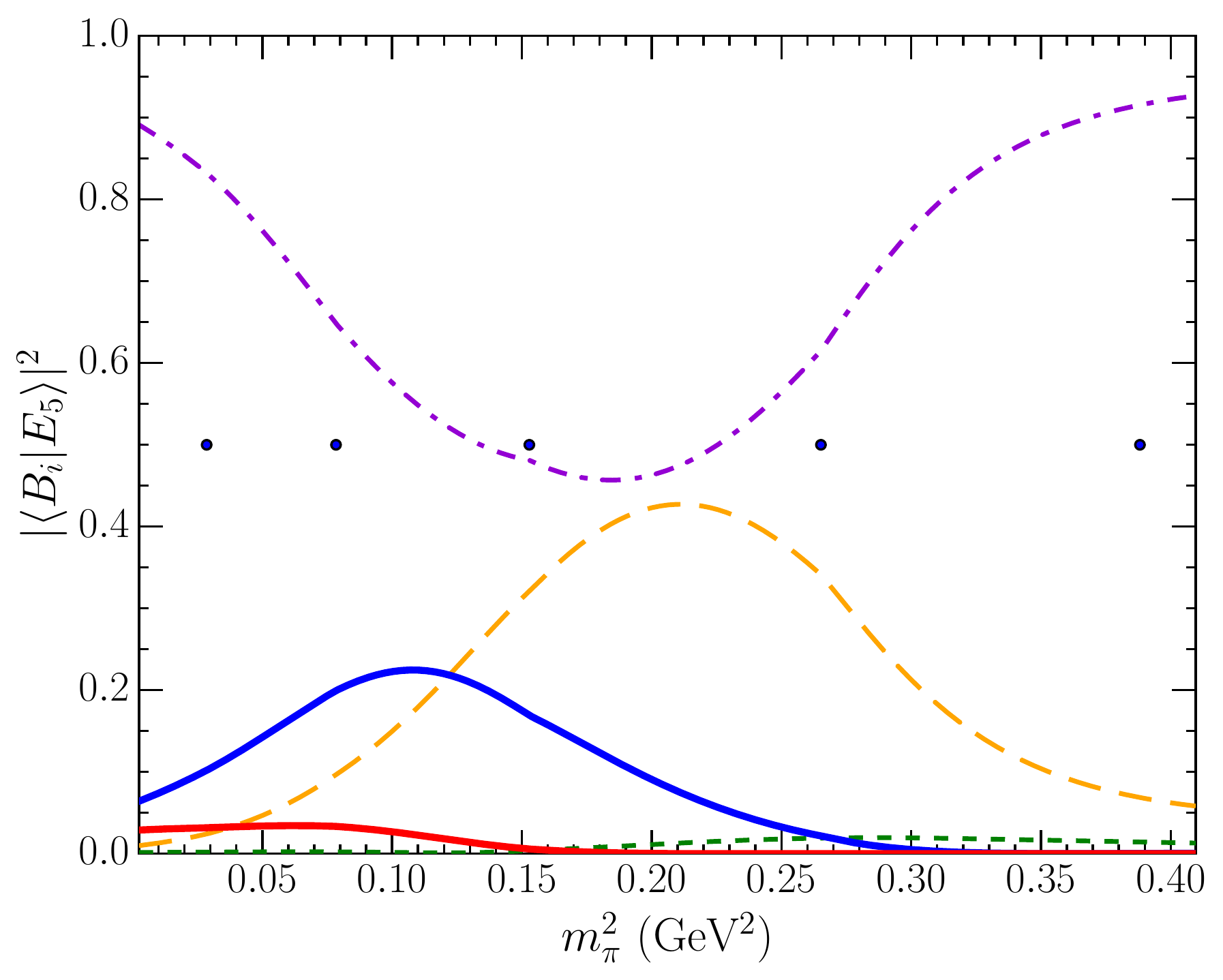}~
  \includegraphics[width=0.32\textwidth]{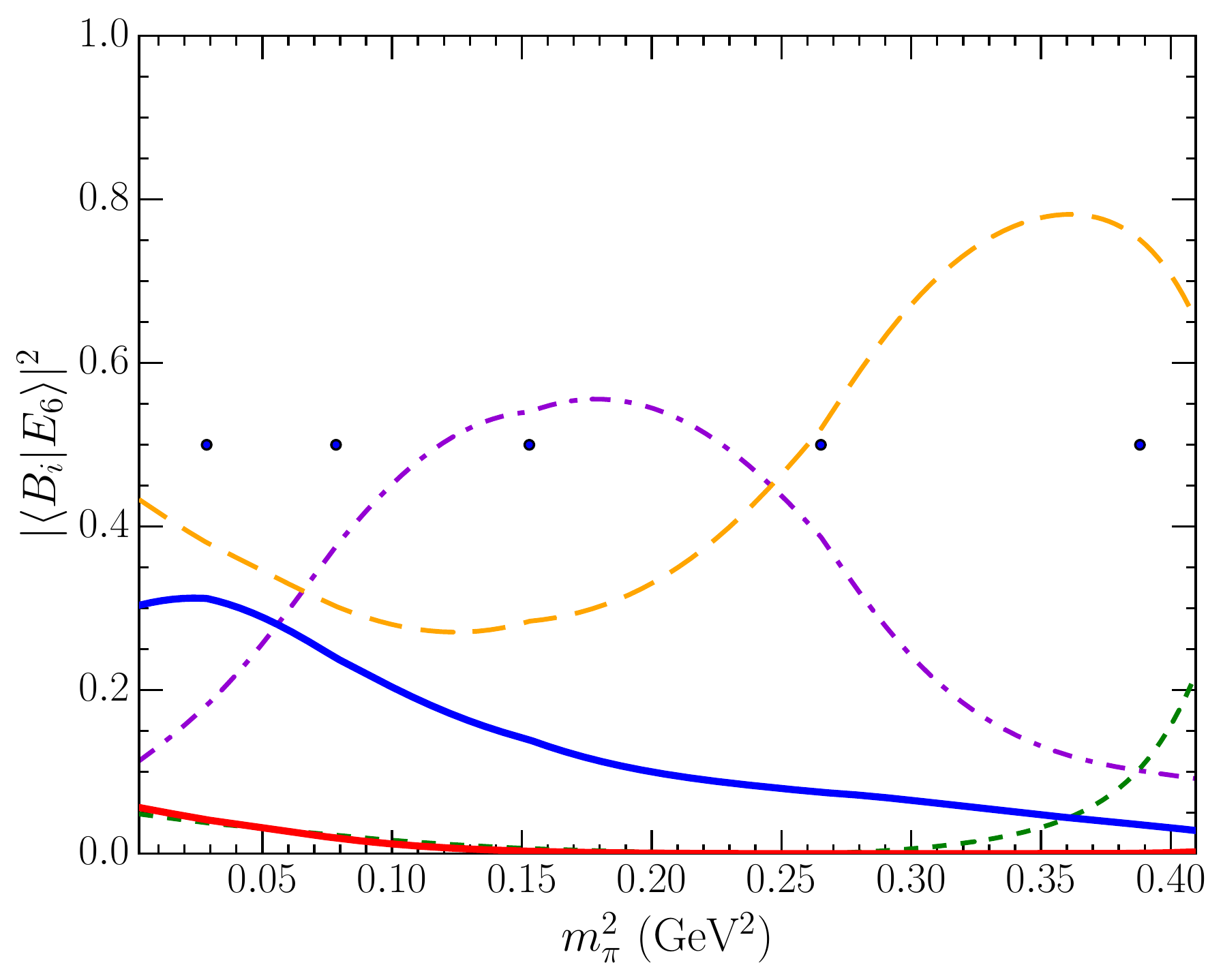}
  \caption{Pion mass dependence of the basis-state contributions for the six lowest eigenstates from the 3 fm spectrum shown in \fref{fig:2b3c_3fm_spectrum}.
    Markers on each plot correspond with the five PACS-CS masses \rref{PACS-CS:2008bkb}.
    Contributions for the sum of all momentum states in the $\pi N$, $\eta N$, and $K\Lambda$ channels are illustrated.}
  \label{fig:2b3c_3fm_eigenvectors}
\end{figure*}
One of the biggest advantages of HEFT however, is the ability to extract the eigenvectors of the Hamiltonian.
The eigenvector $\left|\braket{B_j | E_i}\right|^2$ denotes the contribution from the basis state $\ket{B_j}$ to the eigenstate $\ket{E_i}$, the results of which are shown in \fref{fig:2b3c_3fm_eigenvectors} for the first six eigenvalues at $L\sim 3$ fm.
Here it can be seen that initially at the physical point, is it difficult to interpret a single state as representing one of the odd-parity resonances.
The contributions from the two bare states, denoted by red and blue lines respectively, are instead distributed over the second through to the sixth eigenstates.
However the contributions from the bare states do seem to be concentrated around the masses of the bare states.
At this point, it is only the lowest-lying state which can be definitively interpreted as a $\pi N$ state.

As one moves away from the physical pion mass, contributions from the two bare states seem to become concentrated increasingly in the lower-lying eigenstates, and both the lowest-lying state and next state seem to each contain approximately equal amounts of each bare state.
In other words, the bare states mix to form the energy eigenstates.
The situation is similar to the mixing of the two spin-1/2 negative parity interpolators which mix to form the lattice eigenstates.

In order to better view how the contributions from the bare states are distributed, we overlay coloured lines on the energy spectrum in \fref{fig:2b3c_3fm_spectrum}.
Here, we display the state with the largest and second largest contributions from the first bare basis state as solid and dashed red lines respectively.
The contributions from the second bare basis state are illustrated in the same manner but in blue.
\begin{figure}
  \centering
  \includegraphics[width=0.48\textwidth]{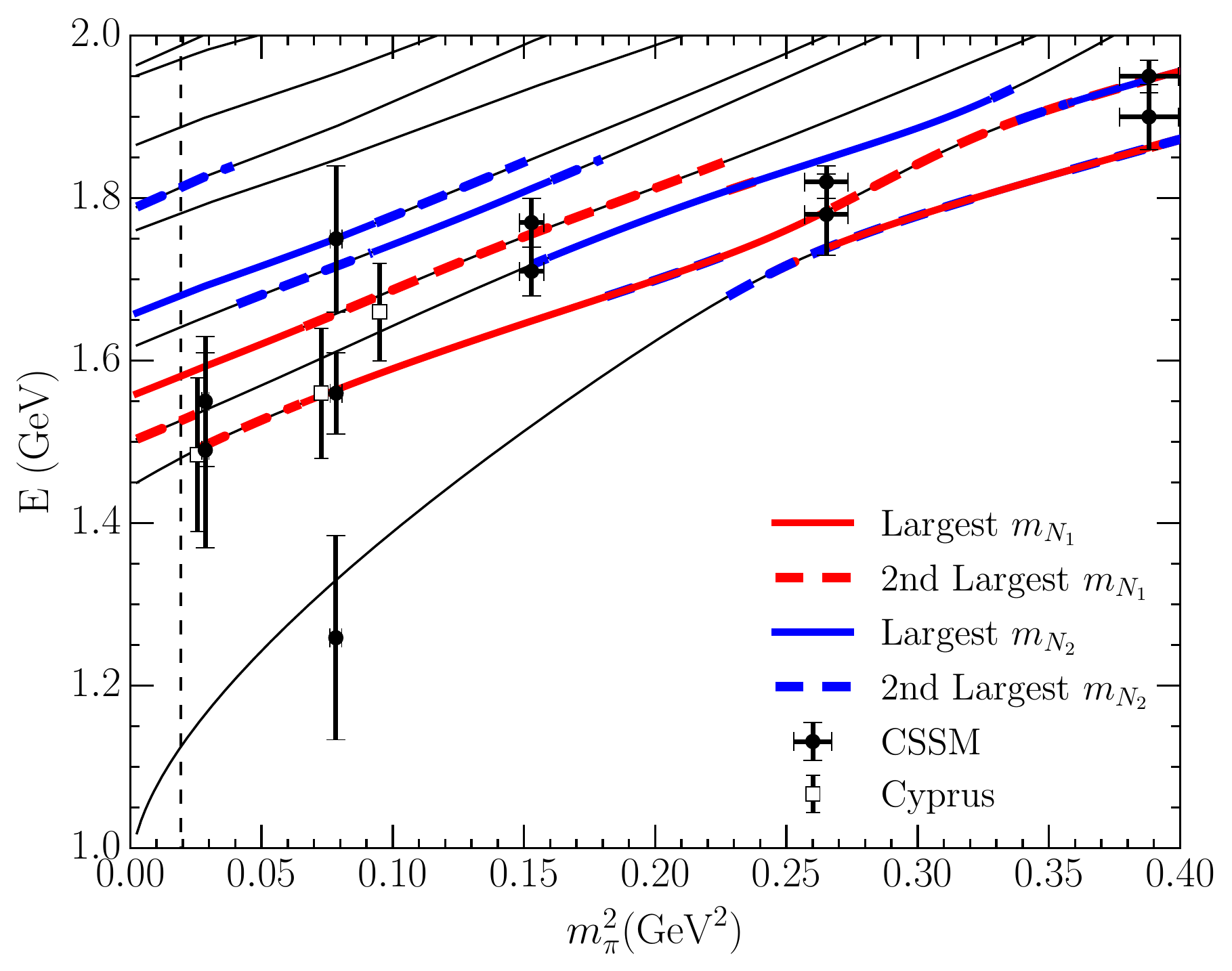}
  \caption{Finite-volume energy spectrum for $L\sim 3$ fm.
    The solid and dashed red lines represent the states with the largest and second largest contributions from the lower bare basis state.
    Similarly the solid and dashed blue liens represent the contributions from the second bare basis state.
    Lattice QCD data from CSSM \rref{Stokes:2019zdd,Mahbub:2013ala,Kiratidis:2015vpa} and the Cyprus collaboration \rref{Alexandrou:2014mka} is overlaid for comparison.}
  \label{fig:2b3c_3fm_spectrum_bare}
\end{figure}
The results of this are illustrated in \fref{fig:2b3c_3fm_spectrum_bare}.

Utilising this method of identifying states with a large bare-basis state component, it becomes easier to understand which states are likely to be observed in the CSSM and Cyprus lattice QCD calculations.
Because they used three-quark operators to form the basis of their correlation matrix, it follows that the states excited in their analysis will contain a large bare basis state component.
Thus we expect each of their lattice QCD results to be associated with a coloured energy eigenstate from HEFT.
The exception to this is the lowest lying CSSM state at $m_{\pi}^2 \sim 0.08$ GeV$^2$, which was obtained from a five-quark operator \rref{Kiratidis:2015vpa}.

Of particular note, as we move to larger quark masses, the contributions from each bare state become primarily concentrated in only two eigenstates, which strongly correspond with the states from lattice QCD.
This is in agreement with the results from \refref{Stokes:2019zdd}, where the magnetic moments of the two resonances become quark-model like as the pion mass increases.
In addition, the three results from the Cyprus Collaboration~\rref{Alexandrou:2014mka} which were constructed using only three-quark operators, correspond with eigenstates dominated by contributions from the lower lying bare state.
%
%
%
%
%
%

\section{Finite-Volume HEFT at 2 \lowercase{fm}} \label{sec:2fm}
Lattice QCD results are available for lattice sizes of approximately 2 fm from Lang \& Verduci \rref{Lang:2012db}, as well as the Hadron Spectrum Collaboration (HSC) \rref{Edwards:2011jj,Edwards:2012fx}.
As Lang \& Verduci’s correlation matrix analysis was not large enough to remove excited-state contaminations from their second and third states, we
focus on their lowest-lying state obtained from a non-local momentum-projected pion-nucleon interpolating field.
While in principle we could use this data for fitting the bare mass slopes, in \refref{Abell:2021awi} it was found that by calculating the bare mass slopes at only one lattice size, the lattice QCD data for other sizes was able to be described.
As such, we continue to use the bare mass slopes from \eref{eq:mass_slopes} for this $L\sim 2$ fm calculation, and thus make predictions for the finite-volume energy eigenvalues at various quark masses.

The HSC collaboration sets their lattice spacing in a scheme
where the physical $\Omega^{-}$ baryon mass is taken as in be
independent of the sea-quark mass. As a result, the lattice
spacing varies with quark mass.
Here, an identical approach is taken to the 3 fm calculation.
At the physical point, the lattice extent is 1.95 fm, corresponding with the lattice size at the Lang \& Verduci lattice QCD mass.
As the pion mass is increased, the lattice size is linearly interpolated between each lattice QCD point, giving a final lattice size of 2.12 fm.
The hadron masses are also varied as described in \eref{eq:hadron_mass}.
\begin{figure}
  \centering
  \includegraphics[width=0.48\textwidth]{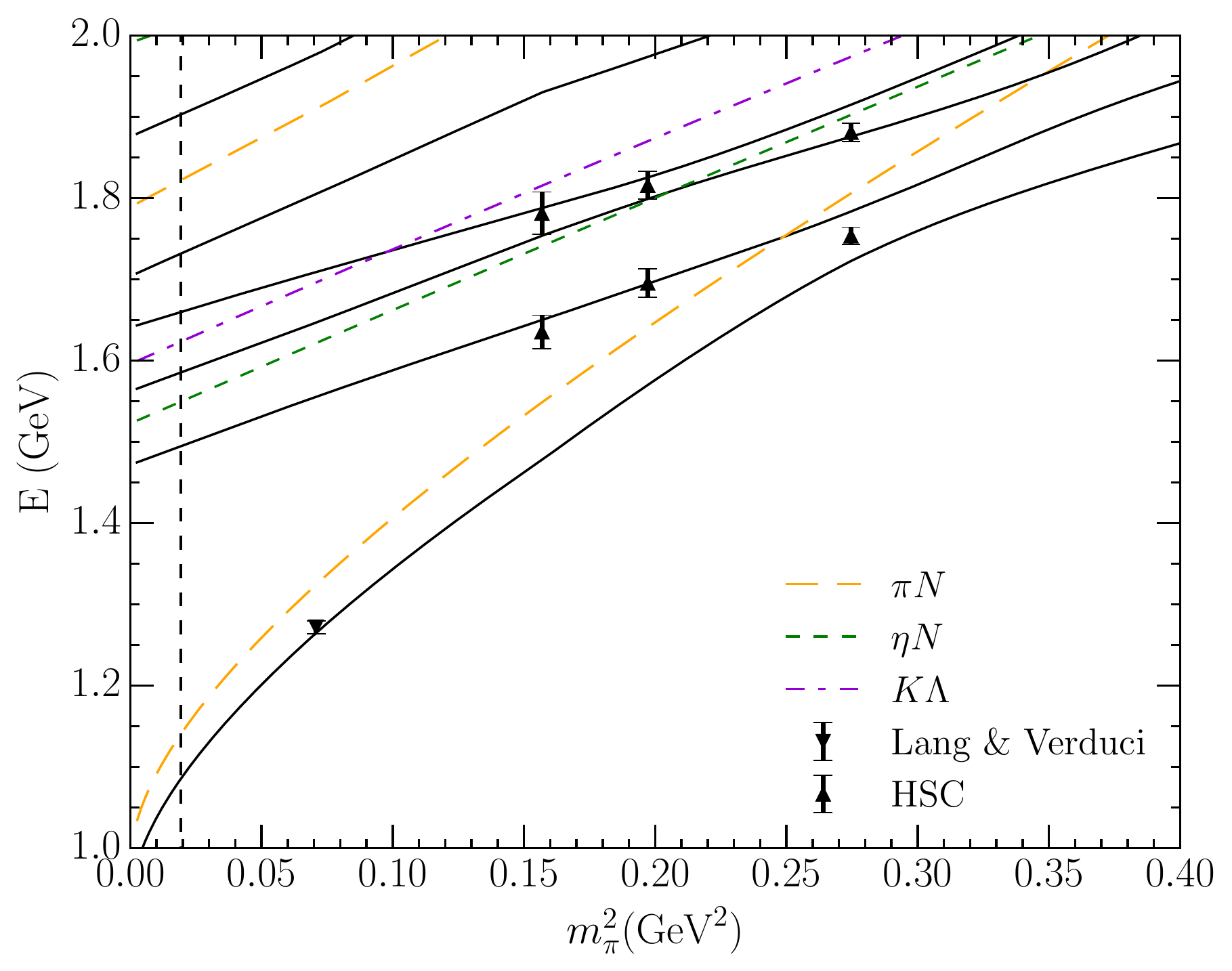}~
  \caption{Finite-volume energy spectrum for $L\sim 2$ fm.
    The vertical dashed line represents the physical point, while the remaining dashed lines correspond with non-interacting basis states for each channel.
    Overlaid is lattice QCD results from Lang \& Verduci \rref{Lang:2012db} using momentum-projected meson-baryon operators, and the Hadron Spectrum Collaboration (HSC) \rref{Edwards:2011jj,Edwards:2012fx} using three-quark operators.}
  \label{fig:2b3c_2fm_spectrum}
\end{figure}
The result for this process is illustrated in \fref{fig:2b3c_2fm_spectrum}, where similarly to the $L\sim3$ fm case, significant shifts in the energy eigenvalues from the non-interacting energies are observed.
Comparing to the lattice QCD data from HSC and Lang \& Verduci, we observe that all data points correspond with an energy eigenvalue, with the exception of a single point from HSC, which sits between the two lowest-lying states we predict.
By investigating the eigenvector composition of these states, we are able to better analyse the consistency of this data with the two bare state analysis.

\begin{figure*}
  \centering
  \includegraphics[width=0.34\textwidth]{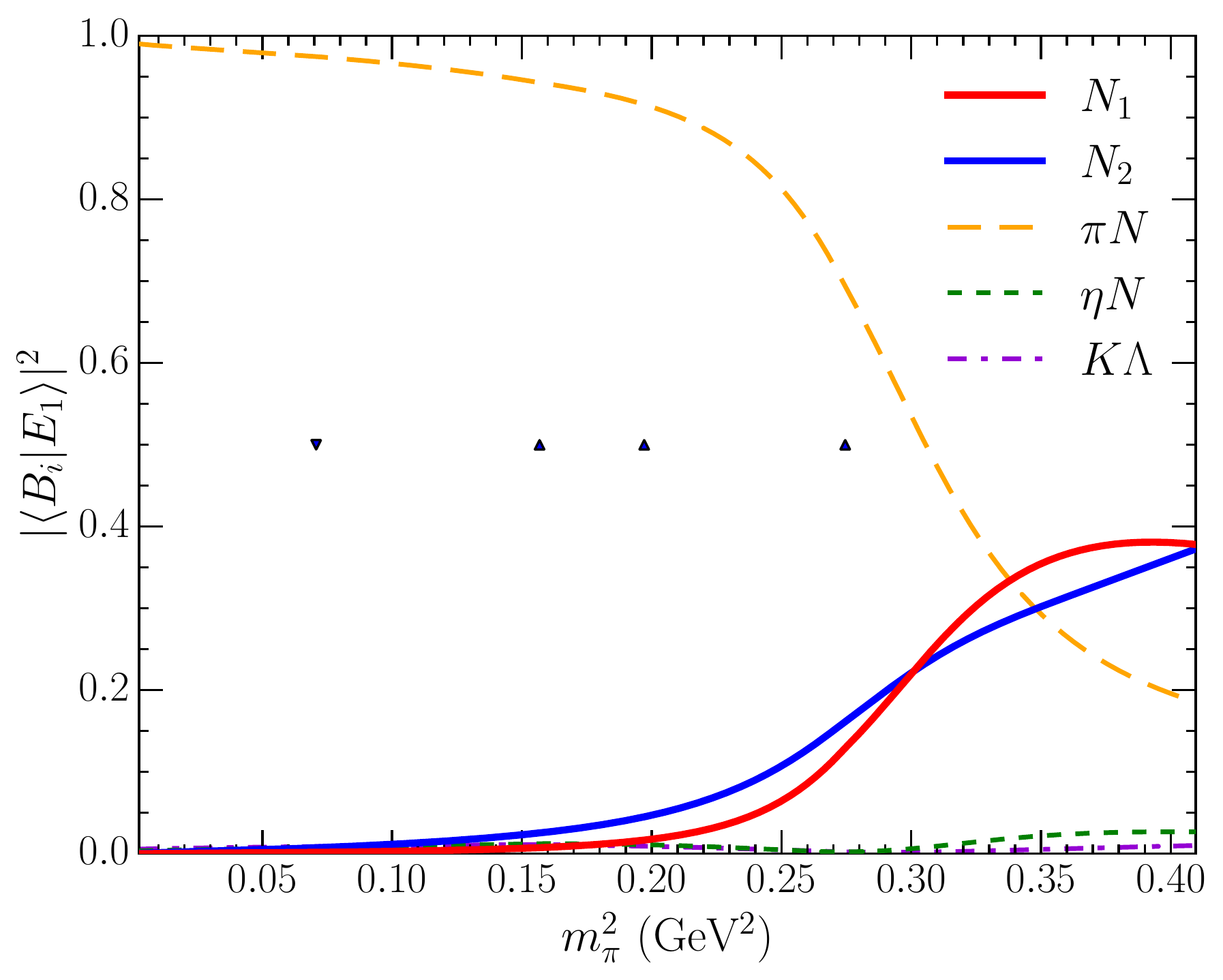}~
  \includegraphics[width=0.34\textwidth]{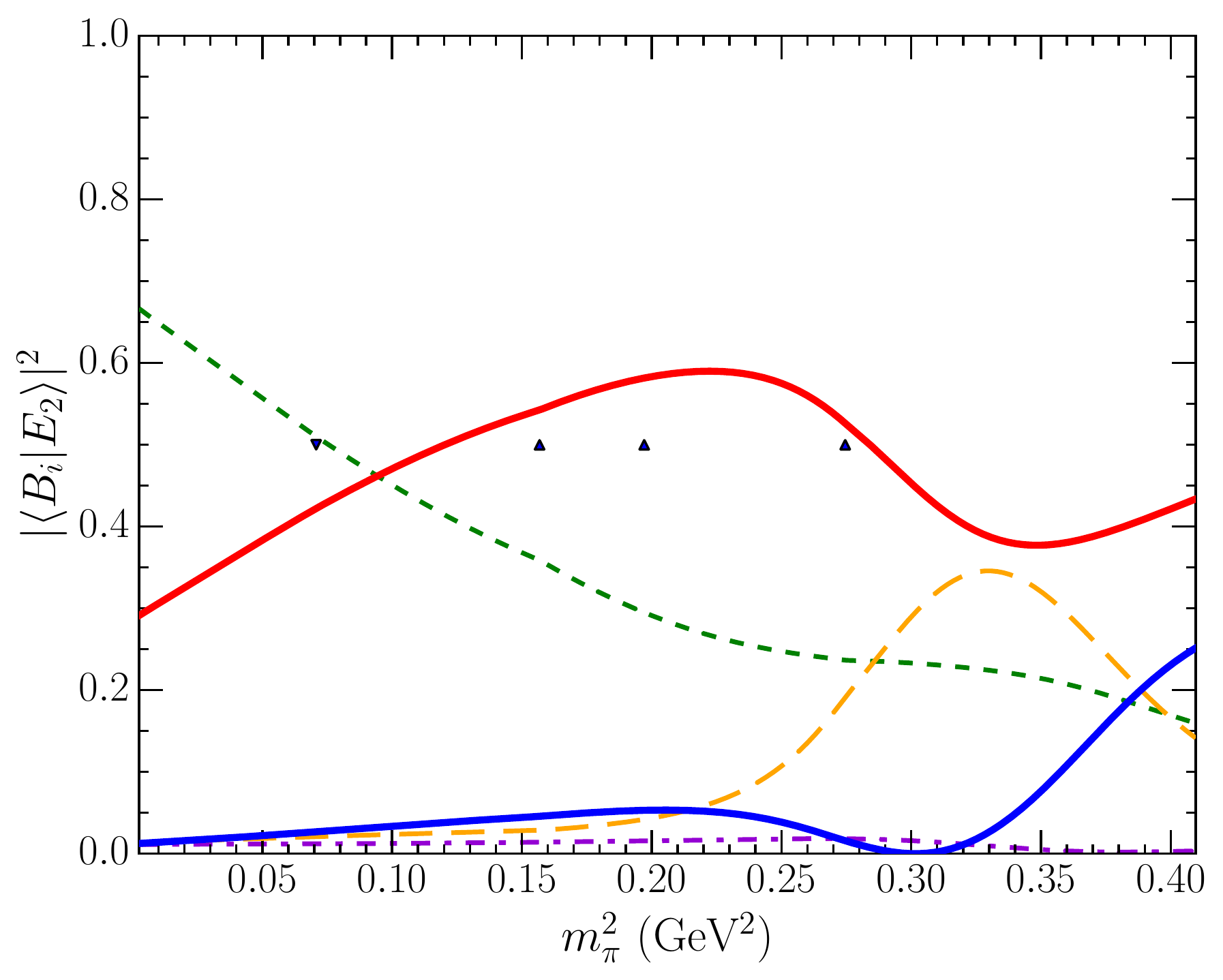}~\\
  \includegraphics[width=0.34\textwidth]{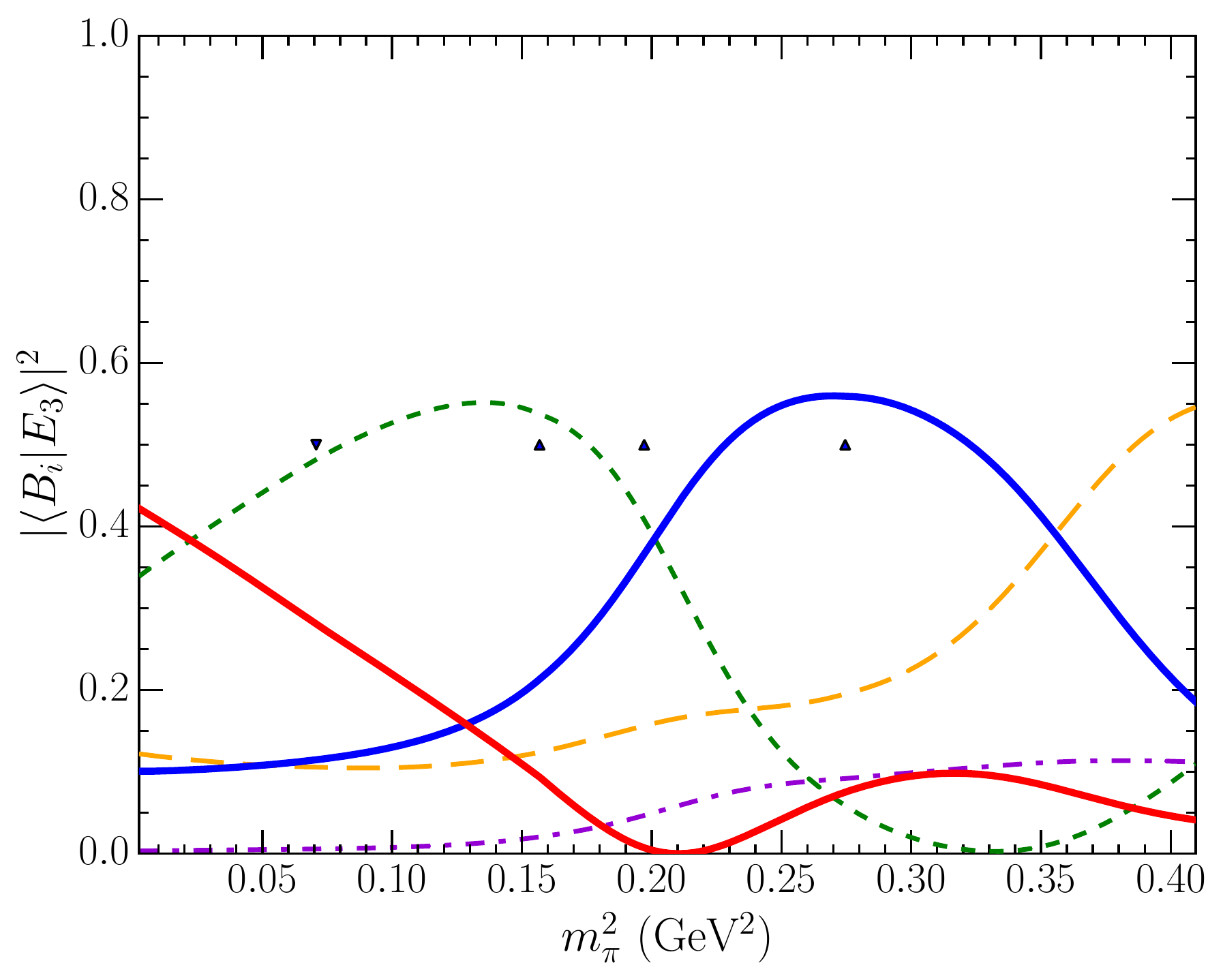}~
  \includegraphics[width=0.34\textwidth]{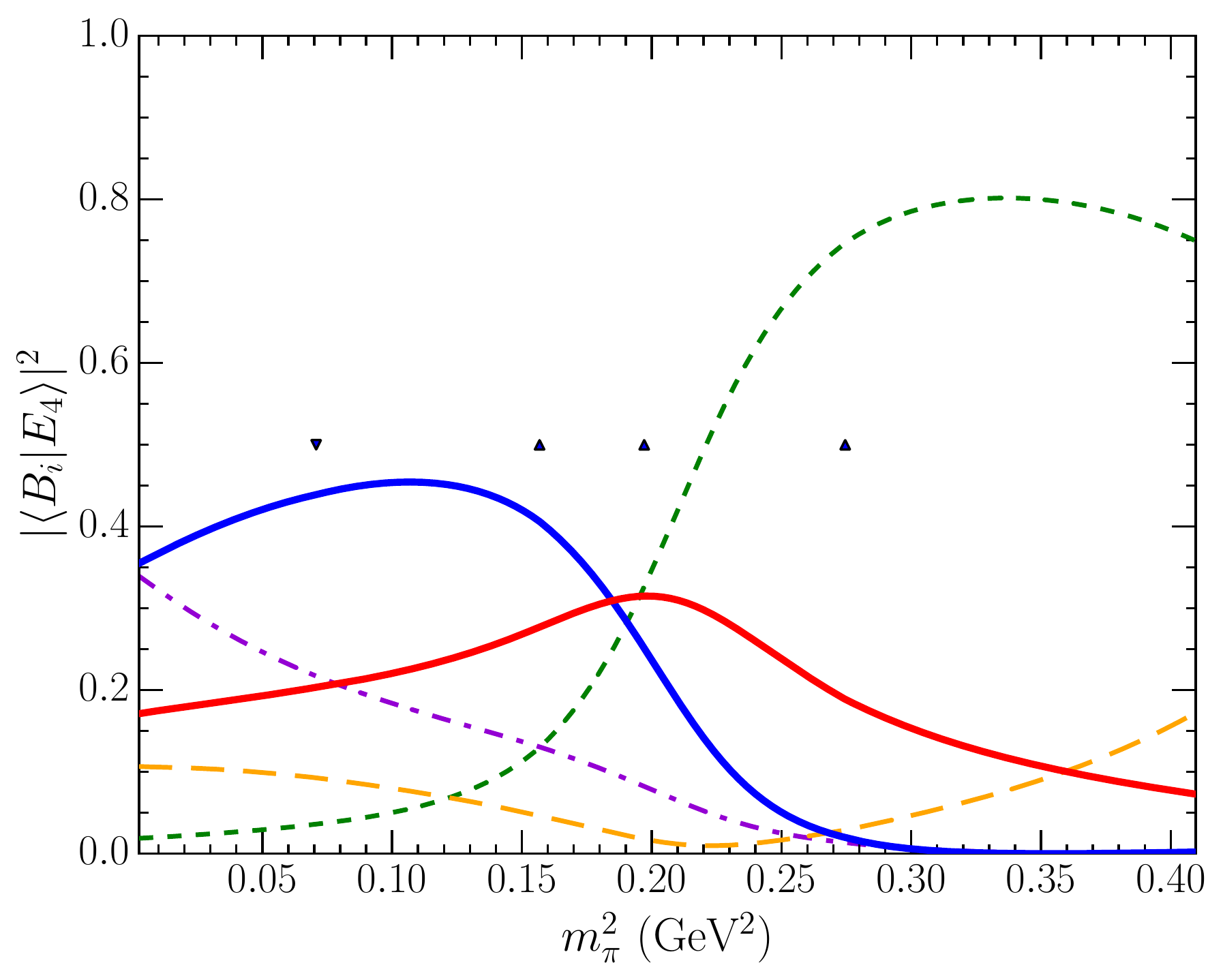}
  \caption{Pion mass dependence of the basis-state contributions for the four lowest eigenstates from the 2 fm spectrum shown in \fref{fig:2b3c_2fm_spectrum}.
    Markers on each plot correspond with the single Lang \& Verduci mass \rref{Lang:2012db} and three HSC masses \rref{Edwards:2011jj,Edwards:2012fx}.
    Contributions for the sum of all momentum states in the $\pi N$, $\eta N$, and $K\Lambda$ channels are illustrated.}
  \label{fig:2b3c_2fm_eigenvectors}
\end{figure*}
Due to the lower density of states, we only consider the eigenvector composition of the four lowest-lying states in \fref{fig:2b3c_2fm_eigenvectors}, as opposed to the six eigenstates in \sref{sec:3fm}.
For the 2 fm spectrum, we observe a similar behaviour in the eigenvectors as in the 3 fm spectrum.
Initially, the lowest-lying state consists almost purely of the $\pi N$ basis state, while the two bare basis states are concentrated in the higher excited states.
As the pion mass increases however, a significantly larger portion of the eigenvectors becomes more concentrated in the lowest-lying state.
As such, at larger pion masses we expect to see lattice QCD states constructed from three-quark operators to correspond with the lower-lying states in the spectrum.

\begin{figure}
  \centering
  \includegraphics[width=0.48\textwidth]{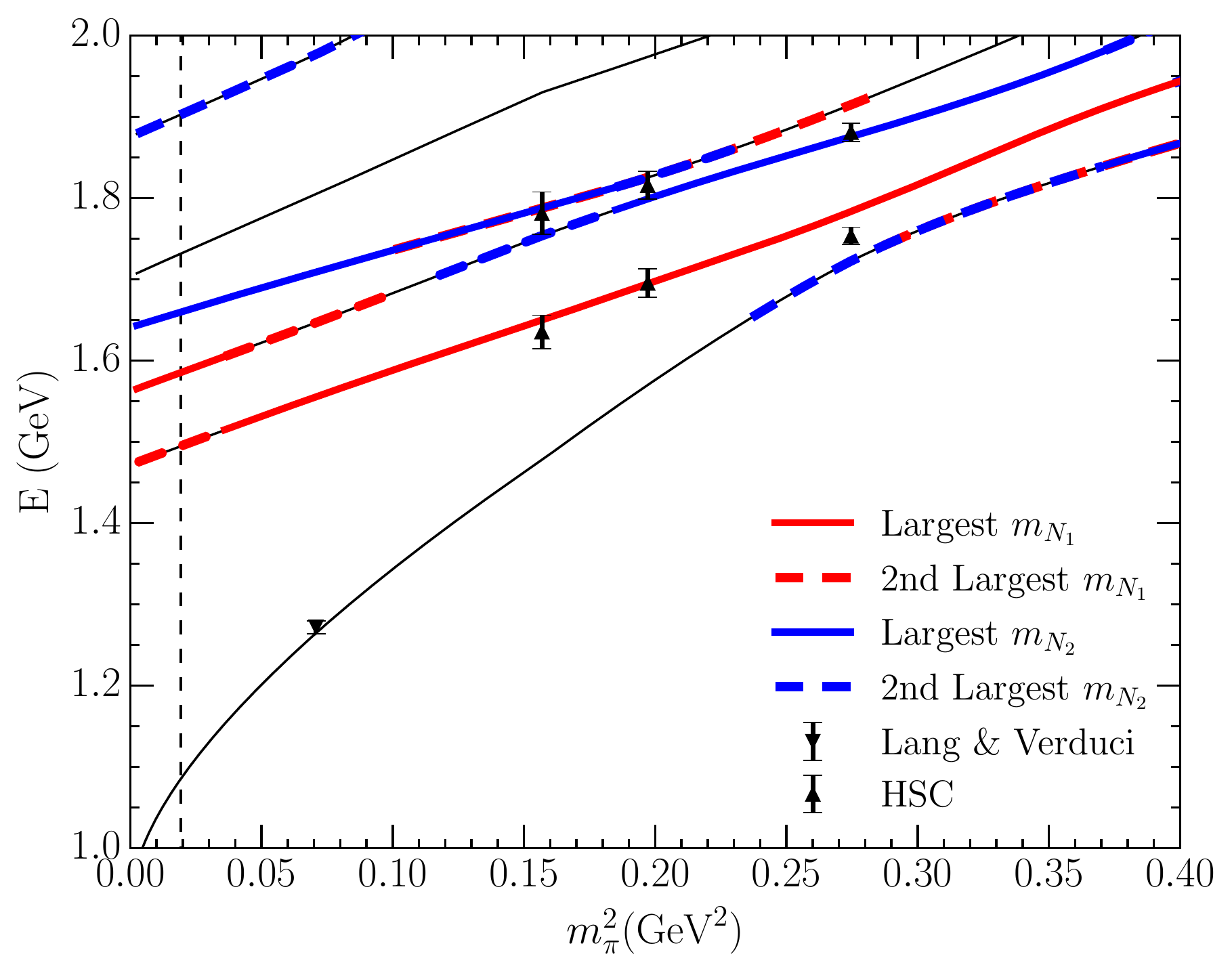}
  \caption{Finite-volume energy spectrum for $L\sim 2$ fm.
    The solid and dashed red lines represent the states with the largest and second largest contributions from the lower bare basis state.
    Similarly the solid and dashed blue liens represent the contributions from the second bare basis state.
    Lattice QCD results from Lang \& Verduci \rref{Lang:2012db} and the Hadron Spectrum Collaboration (HSC) \rref{Edwards:2011jj,Edwards:2012fx}. are overlaid for comparison.}
  \label{fig:2b3c_2fm_spectrum_bare}
\end{figure}
Returning to \fref{fig:2b3c_2fm_spectrum}, it is immediately clear that the lowest lying high-precision point from Lang \& Verduci is very well described by HEFT.
Because this is a low-lying state in the spectrum, it is protected by the L\"uscher relation embedded within the HEFT formalism.
Moreover, because the point is calculated at a relatively small value of the pion mass, it is relatively insensitive to the quark mass interpolation.
In short, this confrontation between lattice QCD and HEFT is also predominantly a confrontation between lattice QCD and experiment.
While this state is composed primarily of the zero momentum $\pi N$ basis state, the other basis state contributions are vital to generating the significant shift in the eigenstate energy down from the non interacting basis-state energy.

To better compare with lattice QCD, we overlay the contributions from these bare states onto the energy spectrum, which can be seen in \fref{fig:2b3c_2fm_spectrum_bare}.
Considering the data from HSC, we observe their six points correspond with states consisting primarily of bare basis states.
The lowest-lying points correspond with the lighter bare state, while their excited states correspond with the eigenstate dominated by the second bare basis state.
This further supports the interpretation of the two odd-parity nucleon resonances as being quark-model like.

It is impressive that five of the six HSC results sit precisely on the HEFT states dominated by bare basis-state components.
It is a testament to the precision of their lattice QCD analysis and the rigour with which HEFT can link different volumes and quark masses within a single formalism.
The notable exception is the lowest-lying state at the largest quark mass where a nearby scattering-state provides a scattering-state contamination in their correlation-matrix analysis.
Of course the authors were completely aware of this possibility and discussed the importance of future calculations including both three-quark interpolators and a complete set of non-local momentum-projected multi-hadron operators.
In \sref{sec:contamination}, a novel HEFT formalism is introduced to quantify the extent of this scattering state contribution.
%
%
%
%
%

\section{Finite-Volume HEFT at 4 \lowercase{fm}} \label{sec:4fm}
Recent lattice QCD calculations of $\pi N$ scattering process were performed by the CLS consortium \rref{Bulava:2022vpq}, and included momentum-projected two-particle interpolating fields.
In particular, we are interested in the zero-momentum $I = 1/2$, $G_{1u}(0)$ results from Fig. 4a of \refref{Bulava:2022vpq}.
These calculations were done for a pion mass of 200 MeV, with a spatial lattice extent of $L = 4.05$ fm.
By altering the nucleon mass at $m_{\pi} = 200$ MeV in the HEFT formalism to $m_{N} = 0.959$ GeV, to match the non-interacting $\pi N(k=0)$ state in Fig. 4a of \refref{Bulava:2022vpq}, we are able to compare the eigenenergies from HEFT with the lattice QCD calculations from the CLS consortium.

\begin{figure}
  \centering
  \includegraphics[width=0.24\textwidth]{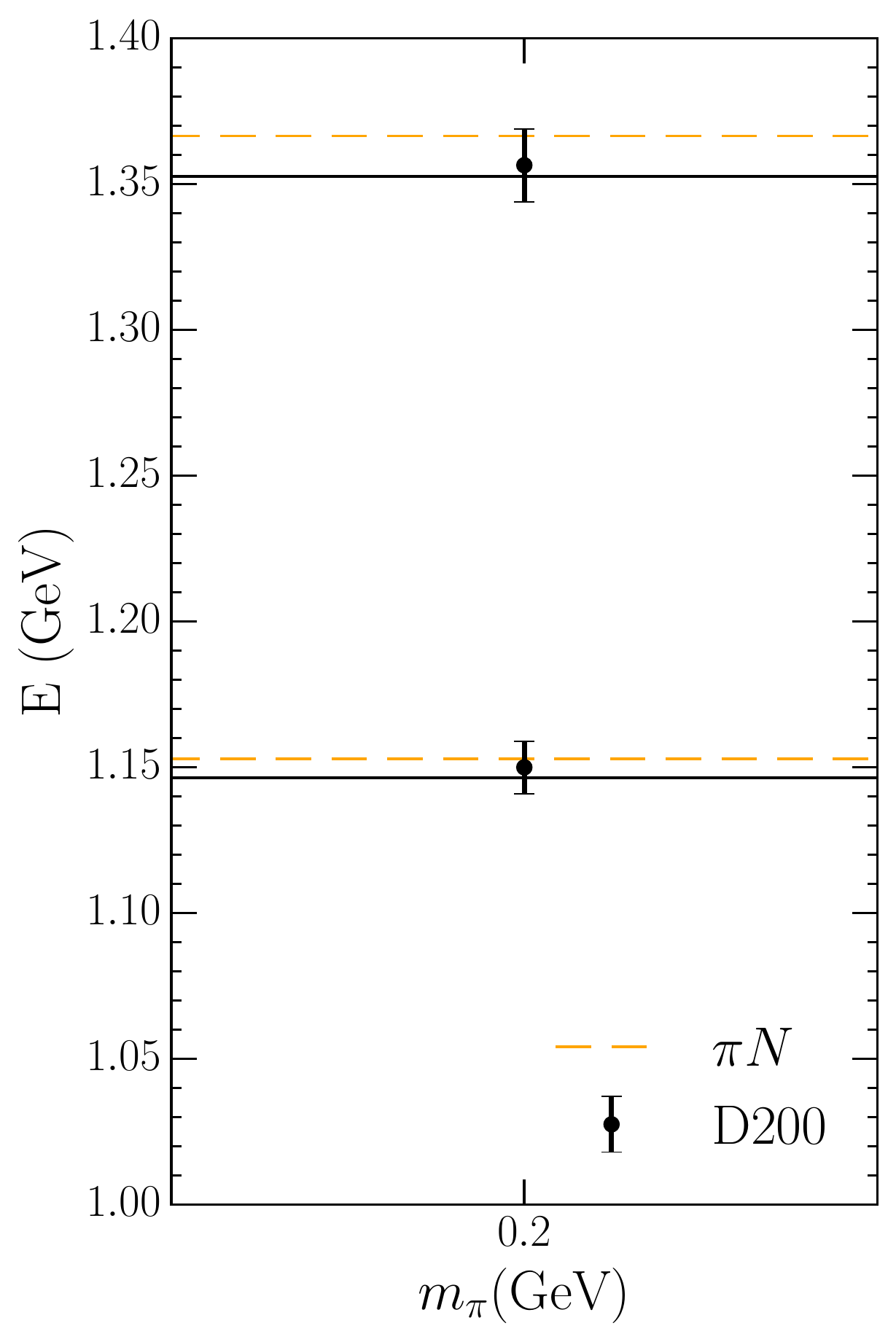}
  \caption{Comparison between the energy eigenvalues calculated in HEFT at a lattice size of $L = 4.05$ fm (solid black lines), and the lattice QCD calculations from the CLS consortium (data points) using the D200 ensemble \rref{Bulava:2022vpq}.
  Dashed lines indicate the non-interacting two-particle $\pi N$ energies for $k=0$ and $k=1$.}
  \label{fig:4fm:EvMpi}
\end{figure}
Using the fit parameters from \sref{sec:inf}, and the bare basis state slopes from \sref{sec:3fm}, this comparison is illustrated in \fref{fig:4fm:EvMpi}.
HEFT predicts a small amount of attraction in the scattering state energies relative to the non-interacting two-particle $\pi N$ basis state energies for $k=0$ and $k=1$.
The CLS results are in excellent agreement with these HEFT predictions, showing effects of a similar magnitude and direction.
Considering the eigenvectors of these two eigenstates from HEFT, the ground state consists of 99.6\% $\pi N(k=0)$, resulting in the minimal shift away from the non-interacting state.
Similarly, 96\% the first excited state is from the $\pi N(k=1)$ state, with a majority of the remaining contributions coming from the two bare states.
As the majority of the lower mass bare state is concentrated in the fourth excited state, the bare basis states have a limited impact on the eigenstates considered in this energy range.
Nonetheless, it is useful to see that the HEFT formalism correctly extends to the $L = 4.05$ fm results from the CLS consortium, as in the $L \sim 2$ fm case in \sref{sec:2fm}

\section{Scattering-State Contaminations in Lattice Correlation Functions} \label{sec:contamination}
\subsection{Contamination Function Formalism} \label{sec:contamination:formalism}
In this section we introduce a novel HEFT formalism for estimating scattering-state contaminations in lattice QCD correlation functions constructed with standard three-quark operators.
The analysis draws on the extensive information available in the finite-volume eigenvectors of the Hamiltonian.

We commence with the consideration of a single bare basis state.
Given a three-quark operator $\chi(\vect x, t)$ with quantum numbers corresponding to a baryonic state of interest, the correlation function \rref{Michael:1985ne,Luscher:1990ck} is given by
\begin{align}
  G_{\chi}(t, \vect p) &= \sum_{\vect x} e^{-i\vect p\cdot \vect x} \braket{\bar{\Omega} | \chi(\vect x,t)\, \bar{\chi}(0,0) | \Omega}\,, \nonumber \\
  G_{\chi}(t) &= \sum_{i} \left| \braket{\Omega | \chi | E_i} \right|^2 e^{-E_i t}\,, \label{eq:lqcd_corr}
\end{align}
where a complete set of energy eigenstates $\mathbb{I} = \sum_{i} \ket{E_{i}}\bra{E_{i}}$ has been introduced, and zero momentum has been taken.

In \refref{Bar:2016jof}, B\"ar and coworkers provided a $\chi$PT estimate of the coupling between a smeared nucleon interpolating field and a non-interacting pion-nucleon basis state as
\begin{equation}
  \frac{3}{16}\, \frac{1}{(f_{\pi}\,L)^2\, E_{\pi}\,L}\left( \frac{E_{N} - m_{N}}{E_{N}} \right) \approx 10^{-3}\,,
\end{equation}
where $E_{\pi}$ and $E_{N}$ are on-shell pion and nucleon energies.
The numerical estimate is based on a 3 fm lattice and the lowest nontrivial momentum contribution where the coupling is largest.
Here the $1/L^{3}$ dependence of the coupling is manifest as the non-interacting two-particle momentum state is spread uniformly throughout the lattice volume.

Noting the small magnitude of the overlap between the local interpolating field and the two-particle basis states, one concludes that the state excited by the local interpolating field is the only local state in the Hamiltonian basis, the bare baryon basis state.
As such, we associate the three-quark nucleon interpolating field $\bar{\chi}$ acting on the nontrivial QCD vacuum, $\ket{\Omega}$, with the bare basis state of HEFT, via $\bar{\chi}(0) \ket{\Omega} = \ket{B_0}$.
Inserting this into \eref{eq:lqcd_corr} gives
\begin{equation}
  G_{B_0}(t) = \sum_{i} \left| \braket{ B_0 | E_i} \right|^2 e^{-E_i t}\,.
\end{equation}

Drawing on the eigenvector components $\braket{B_{0} | E_{i}}$, and eigenenergies $E_{i}$ of HEFT, we can simulate the scattering-state contaminations in lattice QCD correlation functions.
We define the ``contamination function'' $C_{B_0}(t)$ for the bare basis state $\ket{B_0}$ as
\begin{equation}
  C_{B_0}(t) = \frac{1}{ G_{B_0}(t) }\, \sum_{i \neq B_0} \left| \braket{ B_0 | E_i} \right|^2 e^{-E_i t}\,,
\end{equation}
where the sum over all $i \neq B_0$ is considering all energy eigenstates, barring the eigenstate with the largest contribution from the bare state.
We label this eigenstate $\ket{E_{B_0}}$.
If this eigenstate is the ground state, for sufficiently large Euclidean time evolution the contamination function will tend to zero, where all excited states have exponentially decayed through the Euclidean time evolution.
If $\ket{E_{B_0}}$ is not the ground state, we expect a minimum in the contamination function at some time, where the state has the least scattering-state contamination, before becoming completely dominated by the lowest-lying scattering state.

We may extend these definitions to a system with two bare basis states.
This time, states are excited from the vacuum with three-quark operators $\chi_{1}$ and $\chi_{2}$.
For example, the odd-parity proton interpolators
$\chi_1 = \epsilon^{abc}\, \left( u^{Ta}\, C\gamma_5\, d^b \right) \gamma_5\, u^c$ and $\chi_2 = \epsilon^{abc}\, \left ( u^{Ta}\, C\, d^b \right ) \, u^c$ are both ${\mathcal{O}}(p/E)$ in a nonrelativistic reduction and mix strongly in a correlation matrix analysis to isolate the eigenstates.
Each of these interpolating fields acting on the QCD vacuum
will create a bare basis state $\ket{N_{1}}$ and $\ket{N_{2}}$,
\begin{equation}
  \left( \alpha^{*}\,\bar{\chi_{1}} + \beta^{*}\,\bar{\chi_{2}} \right)\ket{\Omega} = \alpha^{*}\ket{N_{1}} + \beta^{*}\ket{N_{2}}\,,
  \label{eq:corr2}
\end{equation}
which are mixed in forming the energy eigenstates, $\ket{E_{i}}$.
Where previously we had a single eigenstate with largest bare state contribution, labelled $\ket{E_{B_0}}$, in the two bare state system there will be a corresponding Hamiltonian eigenstate for each bare state.
We label these states $\ket{E_{N_1}}$ and $\ket{E_{N_2}}$.
As these eigenstates are a mixture of each bare state, they will be constructed for different combinations of $\alpha$ and $\beta$, given by $\alpha_{j}$ and $\beta_{j}$, where $j = 1, 2$ corresponds with $N_1$ and $N_2$ respectively.
With this in mind, correlation functions optimised for these two eigenstates are constructed as
\begin{align}
  G_{j}(\vect{p}, t) &= \sum_{\vect{x}} e^{-i\vect{p}\cdot\vect{x}}\, \bra{ \bar{\Omega} } \left( \alpha_{j}\, \chi_{1}(\vect{x},t) + \beta_{j}\, \chi_{2}(\vect{x},t) \right) \nonumber\\
                    &\qquad\qquad\qquad\qquad \times \left( \alpha_{j}^{*}\, \bar{\chi}_{1}(0) + \beta_{j}^{*}\, \bar{\chi}_{2}(0) \right) \ket{\Omega}\,. \label{eq:contamination:Gt2}
\end{align}
Inserting a complete set of states, setting $\vect{p} = 0$, and applying \eref{eq:corr2},
\begin{align}
  G_{j}(t) &= \sum_{i} \left( \alpha_{j}\bra{N_1} + \beta_{j}\bra{N_2} \right) \ket{E_i}\bra{E_i} \nonumber\\
           &\qquad\qquad\qquad\quad \times \left( \alpha_{j}^{*}\ket{N_1} + \beta_{j}^{*}\ket{N_2} \right)\, e^{-E_{i} t}\,, \nonumber\\
           &= \sum_{i} \left| \alpha_{j}\braket{N_1 | E_i} + \beta_{j}\braket{N_2 | E_i} \right|^{2}\, e^{-E_{i} t}\,. \label{eq:contamination:Gt2_zero_mom}
\end{align}
We note that $\alpha_j$ and $\beta_j$ can be made real \rref{Leinweber:1990dv}, and the eigenvector components $\braket{N_{j} | E_{i}}$ are real.

The mixing parameters for each of the two eigenstates, labelled $\alpha_j$ and $\beta_j$, may be obtained either through the eigenvectors of correlation matrices from lattice QCD, or through the Hamiltonian eigenvectors from HEFT.
Importantly however, the lattice QCD correlation matrix eigenvectors must be normalised to $\mathcal{O}(1)$, as described in \refref{Mahbub:2013ala}.
In the case of HEFT, the eigenvector components are ${\mathcal{O}}(1)$ via the standard normalisation with the sum of the squares of the components equal to one.
Given that strength is localised within the spectrum, the values are insensitive to the size of the Hamiltonian matrix.

The scattering-state contamination to each of the eigenstate-optimised
correlation functions of Eqs. (\ref{eq:contamination:Gt2}) and (\ref{eq:contamination:Gt2_zero_mom}) is obtained by removing the two energy eigenstates whose composition is dominated by the bare basis states (labelled $\ket{E_{N_1}}$ and $\ket{E_{N_2}}$).
The idea is that the lattice correlation matrix will be effective
in isolating two states which couple strongly to the three-quark operators, but lacks the additional information to isolate the scattering states.
While the lattice QCD calculations of \refref{Mahbub:2013ala} isolate states in an $8\times 8$ correlation matrix, appropriate orthogonality is evident in the optimised correlation function for each state, $G_i(t)$.
For example, the contribution of $\ket{E_{N_2}}$ to the optimised correlator $G_{1}(t)$, governed by $\alpha_{1}\braket{N_{1} | E_{N_{2}}} + \beta_{1}\braket{N_{2} | E_{N_{2}}}$, is small.
Similarly the contribution of $E_{N_1}$ to the optimised correlator $G_{2}(t)$, governed by $\alpha_{2}\braket{N_{1} | E_{N_{1}}} + \beta_{2}\braket{N_{2} | E_{N_{1}}}$, is also small.
At most, contributions from $\ket{E_{N_2}}$ and $\ket{E_{N_2}}$ to $G_{1}(t)$ and $G_{2}(t)$ respectively are of order 5\%, though typically take values closer to 1\%.

The optimised contamination functions for these two bare-dominated states are therefore written as
\begin{equation} \label{eq:contamination:Ct}
  C_{j}(t) = \frac{1}{G_{j}(t)} \sum_{i \neq N_1,N_2} \left( \alpha_{j}\braket{N_1 | E_i} + \beta_{j}\braket{N_2 | E_i} \right)^2 e^{-E_{i} t} \,.
\end{equation}
Here, the notation of $i \neq N_{1}, N_{2}$ denotes that we avoid summing over the energy eigenstates labelled $\ket{E_{N_{1}}}$ and $\ket{E_{N_2}}$.
%
%

\subsection{Contamination Function at 3 fm} \label{sec:contamination:3fm}
\subsubsection{Two Particle Scattering-State Contamination}
To determine the degree of scattering-state contamination in the lattice QCD correlation functions of \eref{eq:contamination:Gt2_zero_mom}, which have been optimised for the states $\ket{E_{N_{1}}}$ and $\ket{E_{N_2}}$, we consider the contamination functions as defined in \eref{eq:contamination:Ct}, eliminating the contribution from the states which are identified as corresponding to the lattice QCD results.

As can be seen in \fref{fig:2b3c_3fm_spectrum_bare}, at each lattice QCD mass there is not necessarily only a single corresponding HEFT eigenstate.
Taking the second heaviest mass from \fref{fig:2b3c_3fm_spectrum_bare} as an example, we see that both the first and second states have approximately equal contributions from $m_{N_1}$.
Indeed, there is no single eigenstate corresponding with the single-particle, three-quark core, but rather both eigenstates may be described as quark-model like, and corresponding with the lattice QCD state associated with the $N^{*}(1535)$.
For the $L \sim 3$ fm analysis, this effect can be seen at all but the heaviest lattice QCD masses.
As a result, to remove the bare basis state contributions from the correlation functions as described in the previous section, we must remove not only the contribution from the two eigenstates with largest bare basis state eigenvector components, but also the contribution from the two eigenstates with second largest bare basis state components.
In the context of \fref{fig:2b3c_3fm_spectrum_bare}, we remove the contributions from all highlighted eigenstates from the correlation functions.
This method will allow a proper determination of the level of two-particle dominated scattering-state contributions, having removed all significant sources of single-particle contributions.

In calculating these contamination functions, we compare two sources of values for $\alpha_j$ and $\beta_j$.
From \refref{Mahbub:2013ala}, the eigenvectors of the correlation matrix were calculated in lattice QCD for an $8 \times 8$ correlation matrix, with four sets of smearings at both the source and sink.
Here, we consider the 100 sweep smearings from Fig. 11a of \refref{Mahbub:2013ala}, which dominate the eigenvalue components.
Coefficients for $\alpha_1$ and $\alpha_2$ are taken from the 100-sweep $\chi_1$ ($u_5$) component of the eigenvectors for states 1 and 2 respectively.
Similarly, $\beta_1$ and $\beta_2$ are taken from the 100-sweep $\chi_2$ ($u_6$) component of the eigenvectors for states  1 and 2.
We note the important sign change in $\beta_{i}$ as one moves from state 1 to 2.

We compare these lattice QCD results for $\alpha_j$ and $\beta_j$ with the corresponding quantities calculated from the eigenvectors of the Hamiltonian in HEFT.
In this case, these mixing factors are given by
\begin{align} \label{eq:HEFT_alpha_beta}
  \alpha_1 = \braket{N_1 | E_{N_1}},\quad & \beta_1 = \braket{N_2 | E_{N_1}}, \nonumber\\
  \alpha_2 = \braket{N_1 | E_{N_2}},\quad & \beta_2 = \braket{N_2 | E_{N_2}}.
\end{align}
The eigenstates $\ket{E_{N_1}}$ and $\ket{E_{N_2}}$ correspond with the states illustrated in \fref{fig:2b3c_3fm_spectrum_bare} with solid red and solid blue lines respectively.
\begin{figure*}
  \centering
  \includegraphics[width=0.24\linewidth]{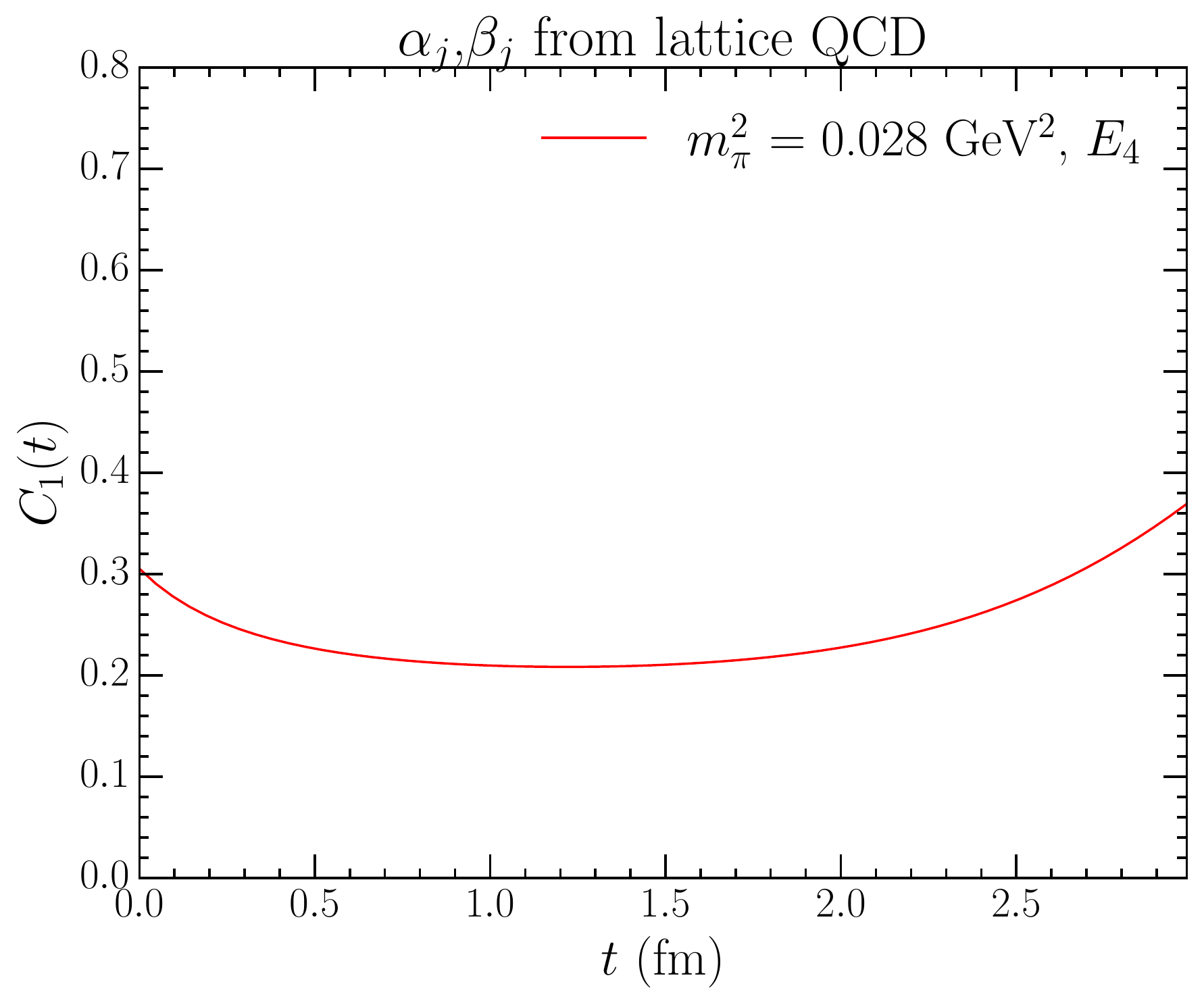}~
  \includegraphics[width=0.24\linewidth]{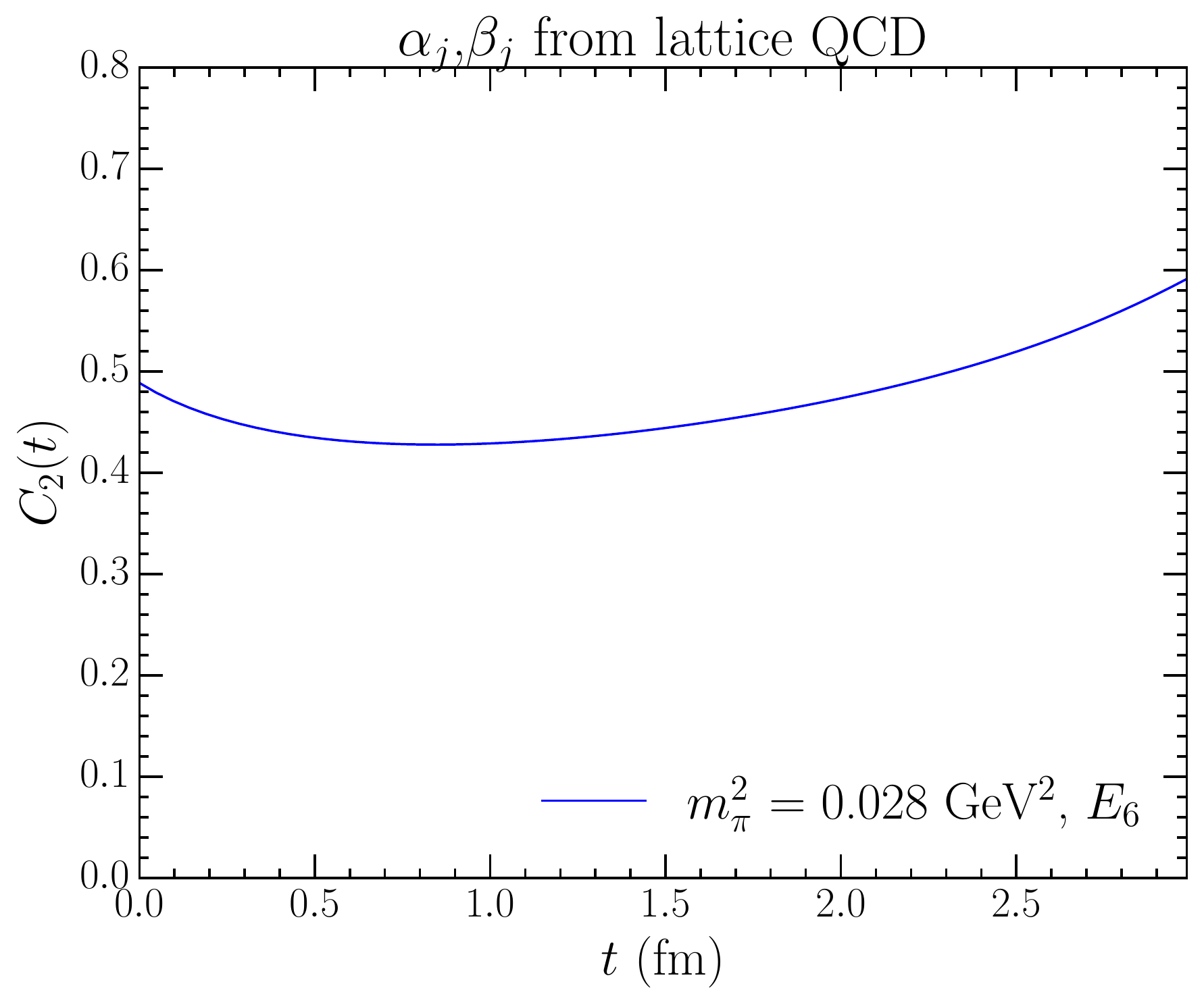}~
  \includegraphics[width=0.24\linewidth]{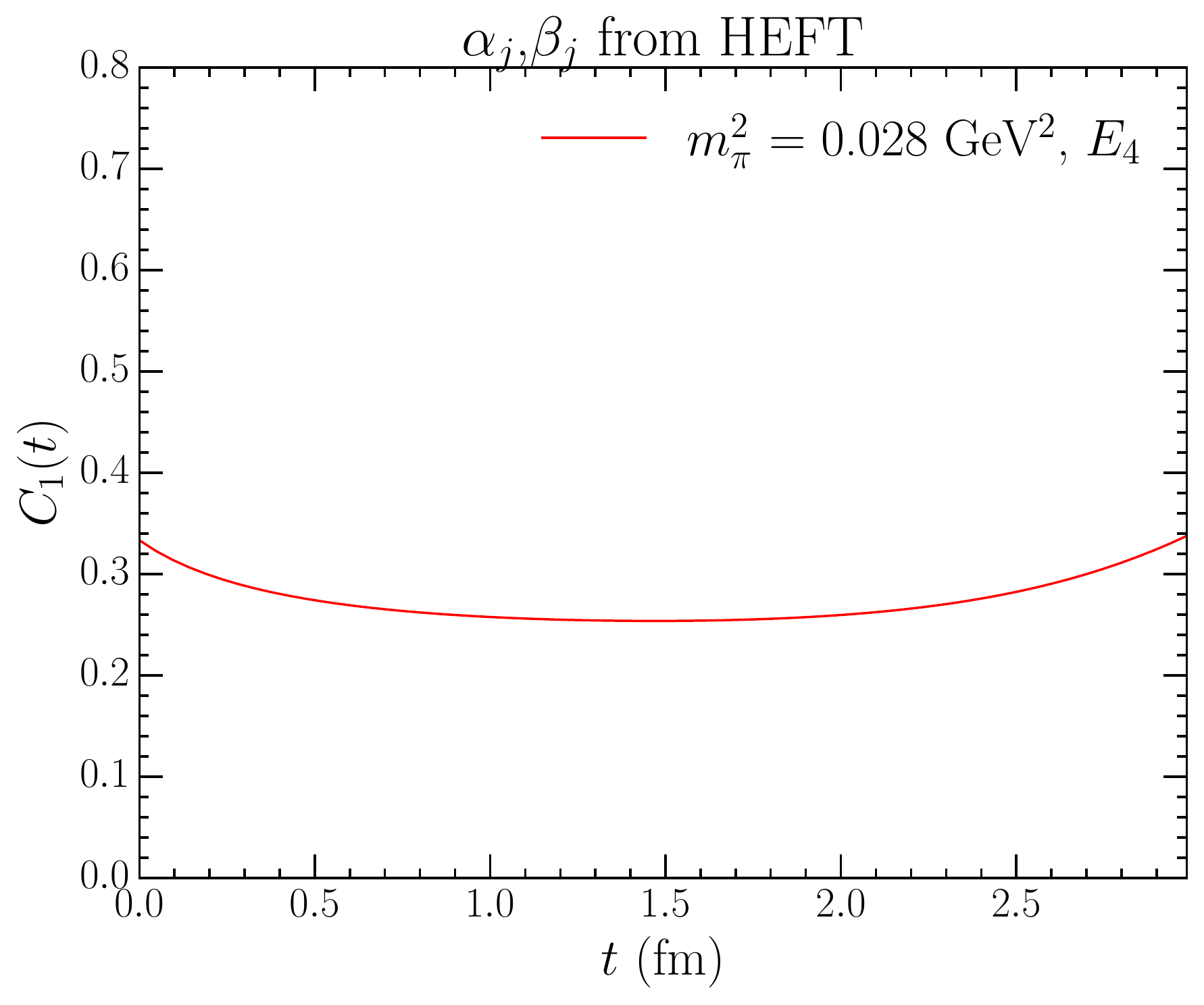}~
  \includegraphics[width=0.24\linewidth]{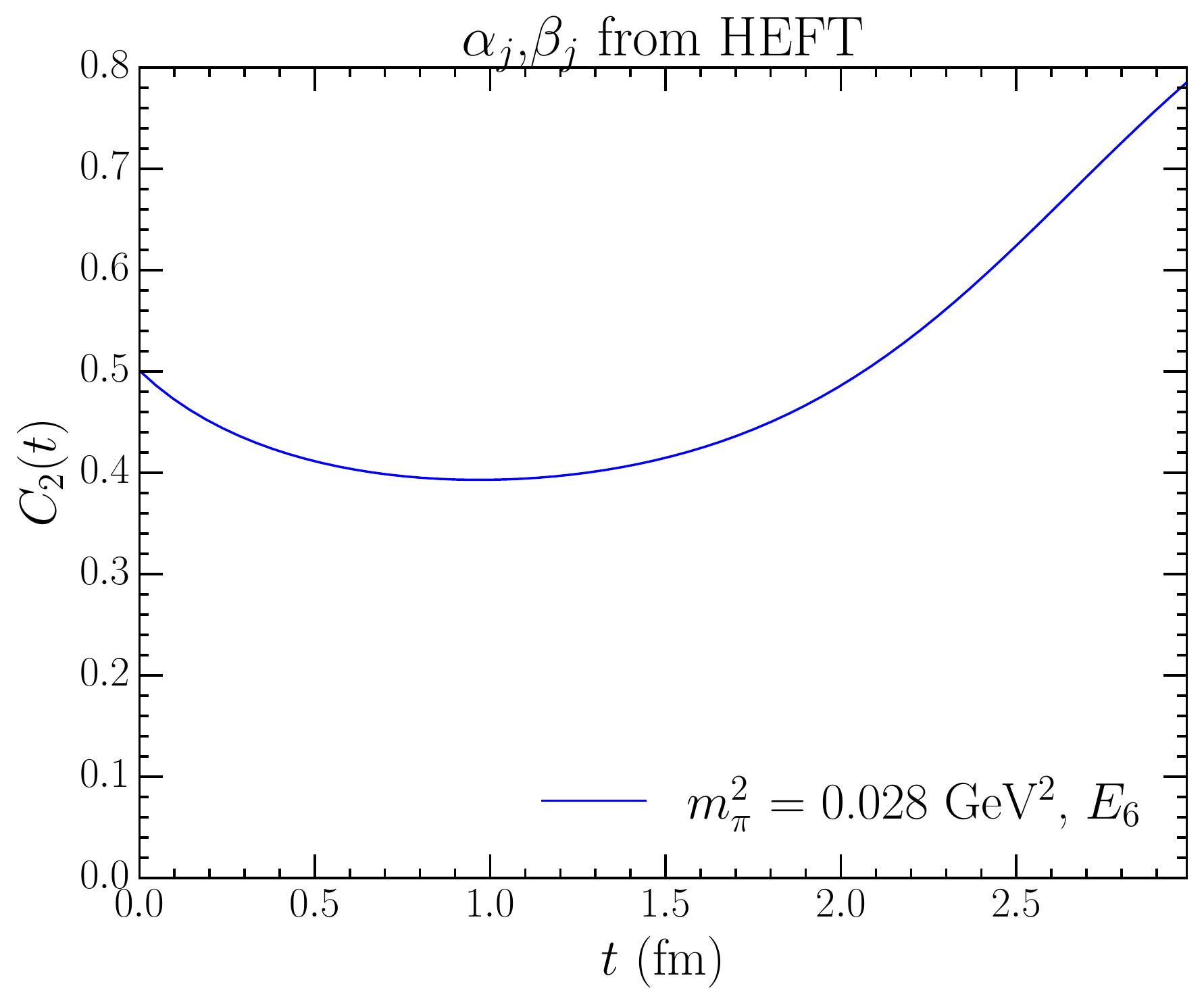}~\\
  \includegraphics[width=0.24\linewidth]{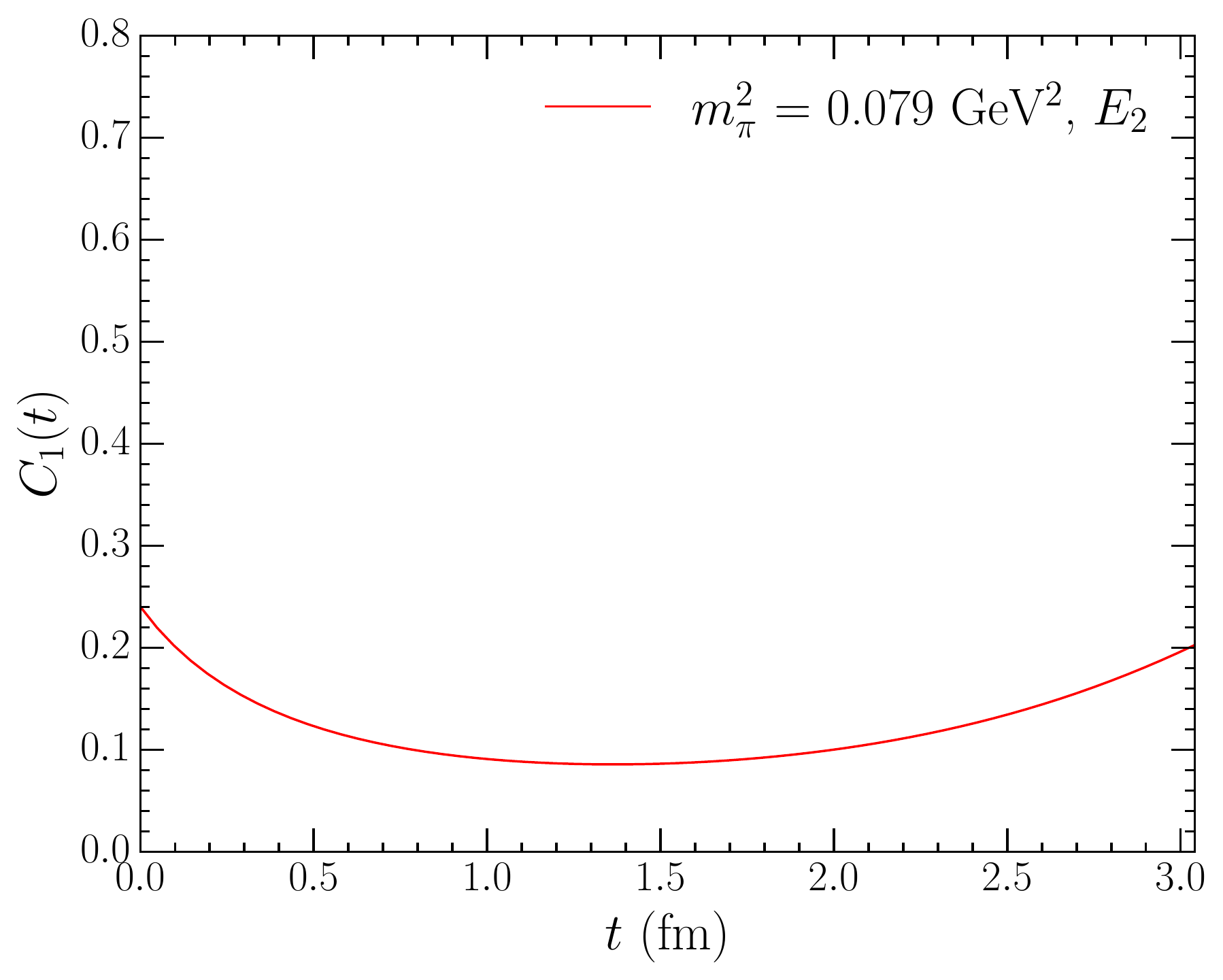}~
  \includegraphics[width=0.24\linewidth]{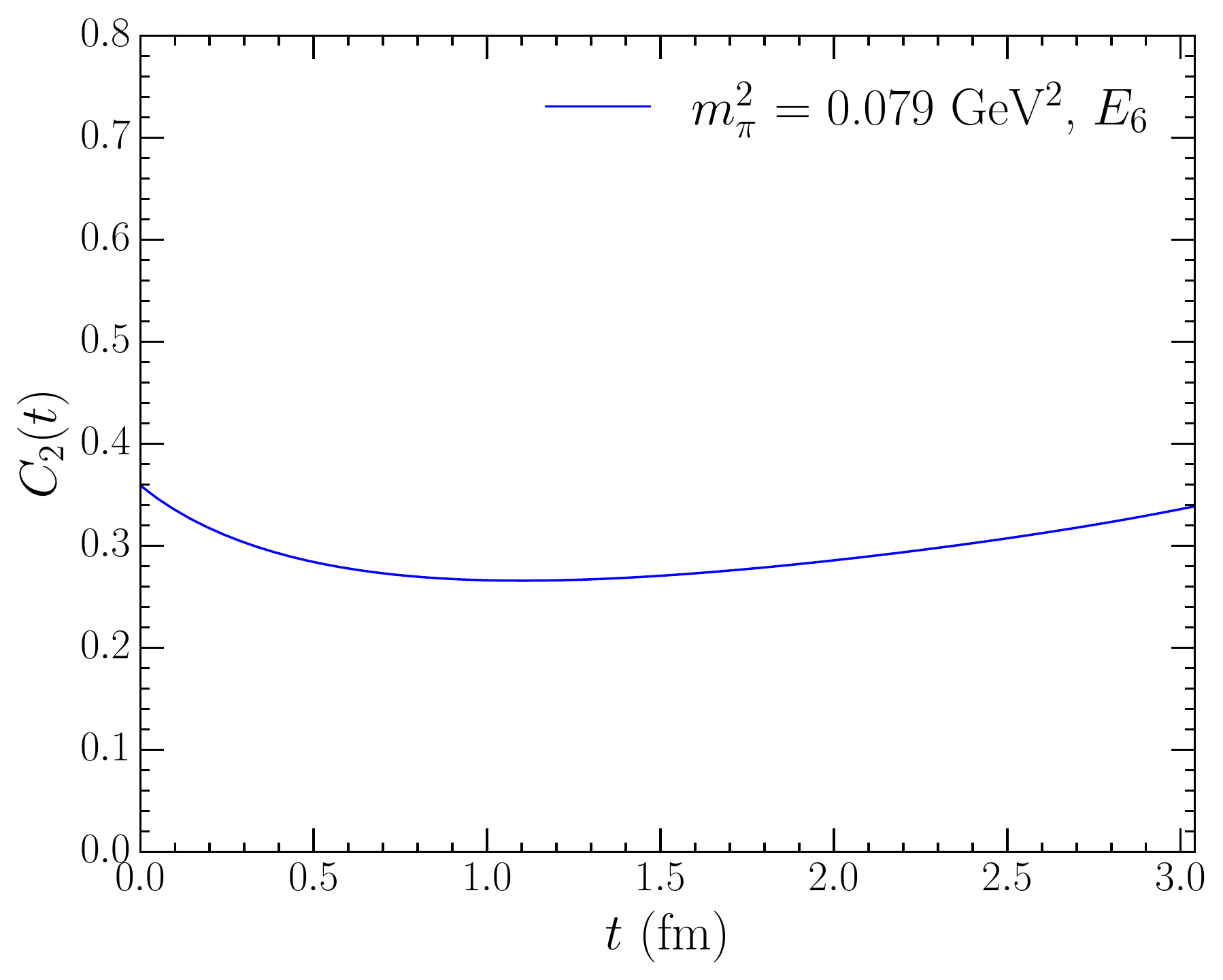}~
  \includegraphics[width=0.24\linewidth]{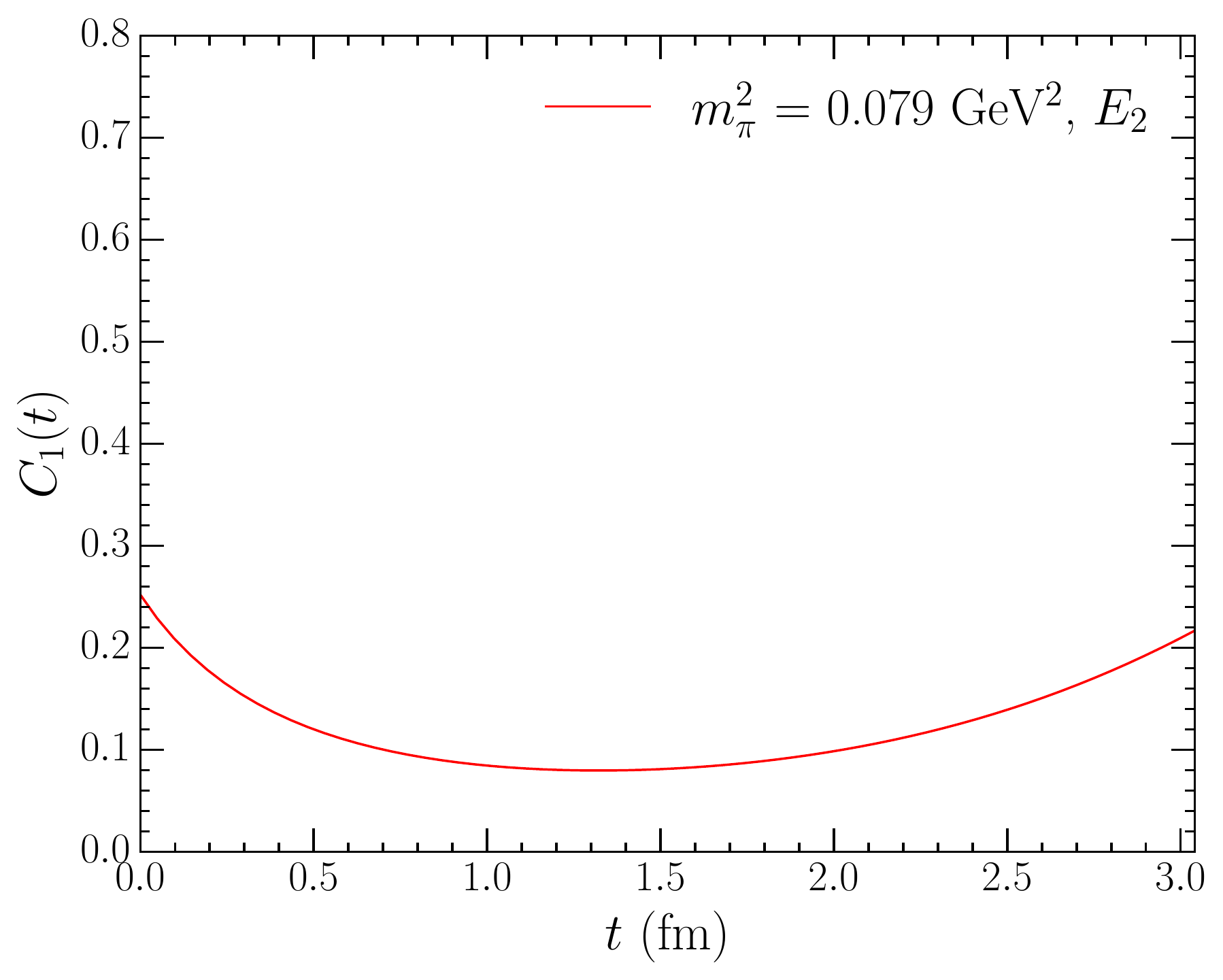}~
  \includegraphics[width=0.24\linewidth]{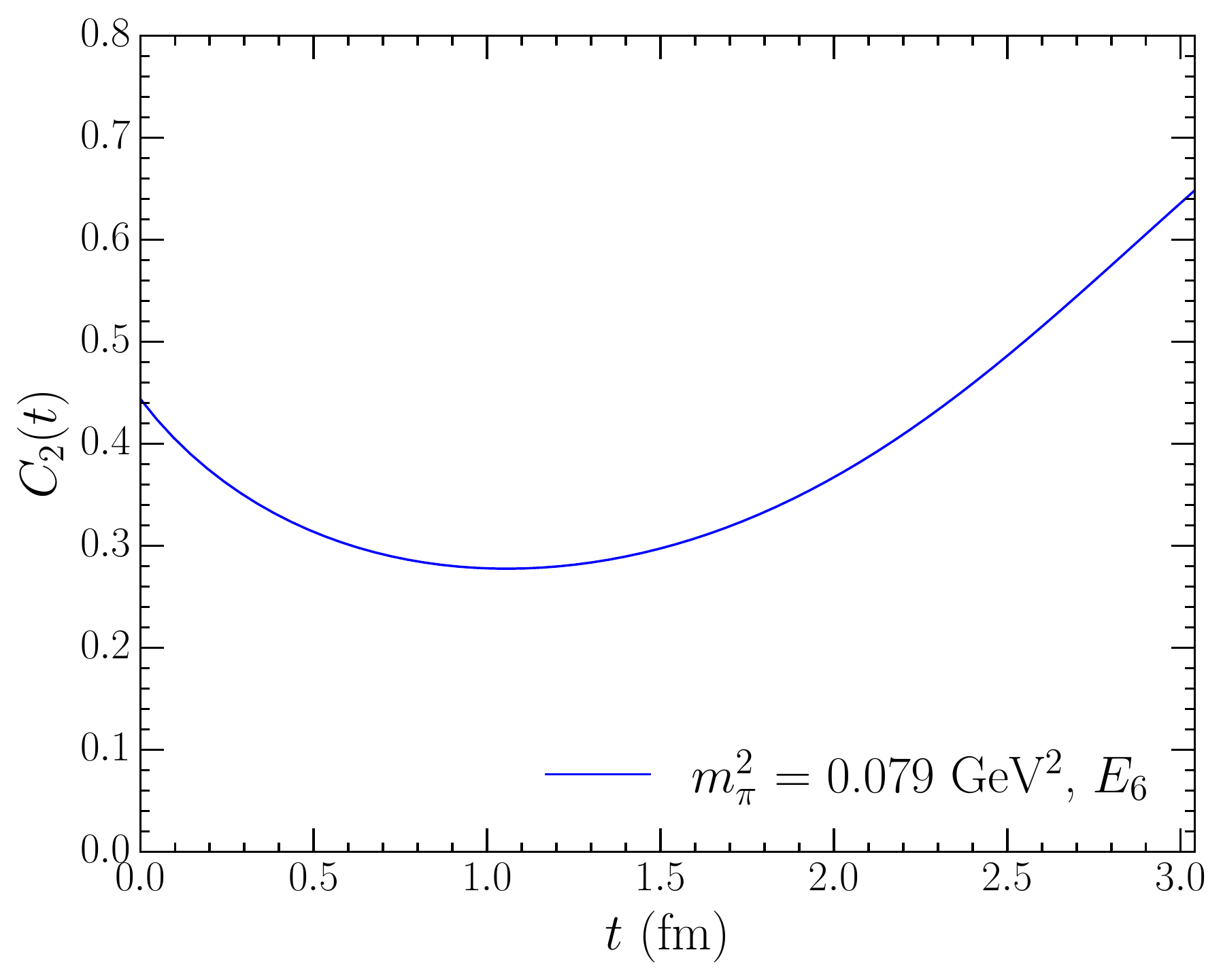}~\\
  \includegraphics[width=0.24\linewidth]{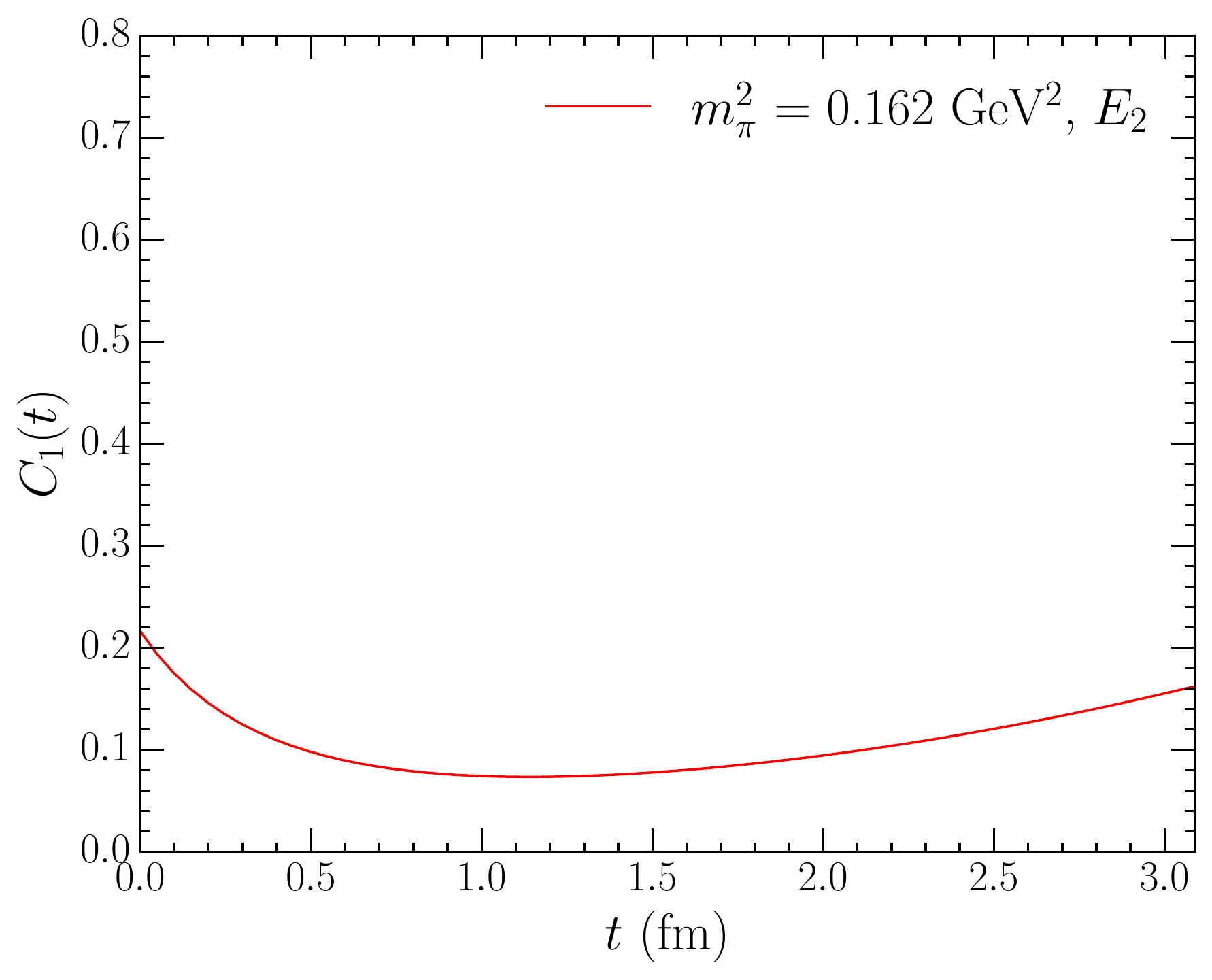}~
  \includegraphics[width=0.24\linewidth]{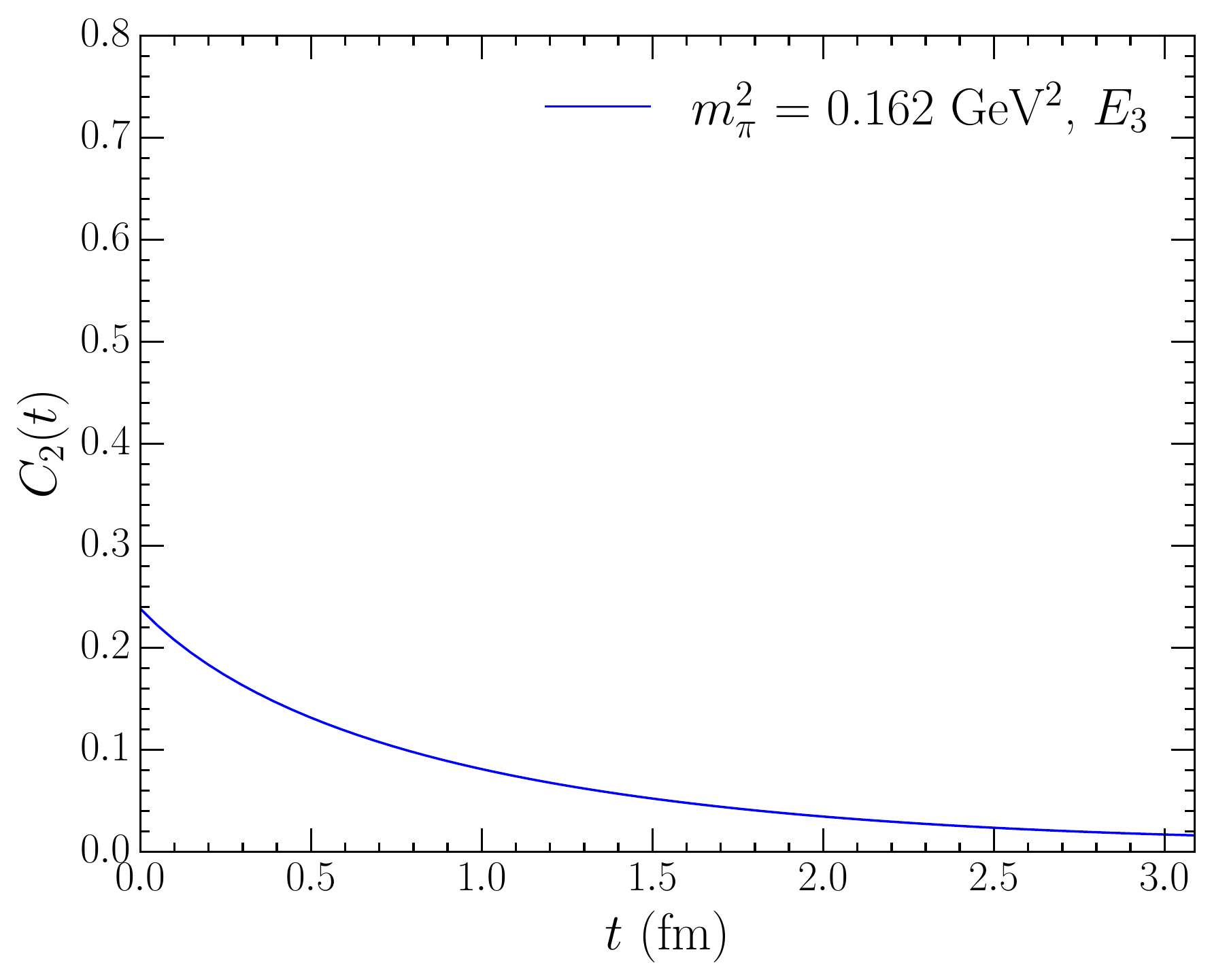}~
  \includegraphics[width=0.24\linewidth]{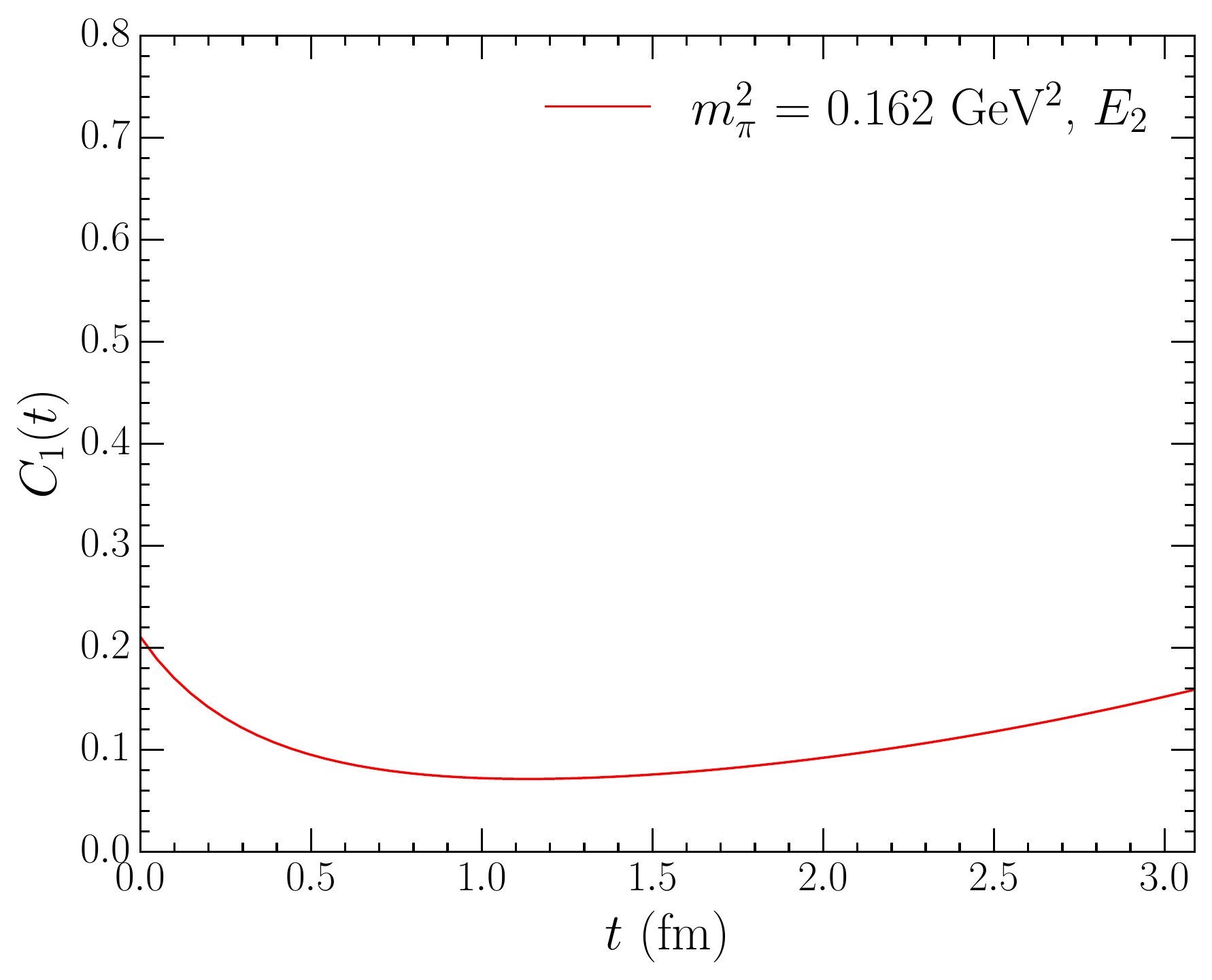}~
  \includegraphics[width=0.24\linewidth]{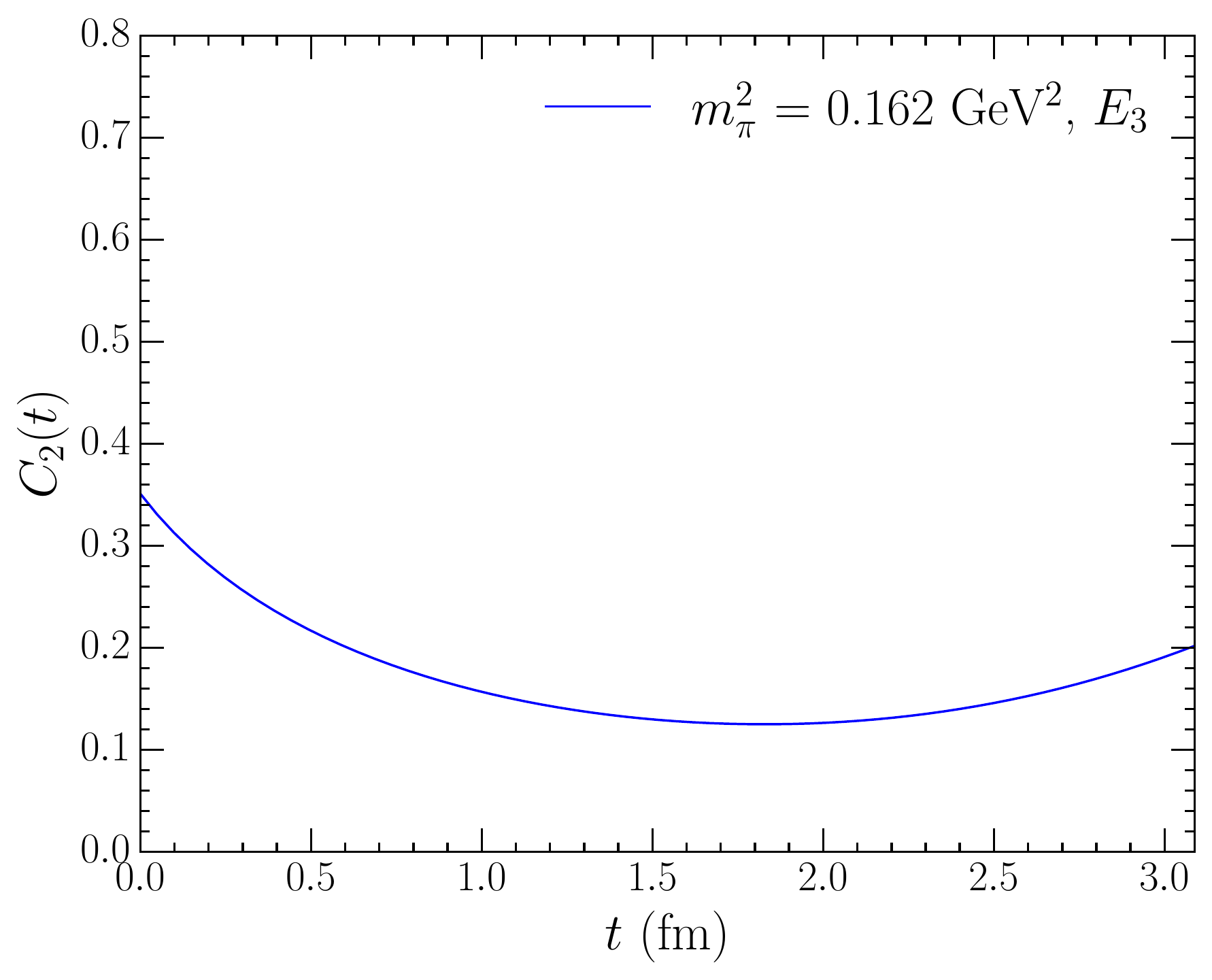}~\\
  \includegraphics[width=0.24\linewidth]{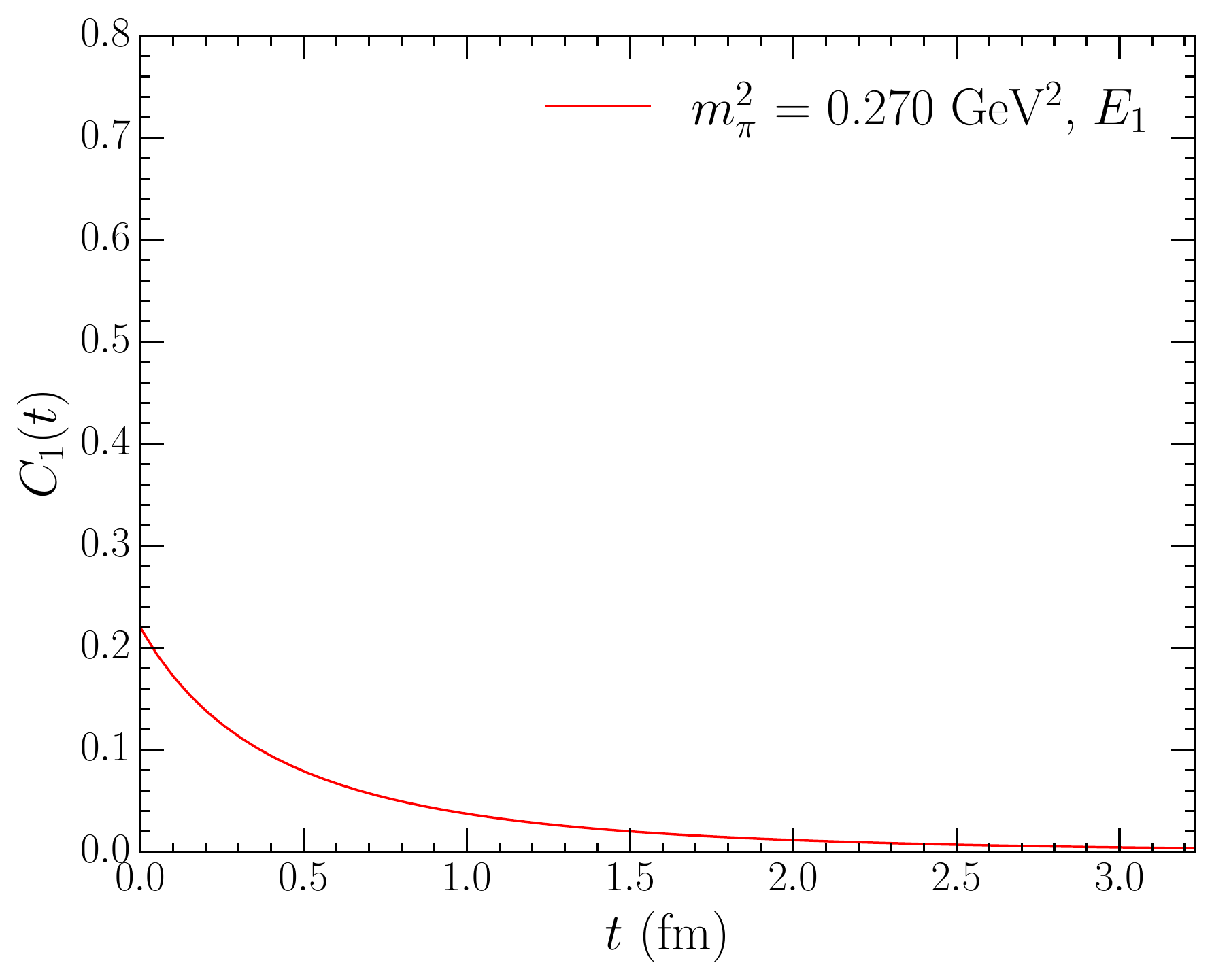}~
  \includegraphics[width=0.24\linewidth]{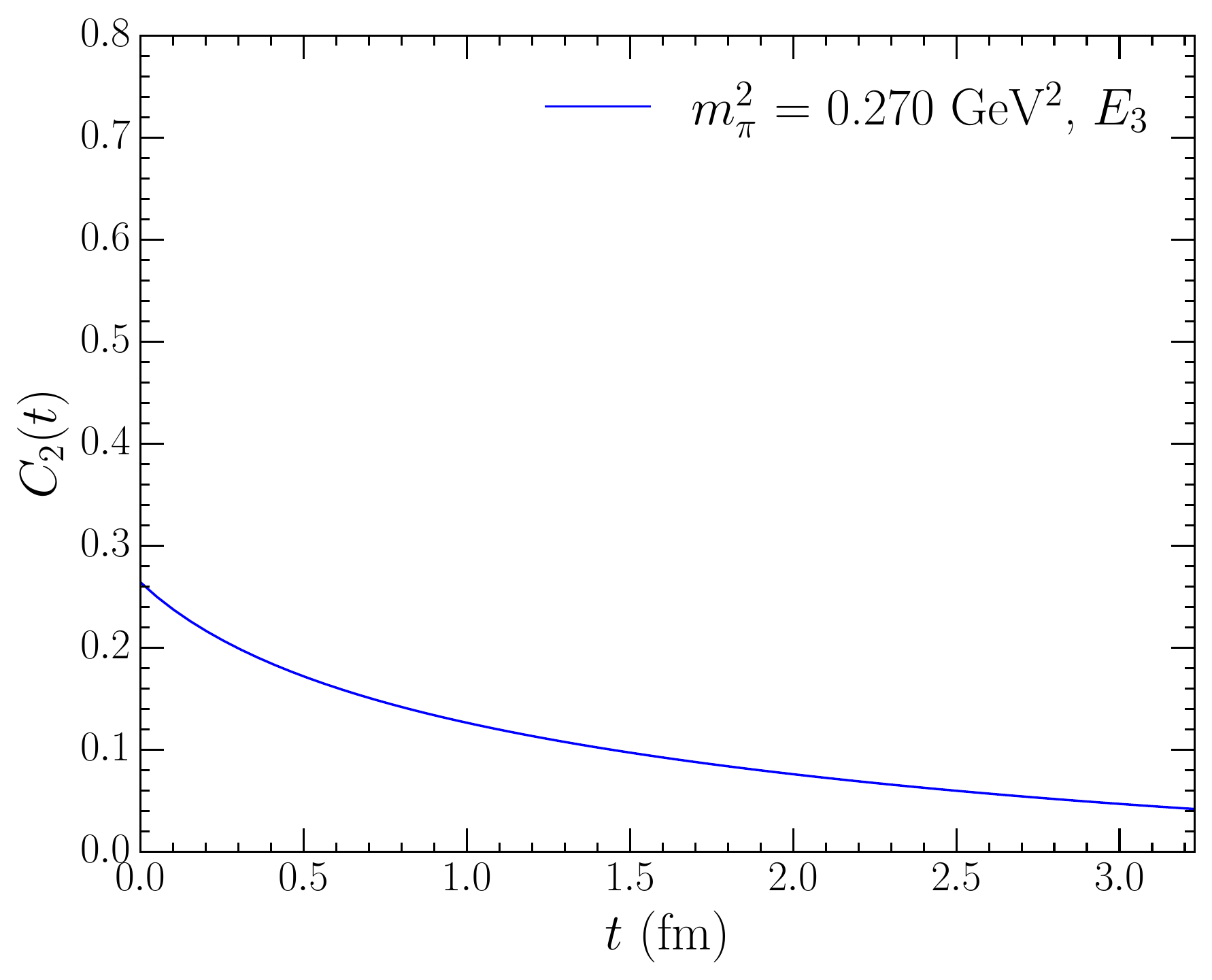}~
  \includegraphics[width=0.24\linewidth]{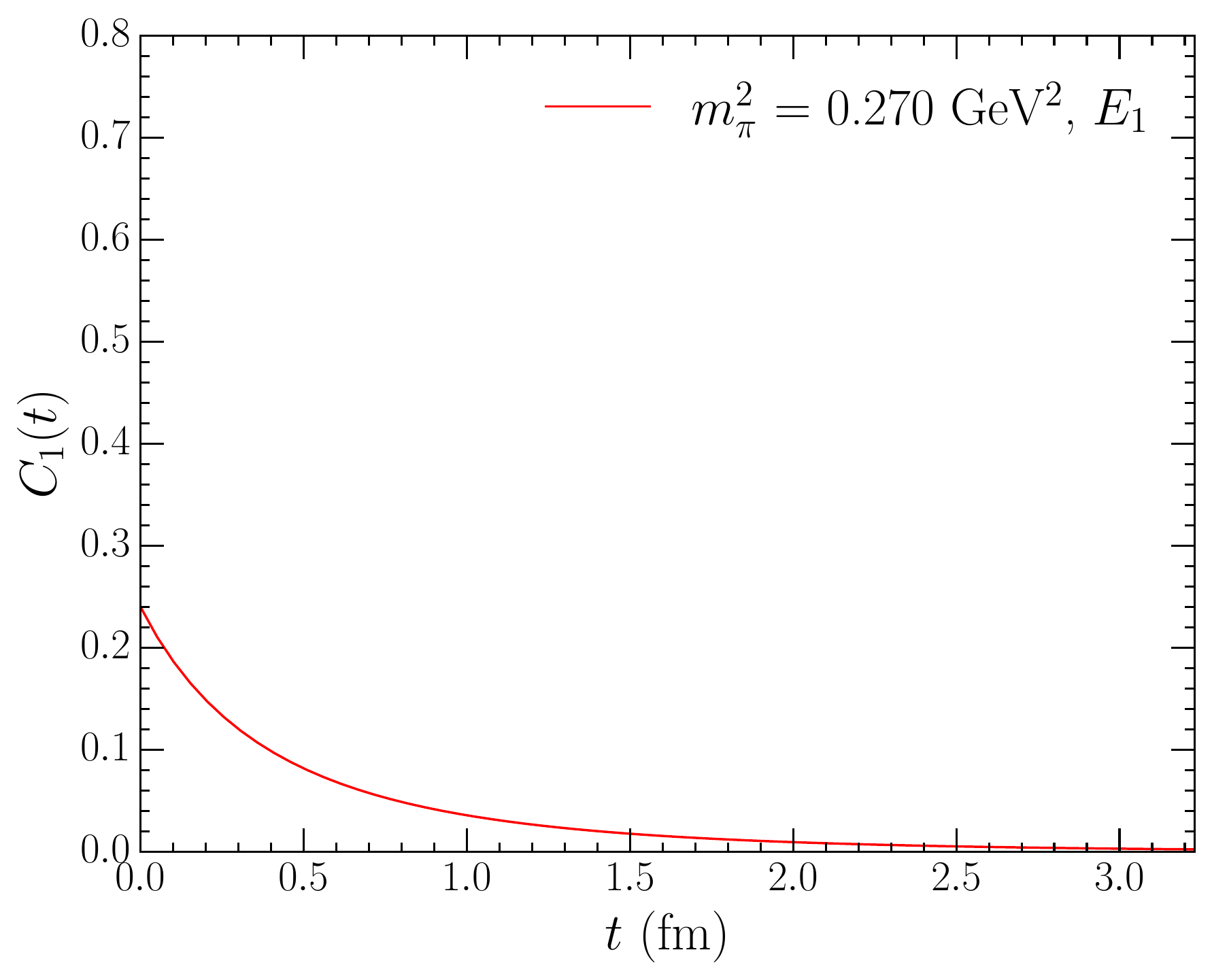}~
  \includegraphics[width=0.24\linewidth]{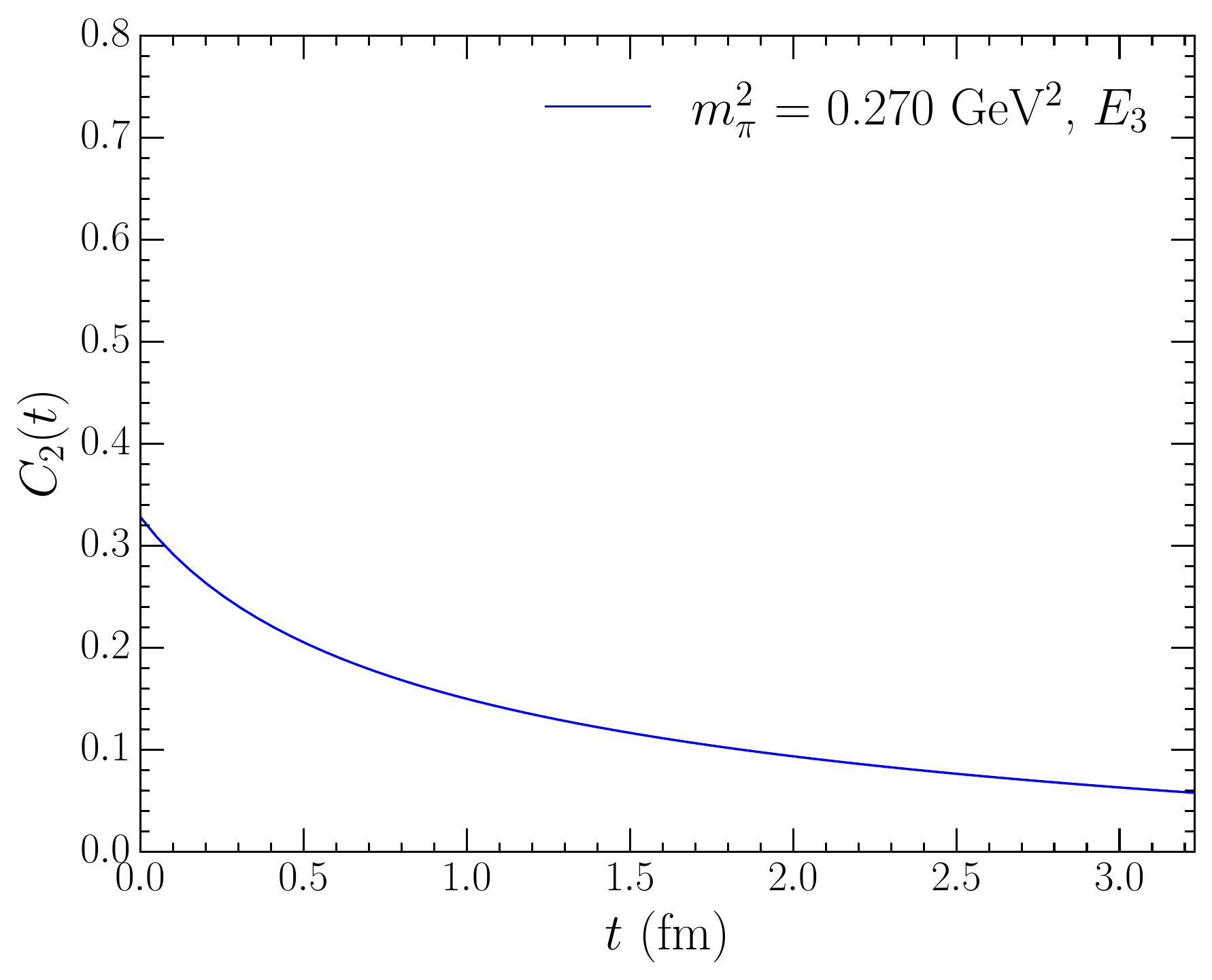}~\\
  \includegraphics[width=0.24\linewidth]{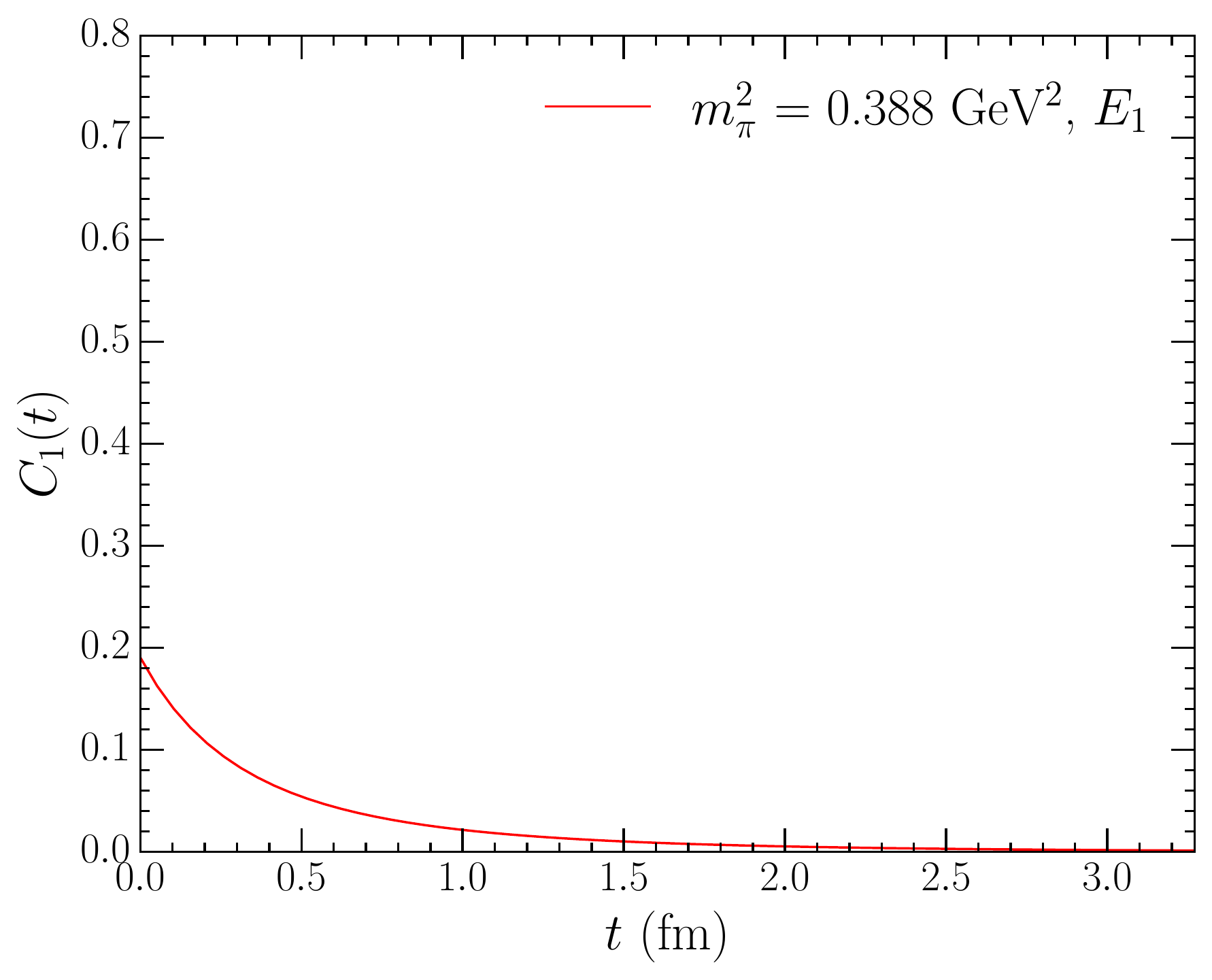}~
  \includegraphics[width=0.24\linewidth]{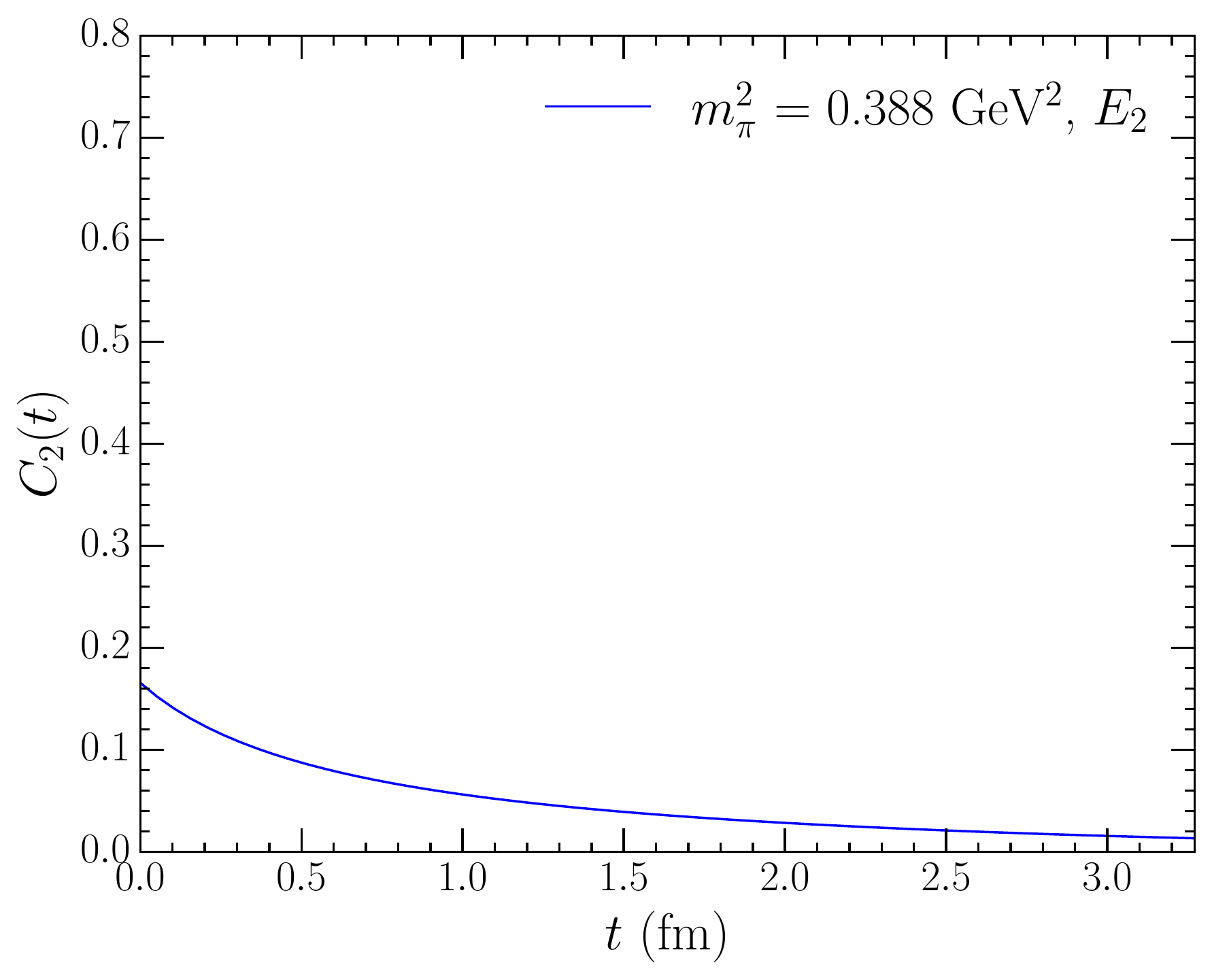}~
  \includegraphics[width=0.24\linewidth]{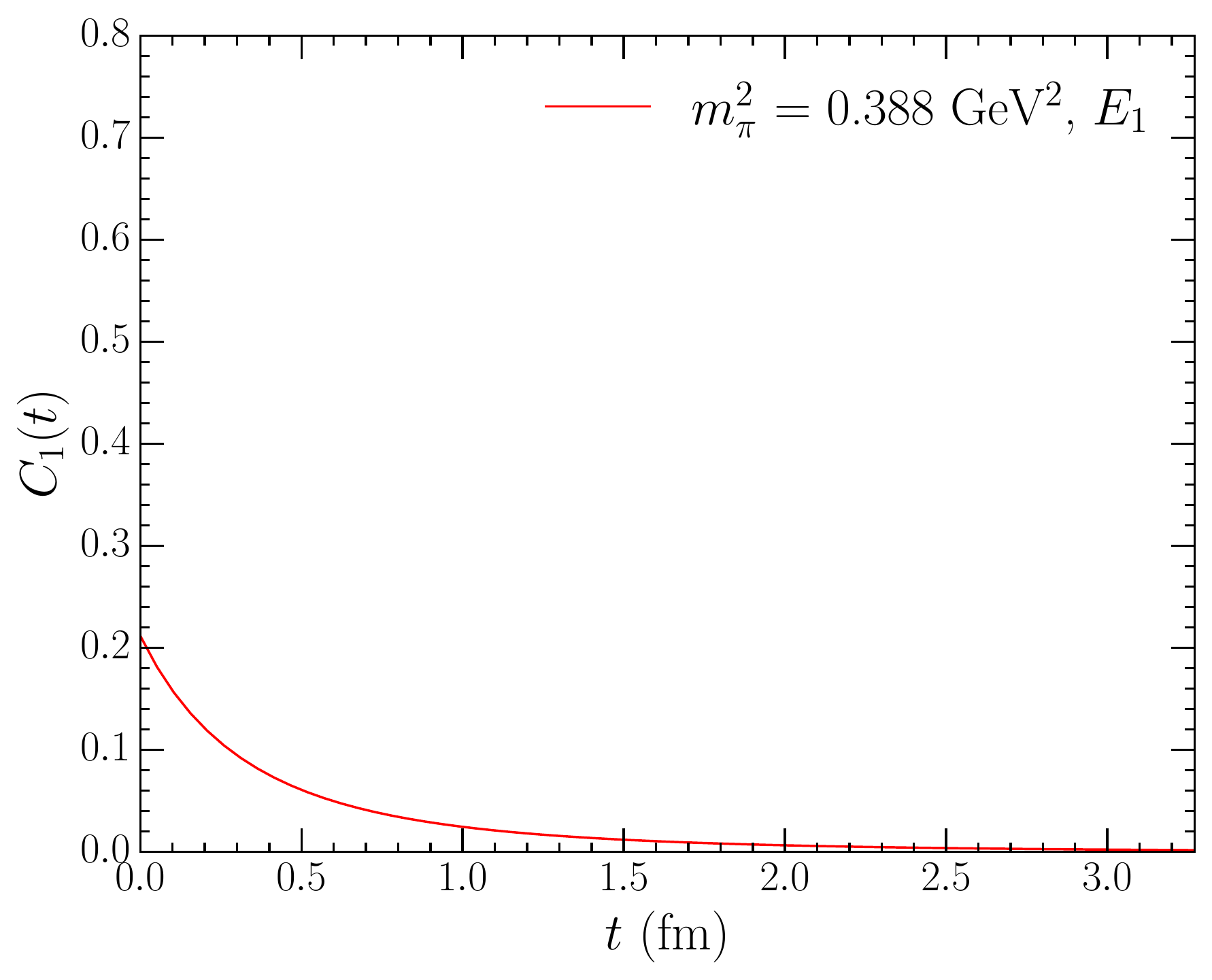}~
  \includegraphics[width=0.24\linewidth]{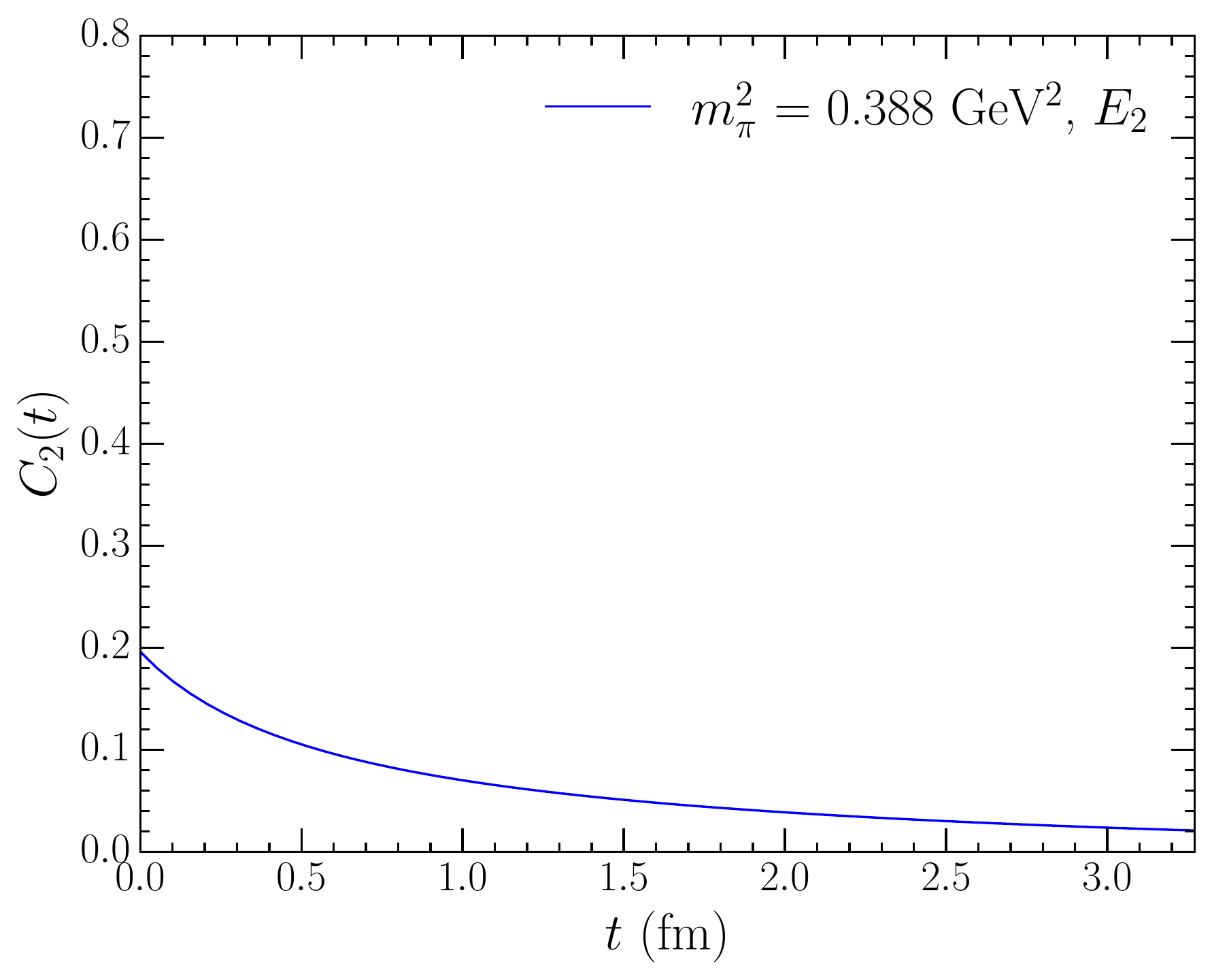}
  \caption{Contamination functions from \eref{eq:contamination:Ct} at the five pion masses considered by the PACS-CS collaboration \rref{PACS-CS:2008bkb}, where the pion mass increases as one moves down the columns.
    Contributions from all eigenstates with large single-particle component have been removed from the correlation functions (all highlighted eigenstates in \fref{fig:2b3c_3fm_spectrum_bare}).
    Values for $\alpha_j$ and $\beta_j$ are taken from lattice QCD correlation matrix eigenvectors for the two left-most columns, and taken from HEFT eigenvectors as defined in \eref{eq:HEFT_alpha_beta} for the two right-most columns.
    The first and third columns (red lines) correspond with the eigenstates dominated by contributions from $\ket{N_1}$, while the second and fourth columns (blue lines) correspond with the eigenstates dominated by contributions from $\ket{N_2}$.
  The relevant eigenstate for each lattice QCD energy level is labelled by $E_{i}$.}
  \label{fig:contamination_3fm}
\end{figure*}

In \fref{fig:contamination_3fm}, these two schemes for determining $\alpha_j$ and $\beta_j$ are compared by calculating contamination functions as defined in \eref{eq:contamination:Ct} at each PACS-CS pion mass.
As described above, we remove not only the contribution from the two eigenstates with largest bare basis state eigenvector component, but also the contribution from the eigenstates with second largest bare basis state component.

Broadly considering these two-particle contamination functions, we observe two situations.
At lighter pion masses, where the bare-dominated states $\ket{E_{N_1}}$ and $\ket{E_{N_2}}$ sit above the lower-lying eigenstates, we observe a scattering-state contamination which has a minimum in the vicinity of $1 - 2$ fm.
At the heavier pion masses, where these bare-dominated states are found in the lower-lying eigenstates, the contamination tends to zero as time increases, as all excited states become exponentially suppressed.
There is a remarkable similarity between the contamination functions constructed from the correlation matrix eigenvectors from lattice QCD, and the Hamiltonian eigenvectors from HEFT.

Considering specific pion masses, at the two largest masses we observe a strong decay in the contamination, where all scattering state contaminations are at the 5-10\% range at Euclidean times where you'd expect to observe an effective-mass plateau.
At the third heaviest mass, we observe a minimum contamination in the plateau region of 6\% for $N_{1}$, which is in line with the prediction from \refref{Stokes:2019zdd} of approximately 5\%.
As described in \refref{Stokes:2019zdd}, we observe a larger degree of scattering-state contamination in the correlation function corresponding with $N_{2}$ for the second lightest mass.
For the two lightest masses, some degree of scattering-state contamination is to be expected, as they fall near the $\eta N$ and $K\Lambda$ thresholds.

\subsubsection{Single-Particle and Two-Particle Contamination}
In the previous section, we analysed the two-particle scattering-state contamination by removing contributions to the correlation functions from all eigenstates with a significant single-particle bare basis state eigenvector component.
There the two-particle scattering state contaminations for the three
heaviest quark masses considered were found to be typically small,
the order of 10\% in the Euclidean time range where masses and
form factors are extracted.

Here we explore a different problem where the bare basis state
becomes significantly associated with more than one energy
eigenstate. The extent of this distribution over eigenstates is directly
related to the volume of the lattice which governs the number of
eigenstates within a given energy range, {\it i.e.} the density of
energy eigenstates. As the volume increases, the density of
eigenstates increases and the bare basis state becomes spread
over several states

However, lattice QCD aims to isolate a single energy eigenstate. In
the absence of two-particle interpolating fields, this is done via
Euclidean time evolution to allow the higher state to become
exponentially suppressed while the uncertainties in the correlation
function grow to the point that the errors are sufficient to encompass
the behaviour of a single propagating state.

Drawing on the information available in the HEFT eigenvectors, we
are able to quantify the contamination from both the two-particle
scattering states and the distribution of significant single-particle
strength across multiple energy eigenstates. This time only the two
energy eigenstates having the dominant bare basis state
components, $\ket{N_1}$ and $\ket{N_2}$, are
eliminated. In cases where the strength is almost equal, the lower
lying state is considered isolated and eliminated from the
contamination function.
%
%

\begin{figure}
  \centering
  \includegraphics[width=0.24\textwidth]{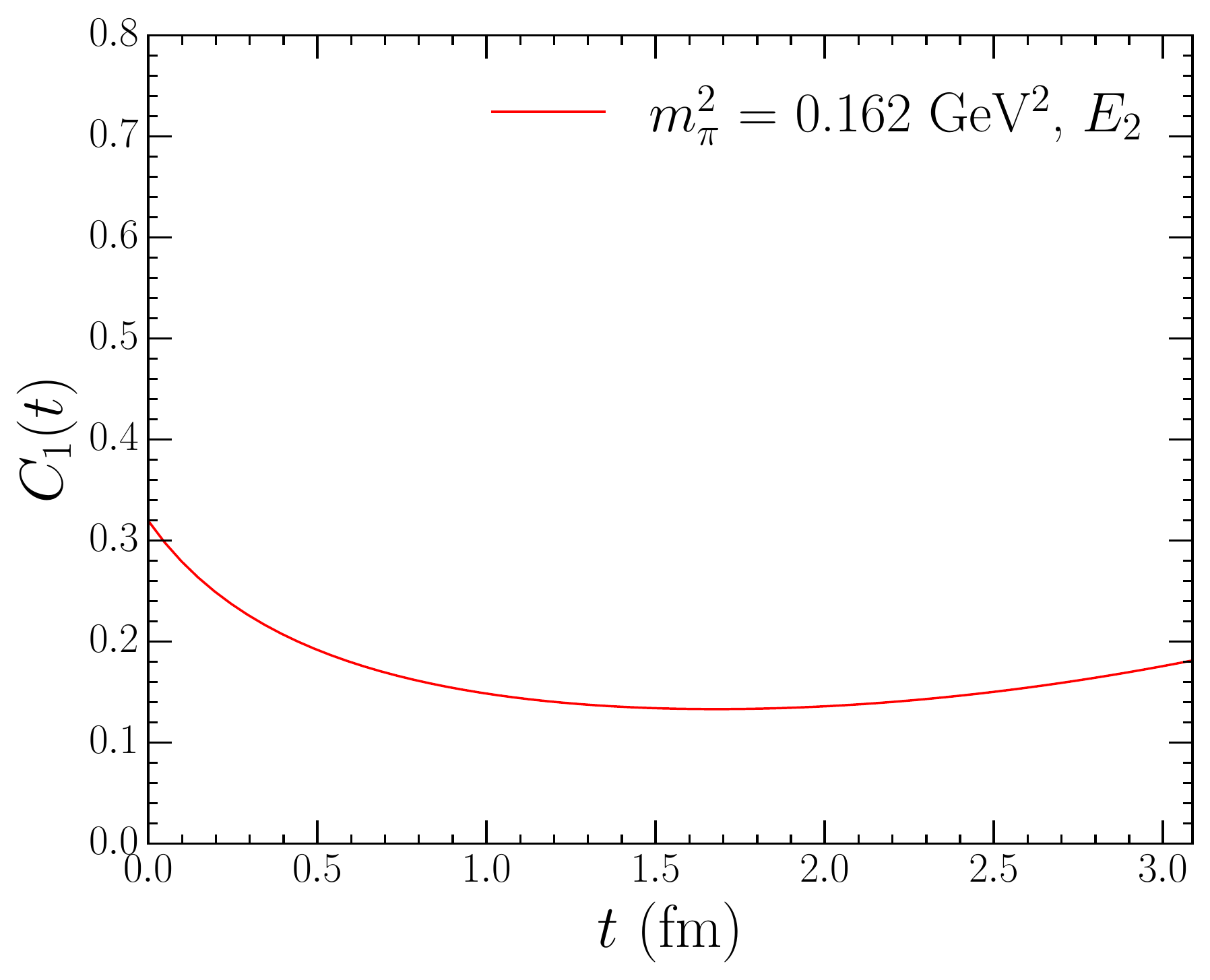}~
  \includegraphics[width=0.24\textwidth]{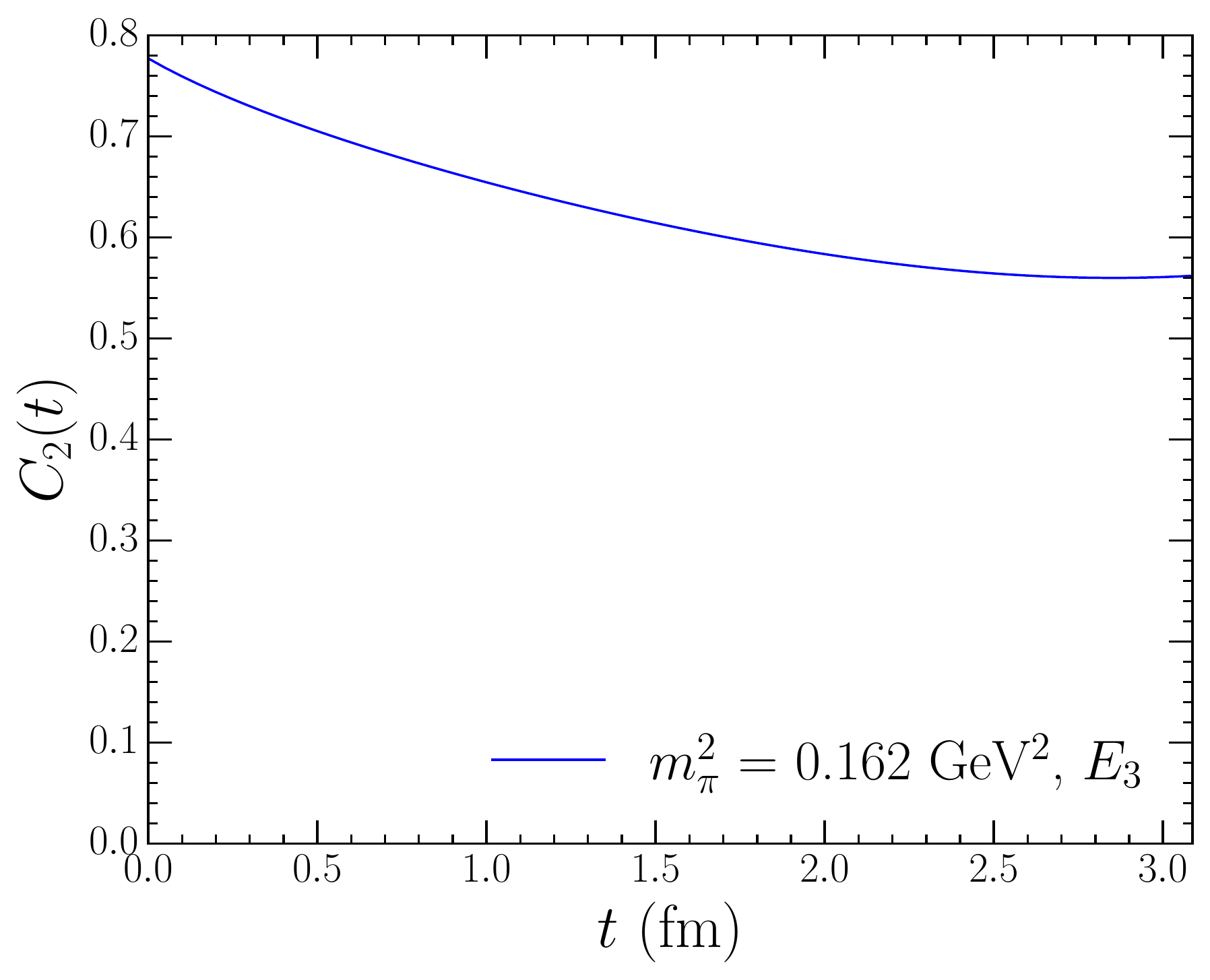}~\\
  \includegraphics[width=0.24\textwidth]{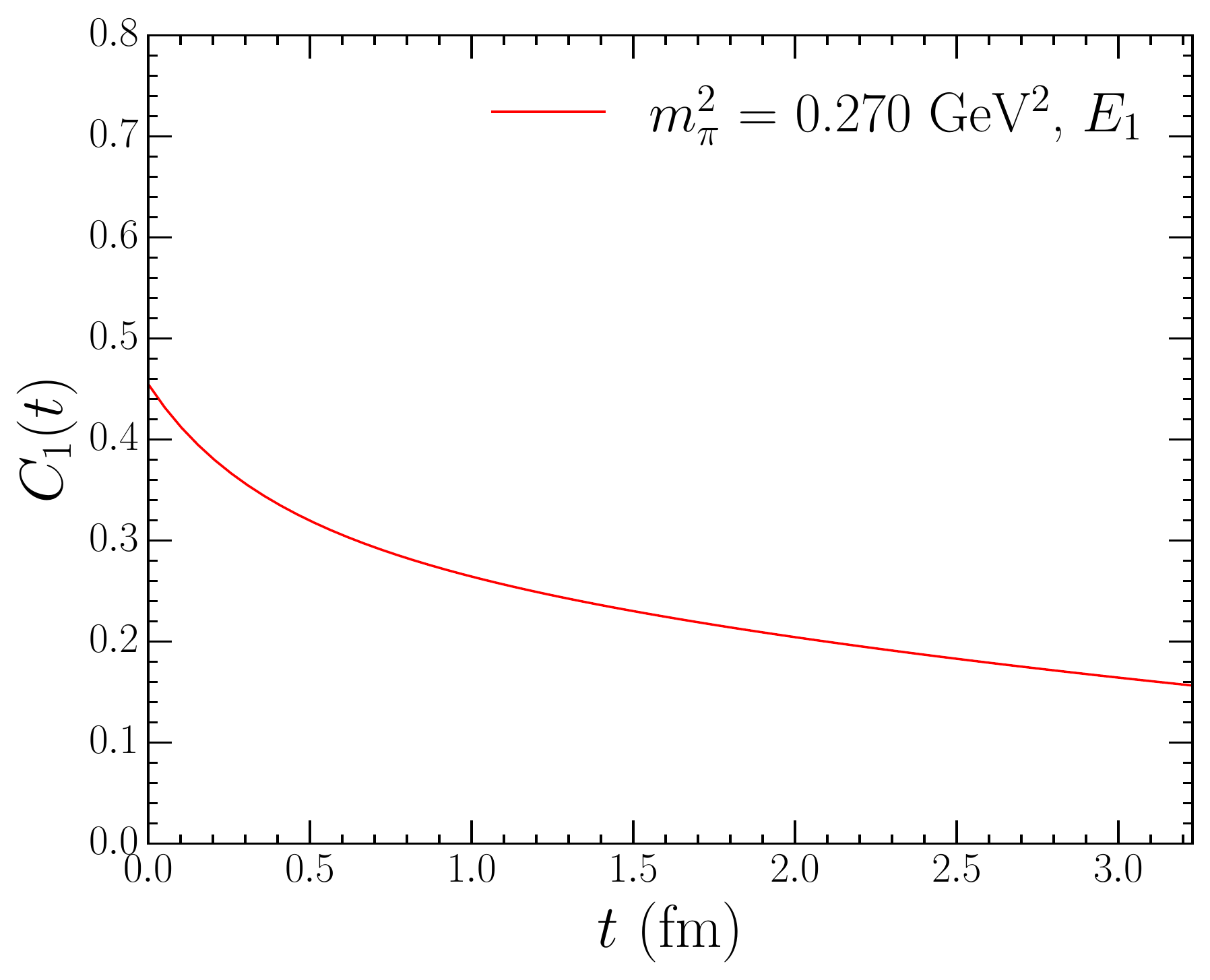}~
  \includegraphics[width=0.24\textwidth]{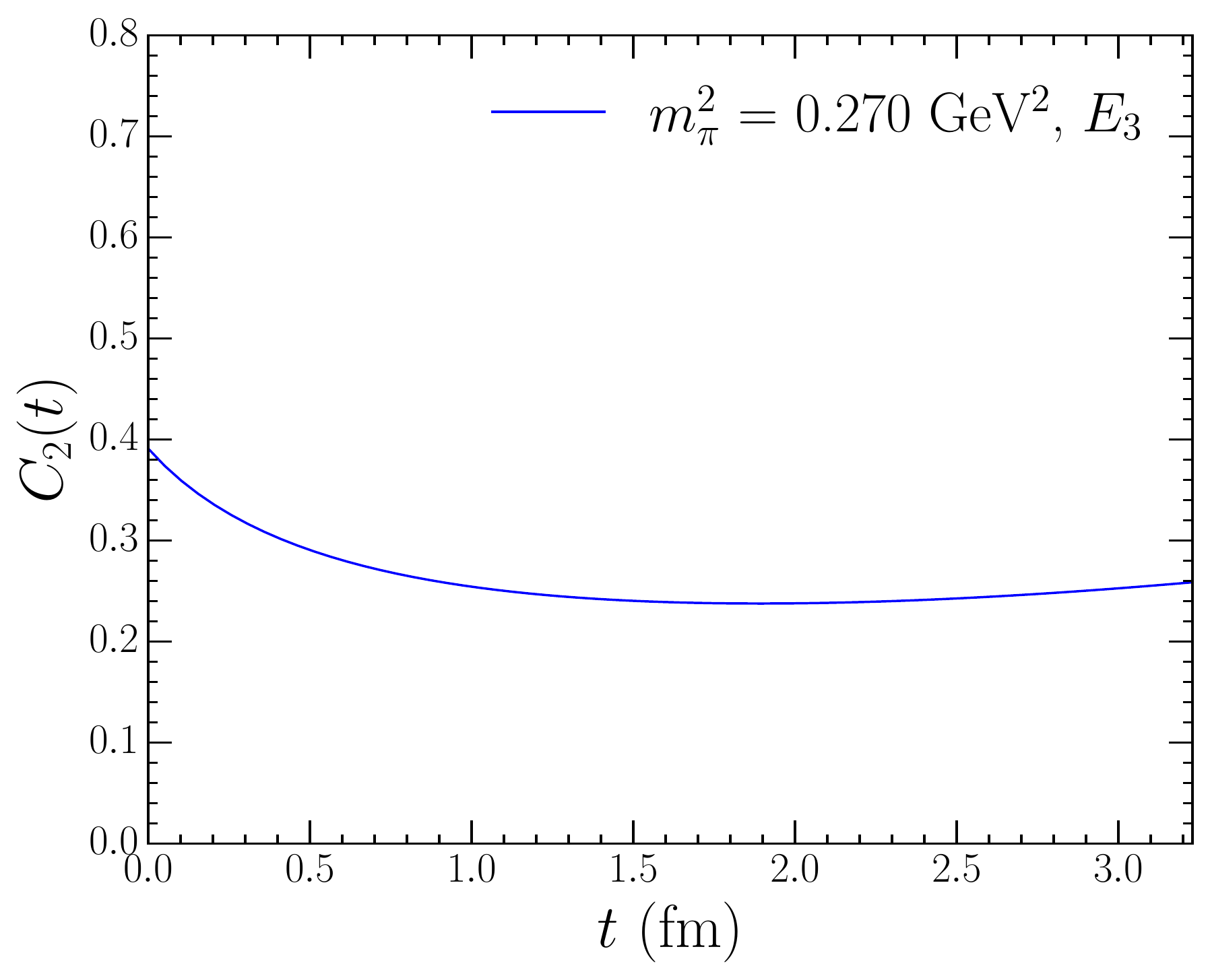}~\\
  \includegraphics[width=0.24\textwidth]{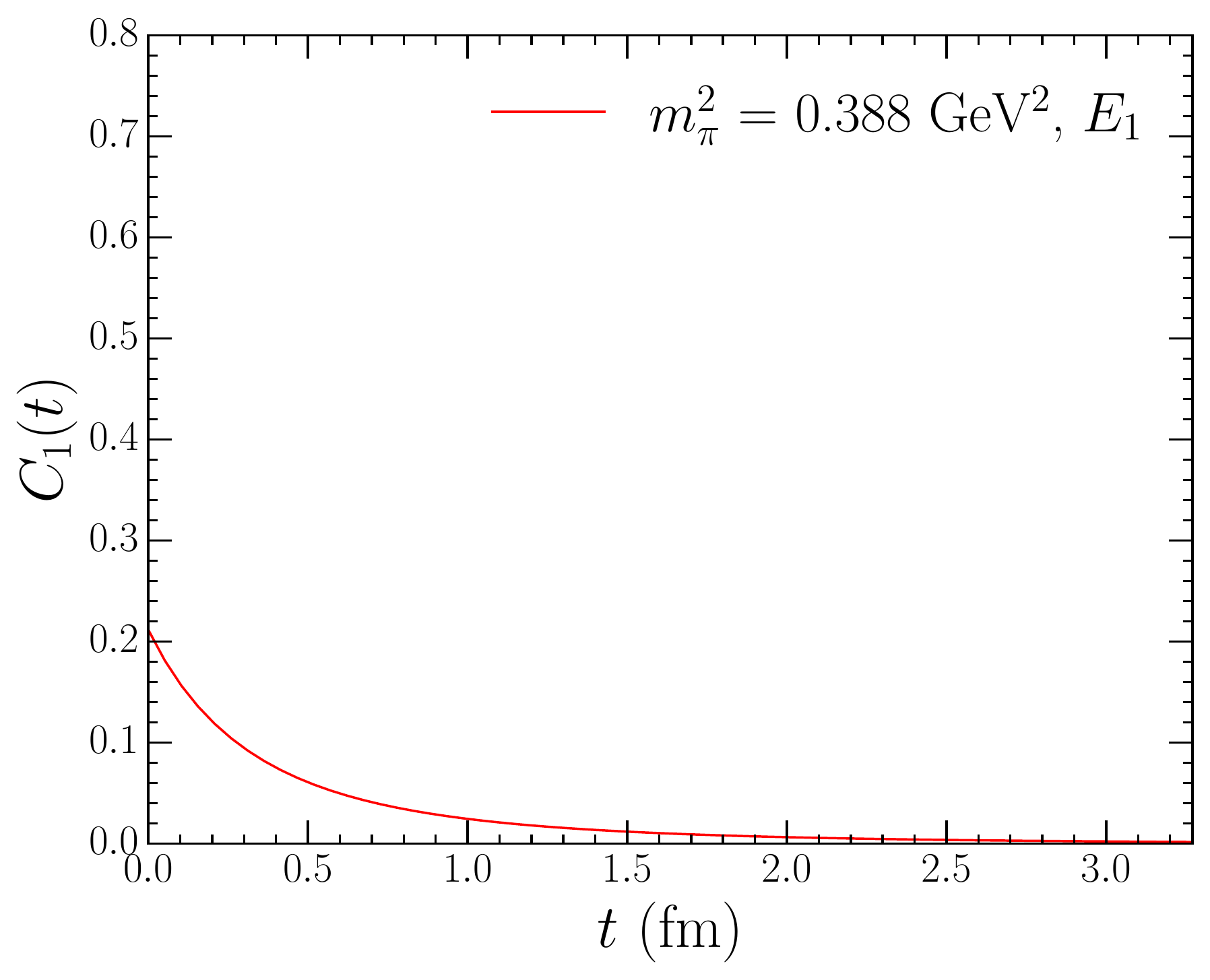}~
  \includegraphics[width=0.24\textwidth]{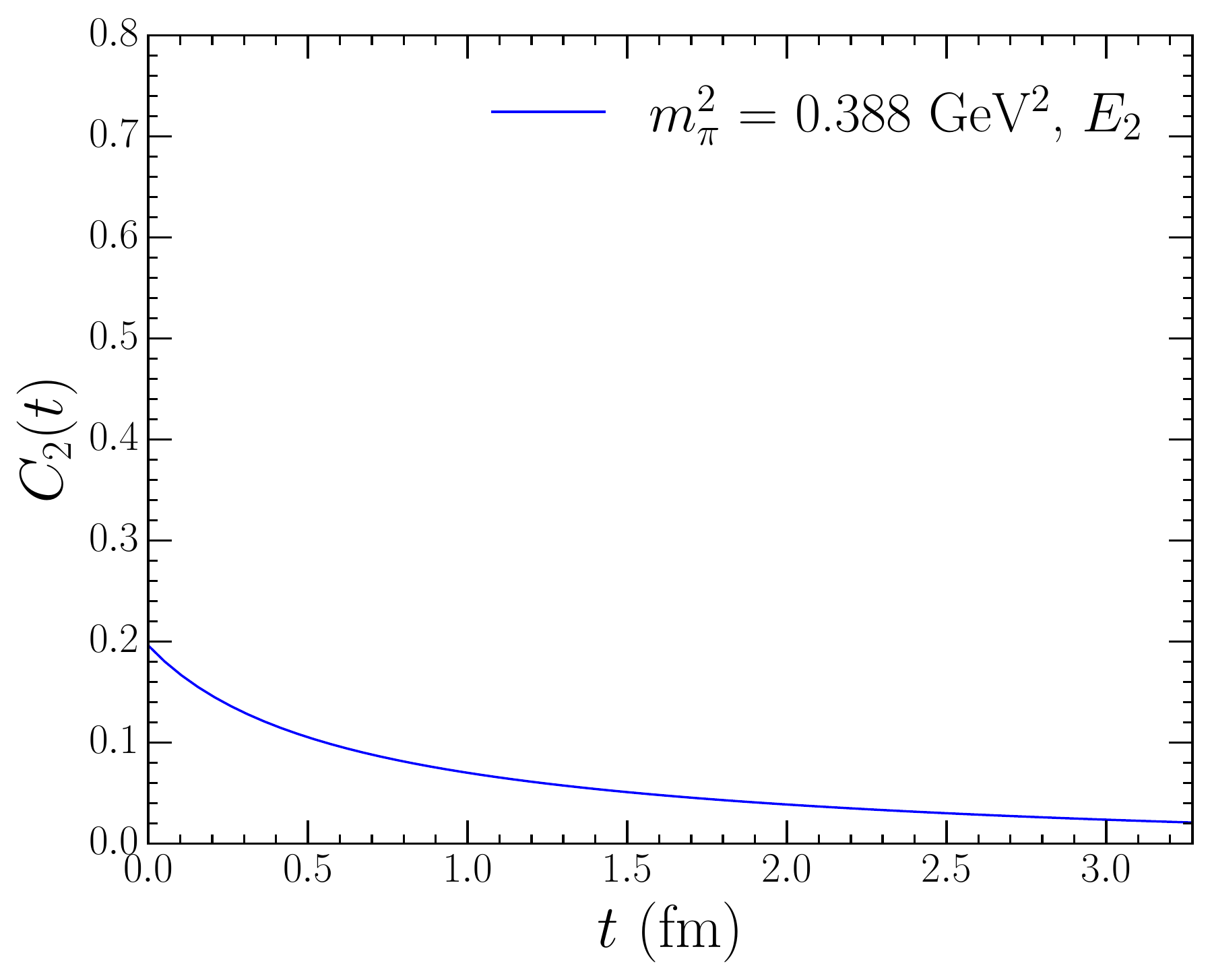}~\\
  \caption{Contamination functions from \eref{eq:contamination:Ct} at the three heaviest pion masses considered by PACS-CS \rref{PACS-CS:2008bkb}.
    The pion mass increases as one moves down the columns.
    Contributions from the eigenstate with the largest eigenvector component for each bare basis state have been removed in calculating the contamination functions (solid highlighted state in \fref{fig:2b3c_3fm_spectrum_bare}).
    Values for $\alpha_{i}$ and $\beta_{i}$ are taken from the HEFT eigenvectors as defined in \eref{eq:HEFT_alpha_beta}.
  The relevant eigenstate for each lattice QCD energy state is labelled by $E_{i}$.}
  \label{fig:contamination_3fm_1eig}
\end{figure}
Using \eref{eq:contamination:Ct} the scattering-state contaminations for the three heaviest PACS-CS masses are illustrated in \fref{fig:contamination_3fm_1eig}.
Here, the label $E_{i}$ on each contamination function refers to the eigenstate associated with each lattice QCD energy level.
In the case where a lattice QCD mass sits on an avoided level crossing, where two different eigenstates have approximately equal large bare basis state eigenvector components, the state with lower eigenenergy is chosen.
Under Euclidean time evolution, excited states in the spectrum decay more quickly, and thus it is expected that the lower eigenenergy is isolated.

Comparing the contamination functions in \fref{fig:contamination_3fm_1eig} with those in the previous section, we observe a significantly higher degree of contamination.
For the heaviest PACS-CS mass, we still observe a decaying contamination for large Euclidean time.
As a vast majority of the bare basis state eigenvector components are concentrated in the two lowest-lying finite-volume eigenstates, we do not expect any scattering-state contamination following Euclidean time evolution.
At the second and third heaviest masses however, we observe a significantly larger level of contamination.

Consider the positions of the second and third heaviest masses on the finite-volume spectrum from \fref{fig:2b3c_3fm_spectrum_bare}.
For the lower-lying lattice QCD mass at the second-heaviest pion mass, this sits directly on an avoided level crossing in the eigenvector component for $N_{1}$.
As a result, whether $E_{1}$ or $E_{2}$ is chosen as the state corresponding with this lattice QCD mass, and removed from the correlation function, a significant single-particle component will remain in the correlation function.
This effect is seen to a greater degree in the larger lattice QCD mass at the third heaviest pion mass.
This mass sits at an avoided level crossing in $N_{2}$, where the eigenvector component for $N_{2}$ is significantly spread over four nearby eigenstates.
In the context of \fref{fig:2b3c_3fm_spectrum_bare}, both the solid blue and dashed blue lines are moving between HEFT eigenstates at this position.
At this position, the eigenstate with largest $N_{2}$ component only contains approximately 15\% of the contribution from $N_{2}$.
As such, removing only a single $N_{2}$-dominated eigenstate from the correlation function will leave a significant degree of single-particle based contamination in the estimate of the scattering-state contamination.
This effect is further exaggerated in the two lightest masses.
Due to the high density of states at this point, the eigenvector components for the two bare basis states are further spread to nearby energy eigenstates.

In the context of exploratory lattice QCD calculations seeking to identify the nature of quark-model like states in the spectrum, the level of scattering-state contaminations illustrated in \fref{fig:contamination_3fm_1eig} is encouraging, in that for five of the six states considered the correlation functions are dominated by the state of interest at the level of 75\% or better where the signal is extracted.
Moreover, \fref{fig:contamination_3fm} illustrates the majority of the contamination comes from a nearby state in the spectrum having the same bare basis state.

On the other hand, it is clear that next generation lattice QCD calculations seeking quantitative comparison with experimental measurements will need to have a complete set of two-particle interpolating fields to complement the single-particle three-quark interpolating fields considered in the leading exploratory calculations.
Only then can one couple to the complete set of energy eigenstates illustrated in Figs. \ref{fig:2b3c_3fm_spectrum} and \ref{fig:2b3c_3fm_spectrum_bare} and isolate them in the solution to the generalised eigenvalue equation for the correlation matrix.
%
%

\subsection{Contamination Functions at 2 fm} \label{sec:contamination:2fm}
\begin{figure}
  \centering
  \includegraphics[width=0.24\textwidth]{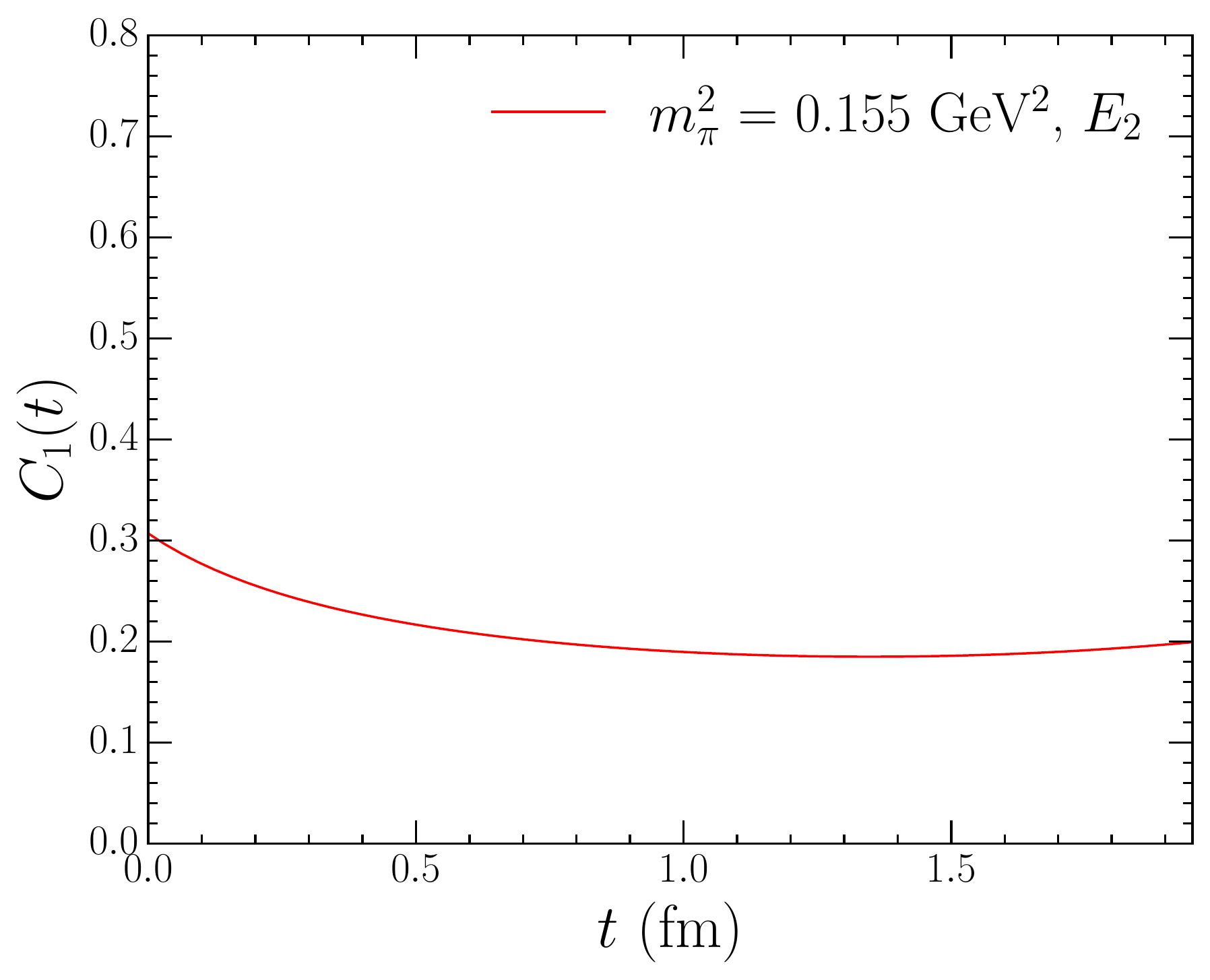}~
  \includegraphics[width=0.24\textwidth]{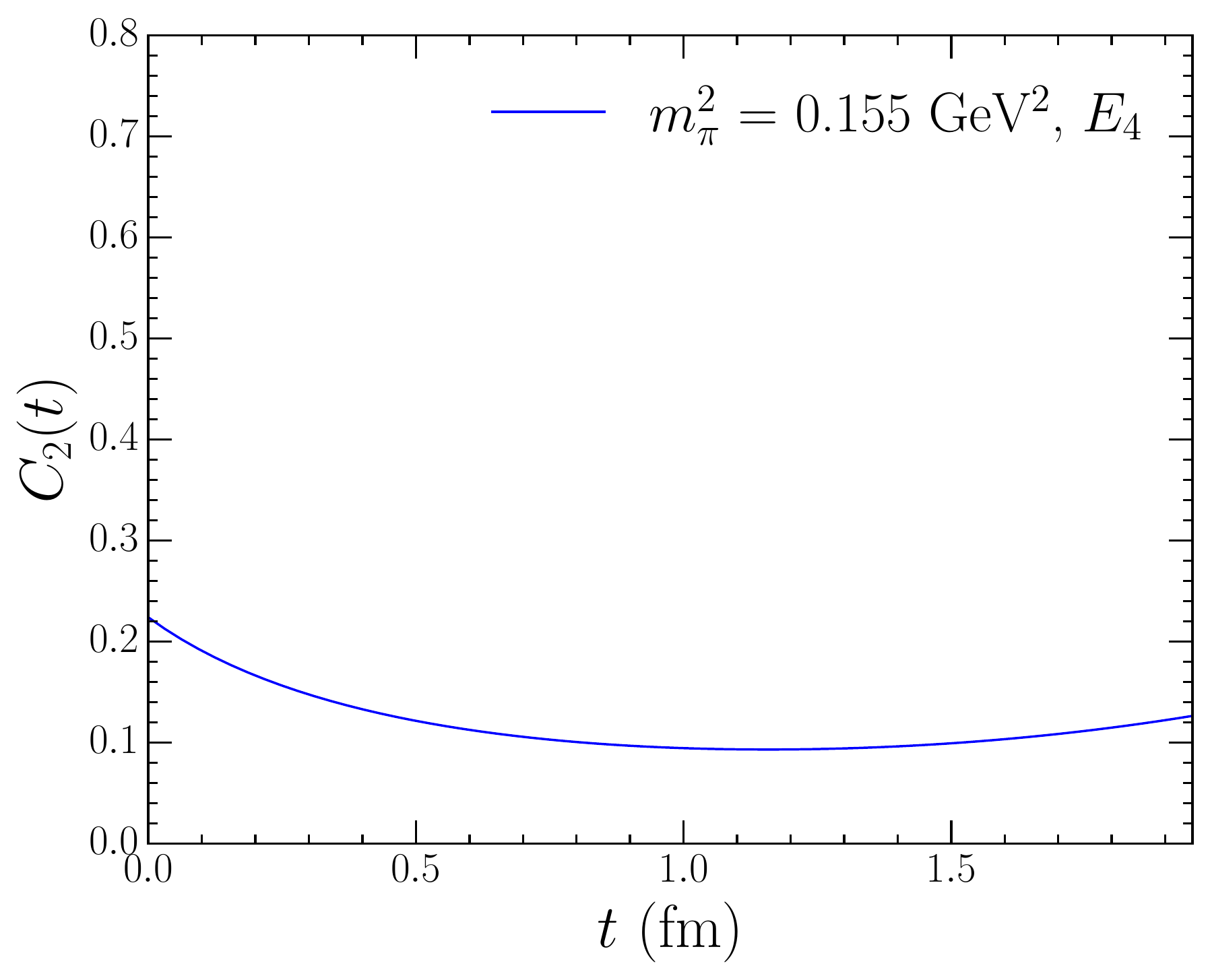}~\\
  \includegraphics[width=0.24\textwidth]{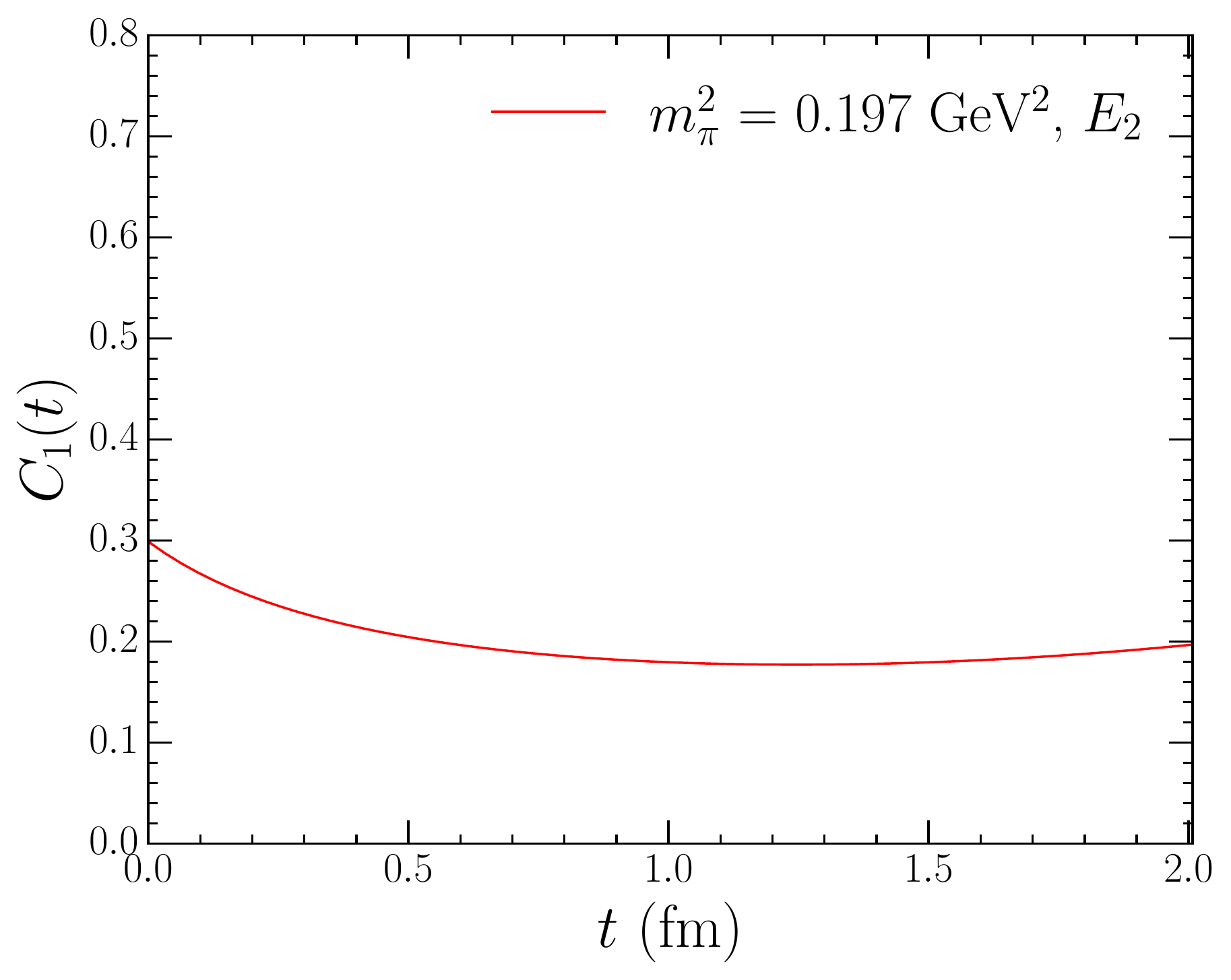}~
  \includegraphics[width=0.24\textwidth]{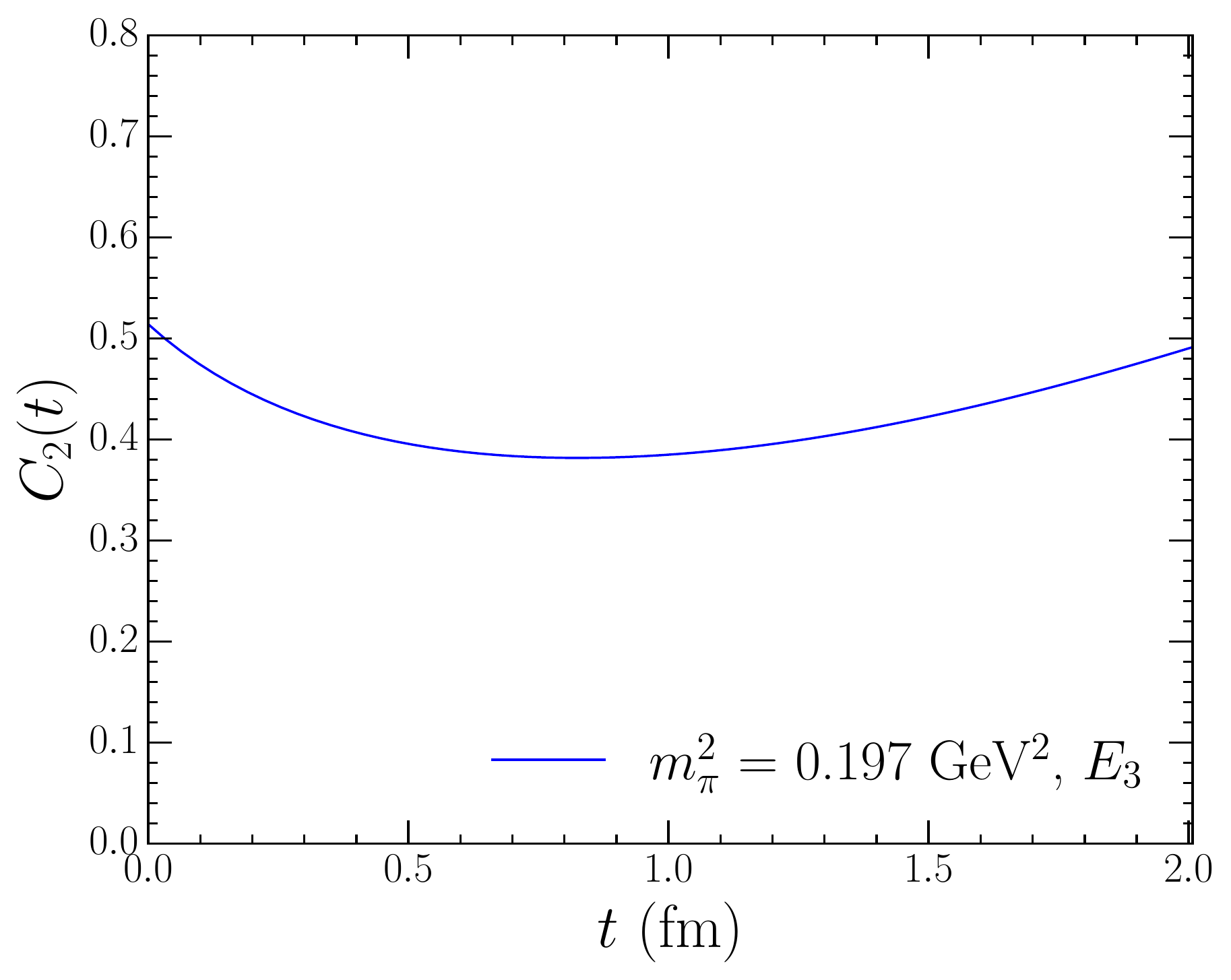}~\\
  \includegraphics[width=0.24\textwidth]{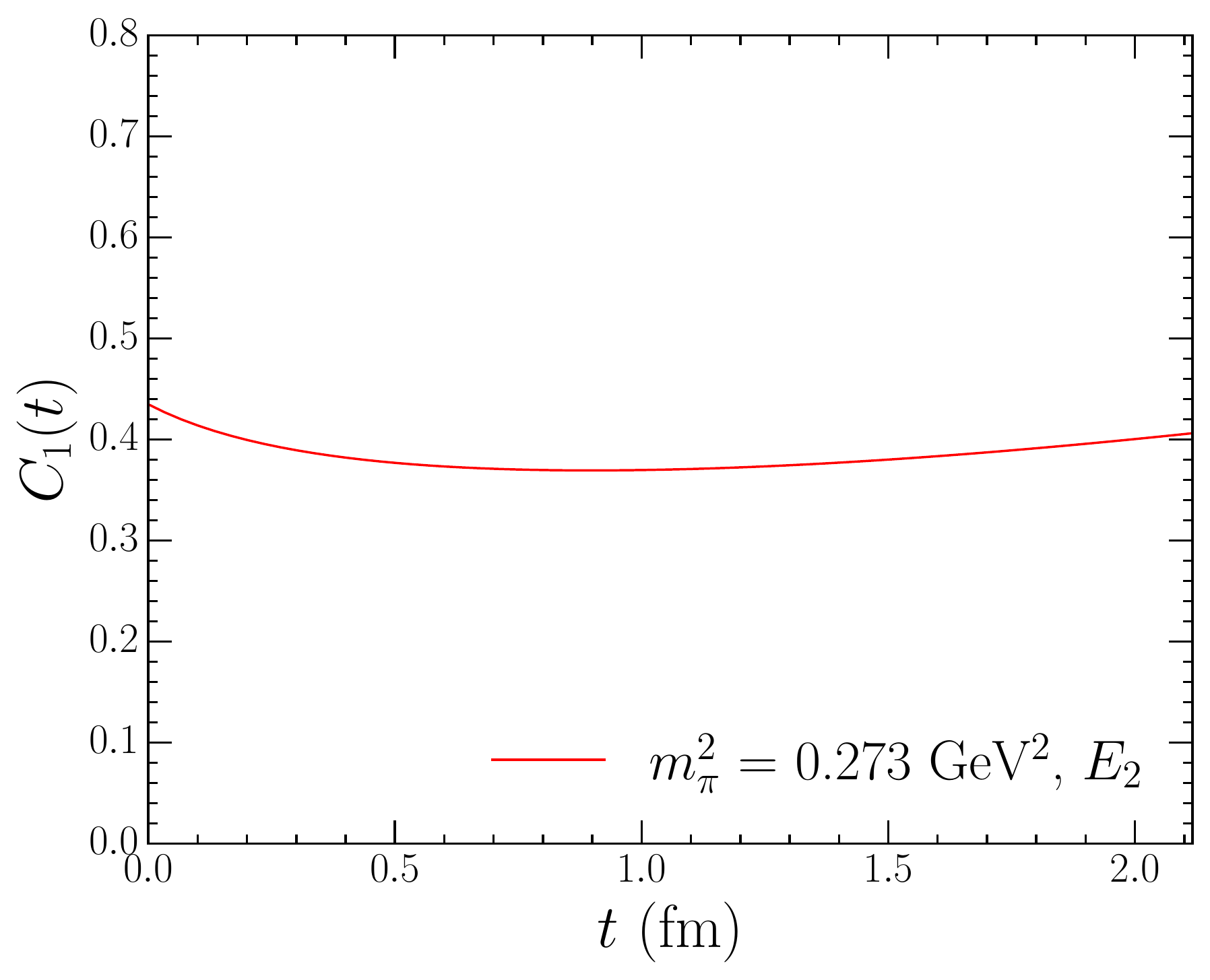}~
  \includegraphics[width=0.24\textwidth]{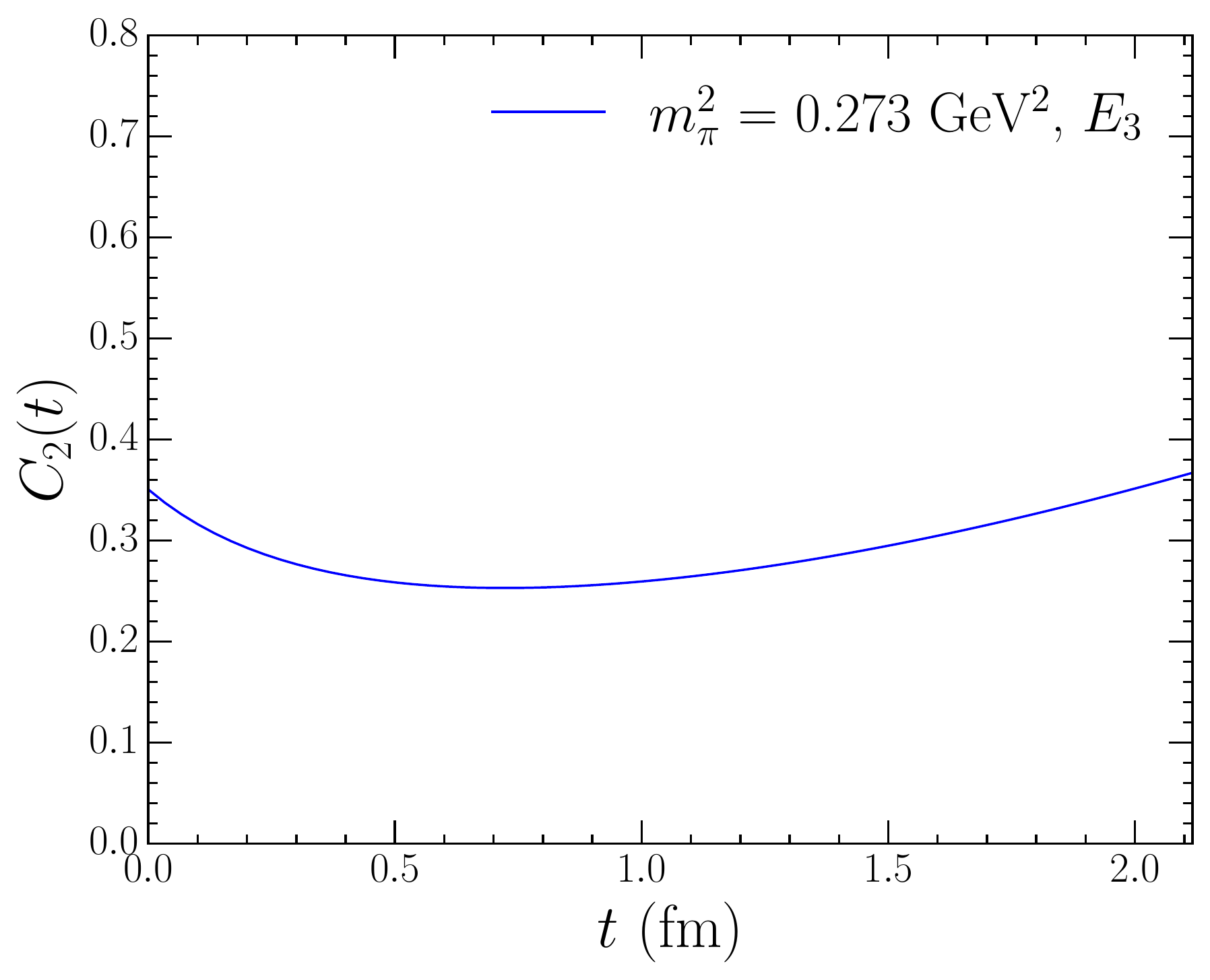}
  \caption{Contamination functions from \eref{eq:contamination:Ct} at the three pion masses considered by the HSC \rref{Edwards:2011jj,Edwards:2012fx}.
    The pion mass increases as one moves down the columns.
    Contributions from the eigenstate with the largest eigenvector component for each bare basis state have been removed from the correlation functions (solid highlighted states in \fref{fig:2b3c_3fm_spectrum_bare}).
    Values for $\alpha_{i}$ and $\beta_{i}$ are taken from the HEFT eigenvectors as defined in \eref{eq:HEFT_alpha_beta}.
  The relevant eigenstate for each lattice QCD energy level is labelled by $E_{i}$.}
  \label{fig:contamination_2fm}
\end{figure}
As for the 3 fm analysis, we can utilise the correlation functions as defined in \sref{sec:contamination:formalism} to calculate the degree of scattering-state contamination in the correlation functions corresponding with the lattice QCD results.
Due to the lower density of states, we explore contamination functions calculated as defined in \eref{eq:contamination:Ct}, where only the eigenstate with largest bare basis state eigenvector component is removed.
In particular, we calculate contamination functions for the six lattice QCD results from the HSC \rref{Edwards:2011jj,Edwards:2012fx}, as these are calculated using three-quark interpolating fields and correspond with bare-dominated states.
As the eigenvectors from both the lattice QCD correlation matrix and the HEFT Hamiltonian were found to produce equivalent contamination functions in \sref{sec:contamination:3fm}, we utilise the HEFT eigenvectors for this section.

In \fref{fig:contamination_2fm}, results for the six contamination functions corresponding to the six lattice QCD results reported by the HSC are illustrated.
These curves can be compared with the first two rows of \fref{fig:contamination_3fm_1eig} reporting results at similar pion masses on a 3 fm lattice.

While the large contamination reported in the top-right plot of Fig. 11 does not appear, broad improvement is not observed.
The second state at the HSC middle mass and the first state at their heaviest mass both show scattering state contamination at 40\%.
As discussed in further detail below, both of these states sit in the midst of avoided level crossings.

Focusing first on the middle mass, the contributions from the second bare state are roughly equally spread between two eigenstates, making it difficult to conclusively comment on the eigenstate to which the lattice QCD energy corresponds.
As only one of these states is removed in calculating the contamination function, there is still an eigenstate containing approximately 40\% of the second bare state in the remaining correlation function.

Interestingly, the $C_{1}(t)$ contamination function for the heaviest pion mass in the lower-left plot of \fref{fig:contamination_2fm} is significantly greater than the contamination functions for the two lighter pion masses.
This may provide an explanation for why the lower-energy HSC state sits at an energy lower than that predicted by HEFT.
The HSC correlation function has a significant contamination from the lower-lying scattering state and the mass obtained in their analysis likely corresponds to a superposition of these two eigenstate energies.

\section{Conclusion} \label{sec:conclusion}
In this study, we consider a Hamiltonian Effective Field Theory (HEFT) analysis of the two low-lying odd-parity nucleon resonances in the $I(J^{P}) = \frac{1}{2}(\frac{1}{2}^{-})$ channel, the $N^{*}(1535)$ and the $N^{*}(1650)$.
This is done using a novel Hamiltonian consisting of two bare basis states, representing a three-quark core for the two resonances.
Two-particle $\pi N$, $\eta N$, and $K\Lambda$ channels are also considered.
In HEFT, the interactions between the basis states are parametrised such that by solving the scattering equations for the system, we are able to obtain a good description of the experimental scattering data and pole positions.

In \sref{sec:inf}, the parameters of the Hamiltonian are constrained to $S_{11}$ scattering data.
These parameters include the masses of the two bare basis states, the coupling strengths of the interactions between the basis states, and the strengths of the dipole regulators for each channel.
In doing so, we are able to obtain a description of the scattering data up to a centre-of-mass energy of 1.75 GeV.
By solving for the pole positions in the $T$-matrix, we obtain a pole for each resonance consistent with the PDG values.

By taking the constrained Hamiltonian and extending it to a finite-volume formalism, we are able to make connection to lattice QCD results.
Using lattice QCD results from a $L\sim 3$ fm lattice, in \sref{sec:3fm} we constrain the mass slopes of the bare basis states, allowing a pion-mass interpolation of the energy eigenvalues.
Here we find that the interpretation of the two resonances as three-quark cores dressed by scattering-state dynamics is consistent with the $L\sim 3$ fm lattice calculations.

Using the parameters constrained by both experimental data and the 3 fm lattice QCD data, we also consider lattice QCD results at $L\sim 2$ fm in \sref{sec:2fm}.
Without any further variation of the Hamiltonian, we find that the HEFT eigenstates with large bare state components are also consistent with these lattice QCD results.
Similarly, in \sref{sec:4fm} we found that HEFT is in agreement with the new $L = 4.05$ fm lattice QCD results from the CLS consortium.
At this lattice size, the two lattice QCD states excited from momentum-projected five-quark operators correspond with HEFT states primarily composed of $\pi N$ basis states, with only small contributions from the bare basis states.

In \sref{sec:contamination}, we create novel HEFT simulations of the correlation functions for the two states observed in lattice QCD.
These correlators are constructed from the eigenvectors of the Hamiltonian and are used to construct two-particle scattering-state contamination functions.
These provide insight into the degree of scattering-state contamination in lattice QCD correlation functions for each lattice QCD energy reported.
Contamination function analysis was also performed for the 2 fm lattice QCD results, where it was found that avoided level crossings induce large scattering state contaminations.

By comparing the Hamiltonian from HEFT with both experimental scattering data, and lattice QCD data at $L\sim 2$, 3, and 4 fm, it is clear that we are able to interpret both the $N^{*}(1535)$ and $N^{*}(1650)$ resonances as three-quark cores dressed by $\pi N$, $\eta N$, and $K\Lambda$ scattering-state contributions.
In addition, by constructing HEFT simulations of the two-particle scattering-state contamination functions at each lattice QCD mass, it becomes clear that two-particle interpolators in lattice QCD are required to gain control over the essential features of the spectrum, particularly as one approaches the physical point.
Future work may be able to apply this multiple bare state formalism to other scattering channels such as the positive-parity nucleon and $\Delta$ systems.
%
%

\section*{Acknowledgements}
This research was supported by the Australian Government Research Training Program Scholarship, and with supercomputing resources provided by the Phoenix HPC service at the University of Adelaide.
This research was undertaken with the assistance of resources from the National Computational
Infrastructure (NCI), provided through the National Computational Merit Allocation Scheme, and
supported by the Australian Government through Grant No.~LE190100021 and the University of Adelaide Partner Share.
This research was supported by the Australian Research Council through ARC Discovery Project Grants Nos. DP190102215 and DP210103706 (D.B.L.).
J.-J. Wu was support by the National Natural Science Foundation of China under Grant Nos. 12175239 and 12221005, and by the National Key R\&D Program of China under Contract No. 2020YFA0406400.
Z.-W. Liu was supported by the National Natural Science Foundation of China under Grant Nos. 12175091, 11965016, 12047501, and 12247101, and the 111 Project under Grant No. B20063.

\newpage
\bibliography{oddParityRefs}
\bibliographystyle{unsrt}
\end{document}